\documentclass[apj]{emulateapj}
\usepackage{lscape}
\usepackage{apjfonts}
\usepackage{graphicx}

\slugcomment{Astrophys.J. 671 (2007) 1640}
\shorttitle{Galactic Cepheid {\it P-L} Relations}
\shortauthors{An et~al.}

\begin{document}

\title{The Distances to Open Clusters from Main-Sequence Fitting. IV.\\
Galactic Cepheids, the LMC, and the Local Distance Scale}

\author{Deokkeun An, Donald M.\ Terndrup, and Marc H.\ Pinsonneault}

\affil{Department of Astronomy, Ohio State University,
140 West 18th Avenue, Columbus, OH 43210;\\
deokkeun,terndrup,pinsono@astronomy.ohio-state.edu}

\begin{abstract}
We derive the basic properties of seven Galactic open clusters
containing Cepheids and construct their period-luminosity ({\it P-L})
relations.  For our cluster main-sequence fitting we extend previous
Hyades-based empirical color-temperature corrections to hotter stars
using the Pleiades as a template.  We use $BVI_CJHK_s$ data to test
the reddening law and include metallicity effects to perform a more
comprehensive study for our clusters than prior efforts.  The ratio of
total to selective extinction $R_V$ that we derive is consistent with
expectations.  Assuming the LMC {\it P-L} slopes, we find
$\langle M_V\rangle=-3.93\pm0.07\ {\rm (statistical)}\pm0.14\ {\rm
(systematic)}$ for 10-day period Cepheids, which is generally fainter
than those in previous studies.  Our results are consistent with
recent {\it HST} and {\it Hipparcos} parallax studies when using the
Wesenheit magnitudes $W(VI)$.  Uncertainties in reddening and
metallicity are the major remaining sources of error in the $V$-band
{\it P-L} relation, but a higher precision could be obtained with deeper
optical and near-infrared cluster photometry.  We derive
distances to NGC~4258, the LMC, and M33 of $(m - M)_0=29.28\pm0.10$,
$18.34\pm0.06$, and $24.55\pm0.28$, respectively, with an additional
systematic error of $0.16$~mag in the {\it P-L} relations.  The distance
to NGC~4258 is in good agreement with the geometric distance derived
from water masers $[\Delta(m - M)_0=0.01\pm0.24]$, our value for M33 is
less consistent with the distance from an eclipsing binary
[$\Delta(m - M)_0=0.37\pm0.34$], and our LMC distance is moderately shorter
than the adopted distance in the {\it HST} Key Project, which formally
implies an increase in the Hubble constant of $7\%\pm8\%$.
\end{abstract}


\section{Introduction}

The Cepheid period-luminosity ({\it P-L}) relation has played a key
role in the determination of distances within the Local Group and to
nearby galaxies.  By extension, it is also crucial for the calibration
of secondary distance indicators used to determine the Hubble constant:
see, for example, the {\it Hubble Space Telescope (HST)} Key Project on the
extragalactic distance scale \citep[][hereafter {\it HST} Key Project]
{freedman01} and the {\it HST} program for the luminosity calibration
of Type Ia supernovae (SNe~Ia) by means of Cepheids \citep{sandage06b}.
The cosmic distance scale was usually established by defining
the {\it P-L} relations for Cepheids in the Large Magellanic Cloud (LMC)
because of its numerous Cepheids, many of which have been discovered
as a result of microlensing campaigns \citep[e.g.,][]{udalski99a}.
Despite decades of effort, however, there have been persistent
differences in the inferred LMC distance from different methods
\citep[see][]{benedict02a}, including ones involving the same basic
calibrators such as RR Lyrae or Cepheid variables.

The Galactic {\it P-L} relationship, on the other hand, was traditionally
established using open clusters and associations containing Cepheids
\citep[see][]{feast87,feast03,sandage06a}.  There is also a steadily
growing body of parallax work
from the {\it Hipparcos} mission \citep{feast97,madore98,lanoix99,
groenewegen00,vanleeuwen07} and the {\it HST} \citep{benedict02b,benedict07},
as well as distances derived from the Baade-Becker-Wesselink moving
atmospheres method \citep{gieren97,gieren98,gieren05} and interferometric
angular diameter measurements \citep{lane00,lane02,kervella04a,kervella04b}.
In the current paper we focus on the Cepheid distance scale as inferred
from Galactic open clusters and its applications for the extragalactic
distance scale.

There are strengths and weaknesses in all of the methods used to establish
the Galactic Cepheid distance scale.  The absolute calibration of the
{\it P-L} relation requires not only accurate distance measurements but
also appropriate accounting for the effects of interstellar extinction and
reddening because Galactic Cepheids are heavily obscured with an average
$E(B - V)$ of order 0.5~mag \citep{fernie90}.  Reddening can be inferred
more precisely for clusters, while field Cepheids are more
numerous.  This difficulty can be partially overcome by the usage of
the ``reddening-free'' or Wesenheit index \citep{freedman91}, but there are
embedded assumptions about the extinction law even in such a system.

We believe that the time is right for a systematic reappraisal of
the Cepheid distance scale as inferred from open clusters.  Recent parallax
work employing the Wesenheit index \citep{benedict07,vanleeuwen07}
has yielded consistent claims of a somewhat smaller LMC distance modulus
($\approx18.40$~mag) than the ones adopted by the {\it HST} Key Project
($18.50$~mag) or \citeauthor{sandage06b} program ($18.54$~mag).  In addition,
there are now independent geometric tests of distances to other galaxies,
which have large numbers of Cepheids.  These include massive eclipsing
binaries in systems such as M31 \citep{ribas05} and M33 \citep{bonanos06}
and astrophysical water masers in NGC 4258 \citep{herrnstein99}.  As
described below, the availability of both more complete data in various
photometric passbands and better theoretical templates permits a more
comprehensive look at the open cluster Cepheid distance scale, which can
in turn be compared
with the alternate methods described above.  We can also use the open
clusters with Cepheids to test our ability to derive precise distances
and reddenings to heavily obscured and poorly studied systems.  We
believe that the net result is a more secure and robust determination
of the extragalactic distance scale.

We have been engaged in a long-term effort to create stellar evolution
models with the latest input physics and to generate isochrones that
are calibrated against photometry in local star clusters with accurate
distances.  After verifying that the models are in agreement with
the physical parameters of binaries and single stars in the Hyades
\citep[][hereafter Papers I and II, respectively]{pinsono03,pinsono04},
we developed a procedure for empirically correcting the color-effective
temperature ($T_{\rm eff}$)
relations to match photometry in the Hyades (Paper II).  Isochrones
generated in this way accurately reproduce the shapes of the main-sequence
(MS) in several colors, allowing the determination of distances to nearby
clusters to an accuracy of $\sim0.04$~mag in distance modulus
(or $2\%$ in distance) \citep[][hereafter Paper III]{an07}.

We extend the Hyades-based empirical corrections on theoretical
isochrones using the Pleiades and apply our techniques to a set of
Cepheid-bearing open clusters that have good multi-color photometry.
Selection of cluster sample and compilation of cluster photometry
are presented in \S~2.  Procedures on the isochrone calibration and
MS-fitting method are described in \S~3.  New cluster distances and
extinctions are presented in \S~4.  In \S~5 we construct multi-wavelength
Galactic {\it P-L} relations and discuss various systematic errors.
In \S~6 we estimate distances and reddenings for the maser galaxy
NGC~4258, the LMC, and M33.  We summarize our results in \S~7 and discuss
the importance of extinction corrections in the Cepheid distance scale.

\section{Cluster and Cepheid Data}\label{sec:data}

\subsection{Cluster Selection, Metallicity, and Age}\label{sec:mehage}

\begin{deluxetable*}{lllcccccccc}
\tabletypesize{\scriptsize}
\tablewidth{6in}
\tablecaption{List of Cepheid Clusters and Cepheid Properties\label{tab:list}}
\tablehead{
  \colhead{Cluster}  &
  \colhead{Other Name\tablenotemark{a}}  &
  \colhead{Cepheid} &
  \colhead{[M/H]\tablenotemark{b}} &
  \colhead{$\log{P}$} &
  \colhead{$\langle B \rangle$} &
  \colhead{$\langle V \rangle$} &
  \colhead{$\langle I_C \rangle$} &
  \colhead{$\langle J \rangle$} &
  \colhead{$\langle H \rangle$} &
  \colhead{$\langle K_s \rangle$}
}
\startdata
NGC~129     & C0027$+$599 & DL~Cas   & $+0.08\pm0.03$ & 0.903 & 10.119 &  8.969 & 7.655 & \nodata & \nodata & \nodata \nl
VDB~1\tablenotemark{c} & C0634$+$031 & CV~Mon & $+0.07\pm0.08$ & 0.731 & 11.607 & 10.304 & 8.646 & 7.310   & 6.783   & 6.475 \nl
NGC~5662    & C1431$-$563 & V~Cen    & $+0.00\pm0.03$ & 0.740 &  7.694 &  6.820 & 5.805 & 5.006   & 4.623   & 4.462 \nl
Lyng{\aa}~6 & C1601$-$517 & TW~Nor   & $+0.12\pm0.05$ & 1.033 & 13.672 & 11.667 & 9.287 & 7.403   & 6.686   & 6.225 \nl
NGC~6067    & C1609$-$540 & V340~Nor & $-0.14\pm0.05$ & 1.053 &  9.526 &  8.370 & 7.168 & 6.192   & 5.724   & 5.515 \nl
NGC~6087    & C1614$-$577 & S~Nor    & $+0.05\pm0.03$ & 0.989 &  7.373 &  6.429 & 5.422 & 4.661   & 4.269   & 4.115 \nl
M25         & C1828$-$192 & U~Sgr    & $+0.10\pm0.06$ & 0.829 &  7.792 &  6.695 & 5.448 & 4.511   & 4.085   & 3.892 \nl
NGC~7790    & C2355$+$609 & CEa~Cas  & \nodata & 0.711 & 12.070 & 10.920 & 9.470\tablenotemark{d} & \nodata & \nodata & \nodata \nl
NGC~7790    & C2355$+$609 & CEb~Cas  & \nodata & 0.651 & 12.220 & 11.050 & 9.690\tablenotemark{d} & \nodata & \nodata & \nodata \nl
NGC~7790    & C2355$+$609 & CF~Cas   & $-0.11\pm0.05$ & 0.688 & 12.335 & 11.136 & 9.754 & \nodata & \nodata & \nodata \nl
\enddata
\tablenotetext{a}{``C'' cluster designation from \citet{lynga87}.}
\tablenotetext{b}{An effective metallicity estimated from [Fe/H] and [$\alpha$/Fe] in 
\citet{fry97} using the \citet{kim02} procedure.}
\tablenotetext{c}{Also named CV~Mon cluster or anonymous VDB cluster \citep{vdbergh57}.}
\tablenotetext{d}{Taken from \citet{tammann03}.}
\end{deluxetable*}

About 30 open clusters and associations in the Galaxy are known to
harbor Cepheid variables \citep{feast87,feast99}, but not all of these
are useful for MS fitting because many are extremely sparse or poorly
studied.  We first excluded systems with overtone or double-mode
Cepheids and then examined the available photoelectric and/or CCD
photometry for each cluster.  From visual inspection of color-magnitude
diagrams (CMDs) we narrowed down the list to 10 promising candidates
for MS fitting.  The populous twin clusters $h\&\chi$~Per were
excluded from the current analysis because their young age introduces
significant uncertainties in
the isochrone calibration, and the membership of UY~Per in the clusters is
also doubtful \citep[see][and references therein]{walker87}.  The eight
remaining clusters are listed in Table~\ref{tab:list} with their
10 Cepheid variables.  We denoted anonymous van den Bergh
\citep[C0634+031;][]{vdbergh57} as VDB~1 for brevity.  Most of
them are close enough that Two Micron All Sky Survey (2MASS) \citep{skrutskie06}
photometry reaches well down the MS.

It is likely that all Cepheids in Table~\ref{tab:list} are
members of their associated clusters.  For the three Cepheids in
NGC~7790 there exist no radial velocity or proper-motion measurements,
although all of them lie near the optical center of the cluster
\citep{sandage58}.  V~Cen is located about $25\arcmin$ away from
the center of NGC~5662, but \citet{baumgardt00} derived a high
membership probability from {\it Hipparcos} proper motions.
The rest of the Cepheids are generally considered to be
members of their clusters since they are located within the optical
radius defined by bright early-type MS stars and red giants.  In
many cases, membership is also supported by radial velocities
\citep[][and references therein]{mermilliod87} and proper
motions \citep[][and references therein]{baumgardt00}.  \citet{orsatti01}
suggested that TW~Nor may not be a member of Lyng{\aa}~6 from
polarization measurements, but we nevertheless included it here.

Previous studies have neglected the effects of cluster metallicity and
age on MS fitting \citep[e.g.,][]{feast87,hoyle03}.  However, the
luminosity of the MS at a fixed color (or $T_{\rm eff}$) is sensitive
to the metal abundance by $\Delta (m - M)_0 / \Delta {\rm [M/H]}
\sim 1$ (Paper~III).  In addition, the cluster age changes the mean
color of the upper MS by $\Delta (B - V) \sim 0.05$ between zero-age
MS (ZAMS) and $\sim100$~Myr isochrone, which is more appropriate for
Cepheid-bearing clusters (see below).  This, in turn, could affect
the reddening estimates from the upper MS by $\Delta E(B - V) \sim 0.05$
and an MS-fitting distance by $\Delta (m - M) \sim 0.1$.

We adopted Cepheid metal abundances from the high-resolution spectroscopy
of \citet{fry97} and assumed the same metallicities for the clusters.
In support of this assumption, \citeauthor{fry97} derived metallicities
for two dwarfs in M25 that are in agreement with the metallicity of U~Sgr.
In addition, \citeauthor{fry97} measured ${\rm \langle [Fe/H] \rangle} =
+0.01\pm0.02$ for two Pleiades dwarfs, which is in good agreement with our
adopted value for the Pleiades, ${\rm [Fe/H]} = +0.04\pm0.02$.

\citeauthor{fry97} also derived $\alpha$-element abundances (Si, Ca, and Ti)
for their sample.  For the stars in this paper, the average enhancement is
$\langle{\rm [\alpha/Fe]}\rangle = 0.14\pm0.02$.  We determined
an effective metallicity [M/H] from the measured [Fe/H] and [$\alpha$/Fe]
using the \citet{kim02} procedure.  These abundances are shown in
the fourth column of Table~\ref{tab:list}.  The mean metal abundance
of our sample is $\langle {\rm [M/H]}\rangle = +0.03\pm0.03$ and
the standard deviation is $0.08$~dex, indicating an intrinsic spread in
metallicity for our clusters.  On average, the rescaled [M/H] are larger
than [Fe/H] by $0.06$ dex.  We computed errors in [M/H] by propagation
of errors in [Fe/H] and [$\alpha$/Fe].

The metallicity scale for the Cepheids is probably robust to $\sim0.1$~dex.
As part of a larger study, \citet{yong06} reanalyzed spectra of 11 Cepheids
in \citeauthor{fry97} and derived lower [Fe/H] by $\approx0.1$~dex and lower
$\alpha$-element enhancement by $\approx0.05$~dex on average.  The mean difference
in the total metallicity is $\Delta \langle {\rm [M/H]} \rangle = -0.15
\pm0.03$, in the sense that the new analysis indicates lower values.
Independently, \citet{andrievsky02} measured abundances of 19 Cepheids in
common with \citeauthor{fry97}.  The average difference between these two
studies is $\approx0.1$~dex in [Fe/H], the \citeauthor{andrievsky02} values
being larger.  They also showed that the alpha abundance ratio ([Si, Ca, Ti/Fe])
is nearly solar, which results in a small difference in the total metal
abundance between \citeauthor{fry97} and \citeauthor{andrievsky02},
$\Delta \langle {\rm [M/H]} \rangle = -0.01\pm0.02$.

We adopted a single representative age (80 Myr) for all clusters, based
on comparing our isochrones (\S~\ref{sec:iso}) to the cluster CMDs
(\S~\ref{sec:phot}), and took its uncertainty as $\Delta \log{t {\rm
(Myr)}} = 0.2$ ($t = 80^{+50}_{-30}$~Myr).  The uncertainty represents
both a range of ages among the clusters and the accuracy of the fitting.
The average age in the Lyng{\aa} Open Clusters Catalog \citep{lynga87}
is 72~Myr for our sample clusters with a standard deviation of 38~Myr.
Previous age estimates based on isochrone fitting are generally within
our adopted age range.  \citet{mermilliod81} found that NGC~7790 and
NGC~6067 are about as old as the Pleiades (78~Myr) from models with
core overshooting.  \citet{meynet93} estimated $\approx70$~Myr for
NGC~5662 and NGC~6087 when the Pleiades is 100~Myr old from core-overshooting
models, while the age of NGC~6067 was estimated to be $\approx170$~Myr.
Cluster ages can also be inferred from the masses of Cepheids.  There are
several classical Cepheids in binary systems, and these have dynamical
masses of about $5 - 6 M_\odot$ \citep{bohm98,evans90,evans97,evans98,evans06}
over the period range similar to those of our sample [$0.65 \leq \log P
({\rm days}) \leq 1.05$].  For our models, stars of this mass range have
an MS lifetime of about $50 - 80$~Myr.

\subsection{Photometry}\label{sec:phot}

For the photometry of the Cepheids, we used intensity-mean apparent
magnitudes in $BVI_C$ from \citet{berdnikov00}, which are on
the Cape (Cousins) system as realized by \citet{landolt83,landolt92}.
For CEa Cas and CEb Cas, the $I_C$-band measurements were taken from \citet{tammann03}.
We adopted $JHK$ photometry from \citet{laney92} on the \citet{carter90}
system and transformed it to the LCO system using the \citet{persson04}
transformation equations.  We assumed $0.025$~mag zero-point error in
the Cepheid photometry and treated this error independently from the cluster
photometry (see below).  Table~\ref{tab:list} lists the Cepheid photometry.

\begin{deluxetable*}{lcccc}
\tabletypesize{\scriptsize}
\tablewidth{4in}
\tablecaption{Photometry References and Zero-Point Differences\label{tab:phot}}
\tablehead{
  \colhead{Reference} &
  \colhead{$\Delta V$} &
  \colhead{$N_{\rm comp}$} &
  \colhead{$\Delta (B - V)$} &
  \colhead{$N_{\rm comp}$}
}
\startdata
\multicolumn{5}{c}{NGC~129} \nl
\hline
\citet{arp59}     & $+0.010\pm0.006$ &26       & $+0.004\pm0.011$ &26       \nl
\citet{hoag61}    & $+0.000\pm0.011$ &18       & $+0.059\pm0.008$ &18       \nl
\citet{turner92}  & $+0.015\pm0.006$ &20       & $+0.048\pm0.008$ &22       \nl
\citet{phelps94}  & standard         & \nodata & standard         & \nodata \nl
\hline
\multicolumn{5}{c}{VDB~1} \nl
\hline
\citet{arp60}    & standard & \nodata & standard & \nodata \nl
\citet{turner98} & tied to \citeauthor{arp60} & \nodata & tied to \citeauthor{arp60} & \nodata \nl
\hline
\multicolumn{5}{c}{NGC~5662} \nl
\hline
\citet{moffat73}  & $+0.009\pm0.005$ &13       & $+0.002\pm0.004$ &28       \nl
\citet{haug78}    & standard         & \nodata & $+0.027\pm0.004$ &20       \nl
\citet{claria91}  & $+0.051\pm0.004$ &19       & standard         & \nodata \nl
\citet{sagar97}   & $-0.021\pm0.015$ &10       & $+0.004\pm0.006$ &14       \nl
\hline
\multicolumn{5}{c}{Lyng{\aa}~6} \nl
\hline
\citet{madore75}  & $+0.009\pm0.015$ &8        & $-0.002\pm0.010$ &9        \nl
\citet{moffat75}  & $-0.097\pm0.018$ &9        & $-0.030\pm0.025$ &10       \nl
\citet{vdbergh76} & $-0.010\pm0.008$ &12       & $+0.024\pm0.014$ &12       \nl
\citet{walker85a} & standard         & \nodata & standard         & \nodata \nl
\hline
\multicolumn{5}{c}{NGC~6067} \nl
\hline
\citet{thackeray62} & $-0.019\pm0.013$ & 15      & $+0.012\pm0.008$    &14       \nl
\citet{walker85b}   & standard         & \nodata & standard            & \nodata \nl
\citet{piatti98}    & $-0.021\pm0.003$\tablenotemark{a} &123 & \nodata & \nodata \nl
\hline
\multicolumn{5}{c}{NGC~6087} \nl
\hline
\citet{fernie61}  & $-0.057\pm0.020$ &9        & $+0.041\pm0.006$ &10       \nl
\citet{breger66}  & $+0.005\pm0.008$ &18       & $-0.019\pm0.011$ &17       \nl
\citet{turner86}  & standard         & \nodata & standard         & \nodata \nl
\citet{sagar97}   & $+0.080\pm0.070$ &2        & $+0.025\pm0.035$ &2        \nl
\hline
\multicolumn{5}{c}{M25} \nl
\hline
\citet{johnson60} & $-0.018\pm0.003$ &74       & standard         & \nodata \nl
\citet{sandage60} & $+0.018\pm0.005$ &51       & $+0.005\pm0.004$ &68       \nl
\citet{wampler61} & standard         & \nodata & $+0.021\pm0.004$ &51       \nl
\hline
\multicolumn{5}{c}{NGC~7790} \nl
\hline
\citet{sandage58} & $-0.010\pm0.013$ & 26      & $-0.018\pm0.008$ & 26 \nl
\citet{romeo89}   & $-0.001\pm0.003$ &186      & $+0.013\pm0.003$ &184 \nl
\citet{phelps94}  & $-0.030\pm0.001$ &360      & $-0.005\pm0.002$ &337 \nl
\citet{lee99}     & $+0.027\pm0.003$ &164      & $-0.032\pm0.003$ &157 \nl
\citet{gupta00}   & $+0.008\pm0.001$ &192      & $+0.009\pm0.002$ &187 \nl
\citet{stetson00} & standard         & \nodata & standard   & \nodata \nl
\enddata
\tablecomments{The mean color and magnitude differences are computed after
$3\sigma$ rejection.  Uncertainties are the standard error of the mean.
The differences are in the sense of individual values minus those adopted
as the local standard.  The photometry in \citet{turner98} are directly tied
to the photometry of \citet{arp60}.}
\tablenotetext{a}{Comparison for stars with $V \leq 15$.}
\end{deluxetable*}

For the cluster photometry, we compiled photoelectric and CCD $BVI_C$ data
on the Johnson-Cousins system from the literature, as well as from WEBDA
\citep{mermilliod03}.\footnote{See http://obswww.unige.ch/webda/webda.html.}
The references for the photometry are shown in Table~\ref{tab:phot},
where we used WEBDA's cross-identification of optical sources among
different references except for a few cases where we found missing
entries and misidentifications.  For each cluster we picked one or two
references to define a local standard in $V$ or $B - V$ by weighing factors
such as the number of observations, number of stars with photometry,
magnitude range in $V$, the photometric calibration procedure, and
whether the photometry was generally in agreement with data in other
studies.  The remaining columns of Table~\ref{tab:phot} show the mean
differences in $V$ and $B - V$ with respect to the local standard, and
the number of stars in common.  The errors shown are standard
errors of the mean difference.  For VDB~1 we combined the photoelectric
photometry by \citet{arp60} and CCD photometry by \citet[][their Table~2]
{turner98}, whose photometry was tied to the former study.

In most cases the differences between one study and another showed no
trends with magnitude.  In M25, however, the differences in $B - V$
between \citet{johnson60} and \citet{sandage60} were statistically
significant ($>2.5\sigma$):
\[
\Delta (B - V) ({\rm Sandage - Johnson}) = +0.009 (V - 10) - 0.010.\nonumber
\]
Similarly, in NGC~6087 there were significant trends with magnitude for
\citet{fernie61} and \citet{breger66} with respect to \citet{turner86}:
\begin{eqnarray}
\Delta V ({\rm Fernie - Turner}) = -0.033 (V - 10) - 0.052,\nonumber\\
\Delta V ({\rm Breger - Turner}) = -0.014 (V - 10) + 0.002.\nonumber
\end{eqnarray}
We applied these corrections to put them on
the same scale as the reference photometry.  For NGC~7790 we only used
the photometry by \citet{stetson00}, although we computed the differences
with respect to other studies and included these in Table~\ref{tab:phot}.
We applied zero-point corrections to individual values if the average
difference was significant at the $2.5\sigma$ level.  After shifting to
a common scale, magnitudes and colors in multiple studies were averaged together.
The weighted rms differences between the studies in Table~\ref{tab:phot}
are 0.026~mag in $V$ and 0.024~mag in $B - V$.  We therefore adopted
0.025~mag as the characteristic size of systematic errors in the photometry.

Useful $V - I_C$ photometry for MS fitting is limited only to four
clusters: VDB~1 \citep{turner98}, Lyng{\aa}~6 \citep{walker85a},
NGC~6067 \citep{piatti98}, and NGC~7790 \citep{romeo89,gupta00,stetson00}.
As with $B - V$ we adopted the \citet{stetson00} photometry for NGC~7790.
Compared to Stetson's data, the photometry in \citet{romeo89}
is bluer by $0.026\pm0.003$~mag, and the photometry in \citet{gupta00}
is redder by $0.023\pm0.001$.  Assuming these values as a characteristic
size of error, we adopted $0.025$~mag for the photometric zero-point
error in $V - I_C$.

We combined optical photometry with $JHK_s$ measurements from the All
Sky Data Release of the 2MASS Point Source
Catalog (PSC)\footnote{See http://www.ipac.caltech.edu/2mass/.}.
WEBDA provides celestial coordinates for only a small fraction of stars in
this study.  For the others we identified each source on the images
from the Digitized Sky Survey,\footnote{The Digitized Sky Surveys were
produced at the Space Telescope Science Institute under US Government
grant NAG W-2166.  The images of these surveys are based on photographic
data obtained using the Oschin Schmidt Telescope on Palomar Mountain and
the UK Schmidt Telescope. The plates were processed into the present
compressed digital form with the permission of these institutions.} or
we computed the celestial coordinates from the plate position information
in WEBDA.  The rms difference between the retrieved and the 2MASS
coordinates was typically $0.5\arcsec$.  For NGC~6067 we matched 2MASS
sources with optical photometry only for those stars with good positional
accuracy.  The validation of the 2MASS source matches was confirmed from the
resulting tight optical and near-infrared color-color relations.  Based
on PSC flag parameters, we ignored the infrared data if sources were
undetected, blended, or contaminated.  Calibration errors in $JHK_s$
were taken as the uncertainty specified in the explanatory supplement
to the 2MASS All Sky Data Release: $0.011$ mag in $J$, $0.007$ mag in
$H$, and $0.007$ mag in $K_s$.\footnote{See
http://www.ipac.caltech.edu/2mass/releases/allsky/doc/explsup.html.}

Most studies do not report individual errors in optical magnitude or color,
so we computed the median of $V$ and $B - V$ differences from various studies
(after shifting to a common scale using the $>2.5\sigma$ criterion) and
assigned this value to represent the random photometric errors for all stars
in each cluster.  For $V - I_C$ we assigned 0.02~mag as the random error.
Errors for individual stars in  \citet{stetson00} were adopted without change.
The $V - K_s$ errors were computed as the quadrature sum of $V$ errors and
the catalog's ``total'' photometric uncertainties in $K_s$.

\begin{figure*}
\epsscale{1.0}
\plotone{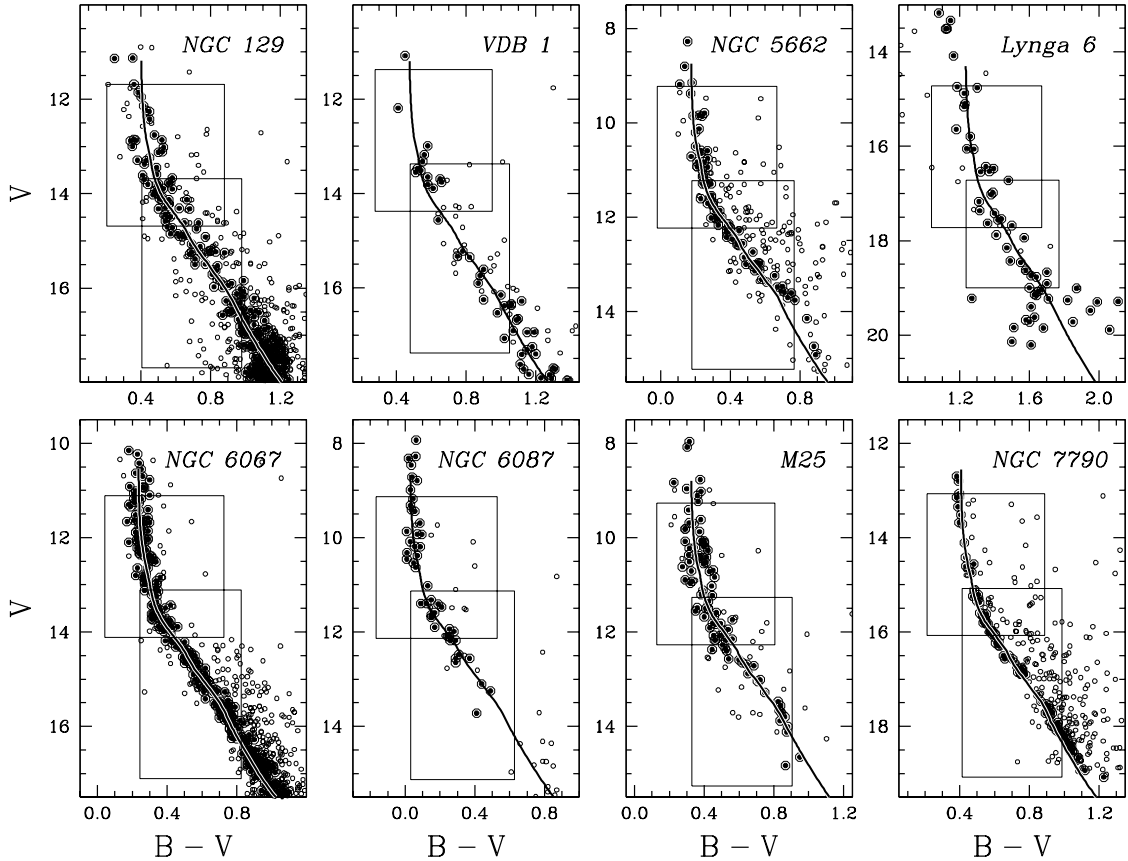}
\caption{Cluster CMDs in $B - V$.  The bull's-eyes are stars remaining
after the photometric filtering, while open circles are stars rejected.  The
solid lines are empirically calibrated isochrones at the best-fitting
distance (Tables~\ref{tab:dist}), reddening (Tables~\ref{tab:ebv}), and
$R_{V,0}$ (Tables~\ref{tab:cluster}).  The boxes on the upper MS and
lower MS are regions where the color excess and distance were determined,
respectively.\label{fig:bvcmd}}
\end{figure*}

\begin{figure*}
\epsscale{1.0}
\plotone{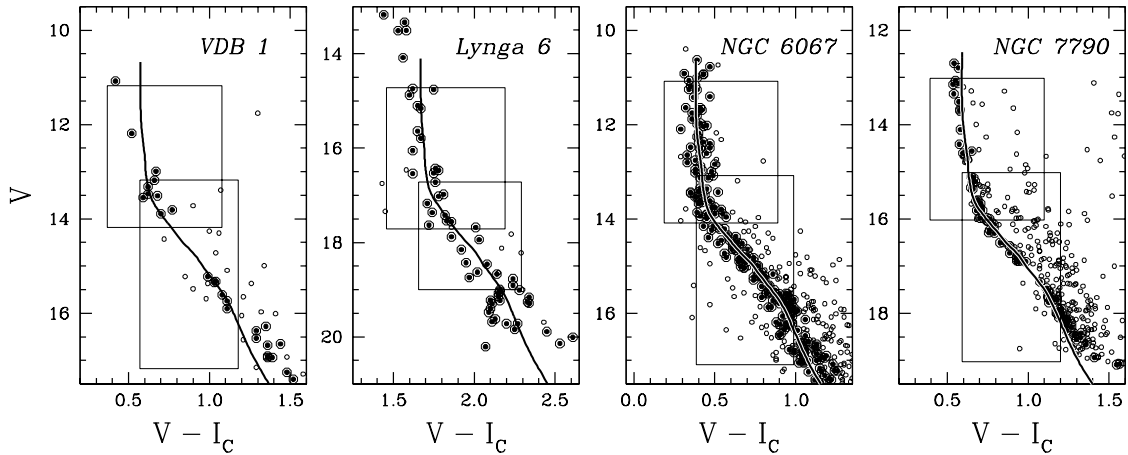}
\caption{Same as Fig.~\ref{fig:bvcmd}, but CMDs in $V - I_C$.
\label{fig:vicmd}}
\end{figure*}

\begin{figure*}
\epsscale{1.0}
\plotone{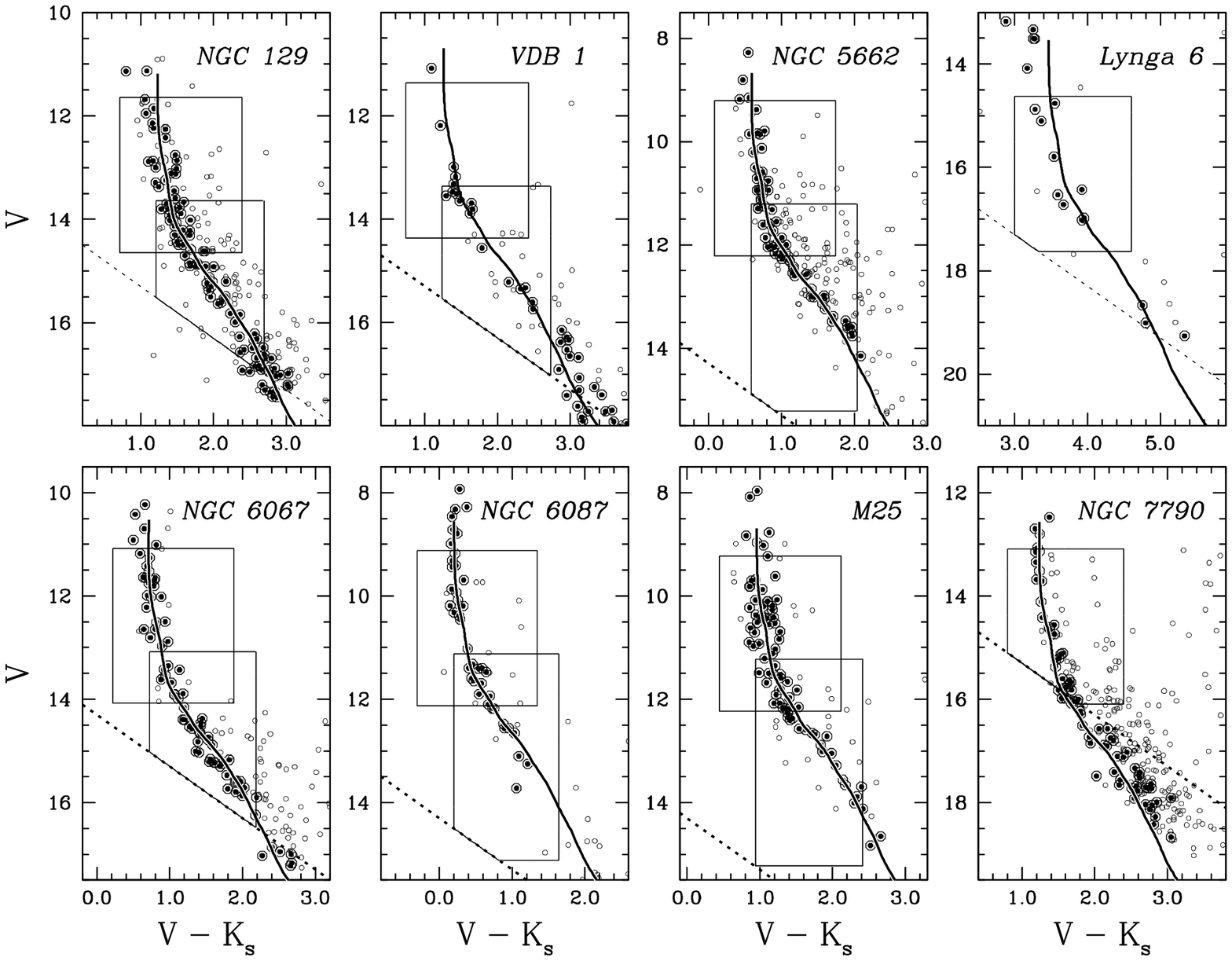}
\caption{Same as Fig.~\ref{fig:bvcmd}, but CMDs in $V - K_s$.  The
dotted line represents the 2MASS completeness limit.\label{fig:vkcmd}}
\end{figure*}

\begin{figure*}
\epsscale{1.0}
\plotone{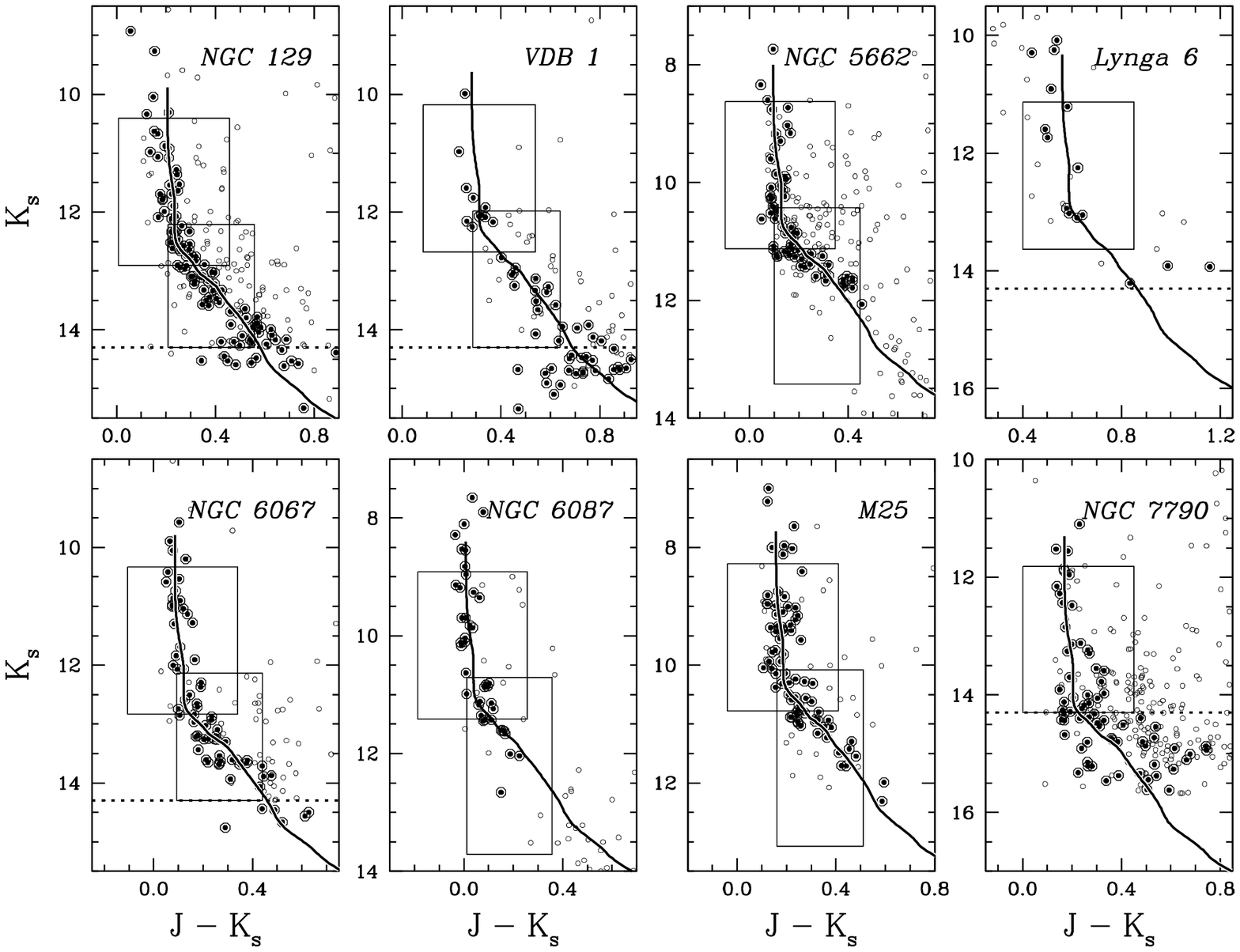}
\caption{Same as Fig.~\ref{fig:bvcmd}, but CMDs in $J - K_s$.  The
dotted line represents the 2MASS completeness limit.\label{fig:jkcmd}}
\end{figure*}

The cluster CMDs in various color combinations are shown in
Figures~\ref{fig:bvcmd} through \ref{fig:jkcmd}.  Stars rejected as being
far from the MS are shown as open circles, while those used in the MS fitting
are shown as bull's-eyes.  The curved line in each panel is the best-fitting
isochrone on each CMD.  Details of the MS fitting and outlier rejection
are given in the next section.  Throughout this paper we denote CMDs by
their color and luminosity indices as $(B - V, V)$, $(V - I_C, V)$,
$(V - K_s, V)$, $(J - K_s, K_s)$, and $(H - K_s, K_s)$.

\section{Main-Sequence Fitting}

\subsection{Extension of Isochrone Calibration}\label{sec:iso}

In our first two papers of this series (Paper~I and Paper~II),
we assessed the accuracy
of distances from MS fitting and examined systematic errors in the
transformation of theoretical to observational quantities.  In Paper~I
we demonstrated that stellar models from the Yale Rotating Evolutionary
Code \citep{sills00} are in  agreement with the masses and luminosities
of the well-studied Hyades eclipsing binary vB~22 \citep{torres02}.
These models also satisfy stringent tests from helioseismology and
predict solar neutrino fluxes in line with observations \citep{basu00,
bahcall01,bahcall04}.  In Paper~II we showed that the models provide
a good match to the spectroscopically determined temperatures
\citep{paulson03} of individual Hyades members with good parallaxes
\citep{debruijne01}.  However, we found that any of the widely-used
color-$T_{\rm eff}$ relations \citep[e.g.,][hereafter LCB]{alonso95,alonso96,
lejeune97,lejeune98} fail to reproduce the observed shapes of the MS
in the Hyades, having differences in broadband colors as large as
$0.1$~mag.  The existence of these systematic errors in the colors, in
the presence of agreement between the spectroscopic and theoretical
luminosity-$T_{\rm eff}$ scales, led us to argue that the problems lie
with the adopted color-$T_{\rm eff}$ relations instead of errors in
the theoretical $T_{\rm eff}$ scale.  We then adopted the LCB
color-$T_{\rm eff}$ relations and computed empirical corrections to the
LCB relations to match photometry in the Hyades.  In Paper~III we showed
that isochrones with the empirical color-$T_{\rm eff}$ corrections
accurately match the MS shapes of other nearby clusters in three color
indices ($B - V$, $V - I_C$, and $V - K_s$), yielding distance estimates
with errors as small as 0.04~mag in distance modulus (or 2\% in distance).

However, the Hyades-based calibration is of limited use for this study.
The calibration is reliable only at $M_V \ga 3$ because the number of
stars in the upper MS of the Hyades is small in this relatively old open cluster
\citep[550~Myr from models excluding convective overshoot;][]{perryman98}.
On the other hand, photometry of our sample clusters is quite sparse or
absent in this lower MS part.  It is therefore desirable to define
an extension of the empirical color-$T_{\rm eff}$ corrections for hotter
and brighter stars.  A good choice for a calibrating cluster is the Pleiades
for several reasons.  First of all, its geometric distance is now known to
a precision comparable to that of the Hyades, $(m - M)_0 = 5.63\pm0.02$
(see Paper~III and references therein for a detailed discussion of
the cluster parameters).  Second, it has low reddening, $E(B - V) = 0.032\pm0.003$.
Third, a precise metal abundance from high-resolution spectra is available,
${\rm [Fe/H]} = +0.04\pm0.02$.
Furthermore, Cepheid clusters tend to be of ages comparable to that of
the Pleiades, making it a good template for distance and reddening
estimates.

\begin{deluxetable*}{lrrrrrrr}
\tabletypesize{\scriptsize}
\tablewidth{6in}
\tablecaption{Merged Photometry in the Pleiades\label{tab:plphot}}
\tablehead{
  \colhead{ID\tablenotemark{a}} &
  \colhead{$M_V$} &
  \colhead{$(B - V)_0$} &
  \colhead{$(V - I_C)_0$} &
  \colhead{$M_{K_s}$} &
  \colhead{$(V - K_s)_0$} &
  \colhead{$(J - K_s)_0$} &
  \colhead{$(H - K_s)_0$}
}
\startdata
Hz~II~153  & $ 1.782$ & $ 0.119$ & $ 0.134$ & $ 1.525\pm0.021$ & $ 0.257\pm0.025$ & $ 0.021\pm0.025$ & $ 0.067\pm0.061$\nl
Hz~II~447  & $-0.278$ & $-0.076$ & $-0.029$ & $-0.147\pm0.008$ & $-0.131\pm0.016$ & $-0.067\pm0.022$ & $-0.026\pm0.036$\nl
Hz~II~531  & $ 2.845$ & $ 0.306$ & $ 0.375$ & $ 2.095\pm0.019$ & $ 0.750\pm0.024$ & $ 0.136\pm0.024$ & $ 0.023\pm0.035$\nl
Hz~II~817  & $ 0.028$ & $-0.066$ & $-0.013$ & $ 0.134\pm0.011$ & $-0.106\pm0.018$ & $-0.041\pm0.025$ & $ 0.043\pm0.046$\nl
Hz~II~859  & $ 0.692$ & $-0.052$ & $-0.021$ & $ 0.790\pm0.019$ & $-0.098\pm0.024$ & $-0.052\pm0.027$ & $ 0.045\pm0.043$\nl
Hz~II~1028 & $ 1.635$ & $ 0.069$ & $ 0.099$ & $ 1.482\pm0.019$ & $ 0.153\pm0.024$ & $-0.024\pm0.020$ & $-0.020\pm0.021$\nl
Hz~II~1139 & $ 3.643$ & $ 0.450$ & $ 0.490$ & $ 2.600\pm0.019$ & $ 1.043\pm0.024$ & $ 0.212\pm0.027$ & $ 0.031\pm0.022$\nl
Hz~II~1234 & $ 1.087$ & $-0.012$ & $ 0.033$ & $ 1.049\pm0.019$ & $ 0.038\pm0.024$ & $-0.011\pm0.030$ & $ 0.035\pm0.060$\nl
Hz~II~1362 & $ 2.528$ & $ 0.230$ & $ 0.274$ & $ 1.996\pm0.013$ & $ 0.532\pm0.019$ & $ 0.110\pm0.026$ & $-0.018\pm0.037$\nl
Hz~II~1375 & $ 0.572$ & $-0.011$ & $-0.006$ & $ 0.614\pm0.016$ & $-0.042\pm0.021$ & $-0.047\pm0.017$ & $-0.015\pm0.029$\nl
Hz~II~1407 & $ 2.393$ & $ 0.220$ & $ 0.259$ & $ 1.877\pm0.016$ & $ 0.516\pm0.021$ & $ 0.080\pm0.020$ & $ 0.017\pm0.019$\nl
Hz~II~1823 & $-0.286$ & $-0.102$ & $-0.043$ & $-0.083\pm0.010$ & $-0.203\pm0.017$ & $-0.075\pm0.025$ & $ 0.024\pm0.027$\nl
Hz~II~1993 & $ 2.645$ & $ 0.242$ & $ 0.313$ & $ 1.989\pm0.010$ & $ 0.656\pm0.017$ & $ 0.109\pm0.019$ & $ 0.063\pm0.017$\nl
Hz~II~2195 & $ 2.387$ & $ 0.190$ & $ 0.220$ & $ 1.933\pm0.011$ & $ 0.454\pm0.018$ & $ 0.079\pm0.019$ & $ 0.017\pm0.029$\nl
Hz~II~2345 & $ 3.375$ & $ 0.407$ & $ 0.466$ & $ 2.393\pm0.017$ & $ 0.982\pm0.022$ & $ 0.204\pm0.021$ & $ 0.055\pm0.025$\nl
Hz~II~2717 & $ 1.682$ & $ 0.101$& \nodata & \nodata & \nodata & \nodata & \nodata \nl
Hz~II~3031 & $ 3.105$ & $ 0.352$ & $ 0.433$ & $ 2.234\pm0.017$ & $ 0.871\pm0.022$ & $ 0.164\pm0.023$ & $ 0.047\pm0.019$\nl
Hz~II~3302 & $ 3.153$ & $ 0.330$ & $ 0.381$ & $ 2.314\pm0.017$ & $ 0.839\pm0.022$ & $ 0.135\pm0.024$ & $ 0.057\pm0.017$\nl
Hz~II~3308 & $ 3.068$ & $ 0.320$ & $ 0.358$ & $ 2.268\pm0.011$ & $ 0.800\pm0.018$ & $ 0.157\pm0.031$ & $ 0.019\pm0.046$\nl
Hz~II~3309 & $ 3.438$ & $ 0.410$ & $ 0.466$ & $ 2.451\pm0.000$ & $ 0.987\pm0.014$ & $ 0.167\pm0.013$ & $ 0.078\pm0.013$\nl
Hz~II~3310 & $ 3.398$ & $ 0.430$ & $ 0.498$ & $ 2.359\pm0.025$ & $ 1.039\pm0.029$ & $ 0.216\pm0.027$ & $ 0.025\pm0.030$\nl
Hz~II~3319 & $ 2.983$ & $ 0.330$ & $ 0.381$ & $ 2.168\pm0.008$ & $ 0.815\pm0.016$ & $ 0.137\pm0.017$ & $ 0.078\pm0.011$\nl
Hz~II~3332 & $ 3.338$ & $ 0.400$ & $ 0.482$ & $ 2.343\pm0.021$ & $ 0.995\pm0.025$ & $ 0.157\pm0.027$ & $ 0.014\pm0.030$\nl
Hz~II~3334 & $ 2.605$ & $ 0.232$ & $ 0.306$ & $ 2.035\pm0.023$ & $ 0.570\pm0.027$ & $ 0.072\pm0.027$ & $ 0.032\pm0.034$\nl
Hz~II~5016 & $ 2.017$ & $ 0.149$ & $ 0.181$ & $ 1.667\pm0.021$ & $ 0.350\pm0.025$ & $ 0.061\pm0.022$ & $ 0.011\pm0.025$\nl
Hz~II~5023 & $ 1.282$ & $-0.001$ & $ 0.010$ & $ 1.239\pm0.017$ & $ 0.043\pm0.022$ & $-0.046\pm0.019$ & $ 0.030\pm0.023$\nl
Hz~II~5025 & $ 0.337$ & $-0.041$ & $ 0.018$ & $ 0.333\pm0.016$ & $ 0.004\pm0.021$ & $-0.027\pm0.017$ & $ 0.069\pm0.061$\nl
TS~23    & $ 1.002$ & $-0.031$ & $-0.006$ & $ 1.029\pm0.016$ & $-0.027\pm0.021$ & $-0.049\pm0.024$ & $-0.003\pm0.017$\nl
TS~25    & $ 1.377$ & $ 0.009$ & $ 0.010$ & $ 1.389\pm0.013$ & $-0.012\pm0.019$ & $-0.071\pm0.021$ & $-0.013\pm0.037$\nl
TS~78    & $ 2.092$ & $ 0.159$ & $ 0.173$ & $ 1.724\pm0.025$ & $ 0.368\pm0.029$ & $ 0.051\pm0.031$ & $ 0.023\pm0.033$\nl
TS~84    & $ 2.272$ & $ 0.190$ & $ 0.228$ & $ 1.894\pm0.010$ & $ 0.378\pm0.017$ & $ 0.049\pm0.016$ & $ 0.044\pm0.052$\nl
TS~115   & $ 2.583$ & $ 0.230$ & $ 0.259$ & $ 2.063\pm0.019$ & $ 0.520\pm0.024$ & $ 0.080\pm0.021$ & $ 0.041\pm0.027$\nl
TS~137   & $ 2.002$ & $ 0.149$ & $ 0.181$ & $ 1.596\pm0.019$ & $ 0.406\pm0.024$ & $ 0.055\pm0.028$ & $ 0.022\pm0.033$\nl
TS~165   & $ 1.907$ & $ 0.129$ & $ 0.095$ & $ 1.641\pm0.010$ & $ 0.266\pm0.017$ & $ 0.007\pm0.013$ & $ 0.042\pm0.011$\nl
TS~194   & $ 1.462$ & $ 0.009$ & $ 0.010$ & $ 1.474\pm0.016$ & $-0.012\pm0.021$ & $-0.032\pm0.017$ & $-0.038\pm0.026$\nl
\enddata
\tablenotetext{a}{The star designation is that of \citet{hertzsprung47}, Hz~II; and
\citet{trumpler21}, TS.}
\end{deluxetable*}

\begin{figure*}
\epsscale{0.9}
\plotone{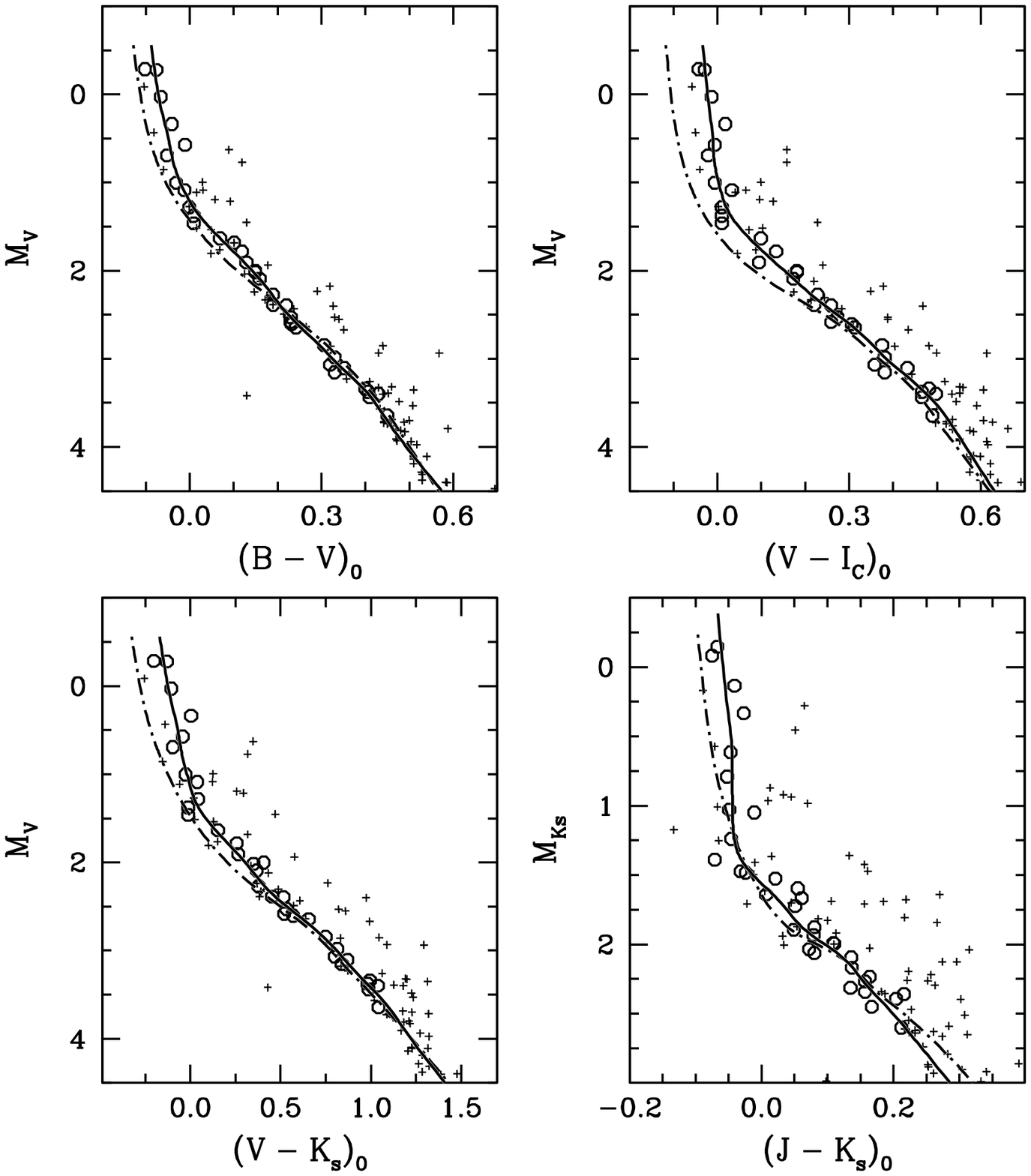}
\caption{Empirically calibrated isochrones for the Pleiades ({\it solid
lines}).  The dot-dashed lines show the isochrones without the color
corrections.  The open circles are stars retained for use in the
calibration process, and the small plus signs show stars rejected
(see text).\label{fig:plcmd}}
\end{figure*}

The sources of our Pleiades photometry are described in Paper~III.  We
began by removing a few known binaries and then rejected stars that are
more than 0.1 mag in color away from a 15 point median of the sample
sorted in $V$; this step was independent of the theoretical isochrones.
Table~\ref{tab:plphot} lists photometry of the hot Pleiades stars that
remained after this selection.  The CMDs of these stars are shown in
Figure~\ref{fig:plcmd}.  Small plus signs denote stars rejected as far
from the MS, while open circles show stars retained for use in the
following calibration process.

\begin{figure}
\epsscale{1.1}
\plotone{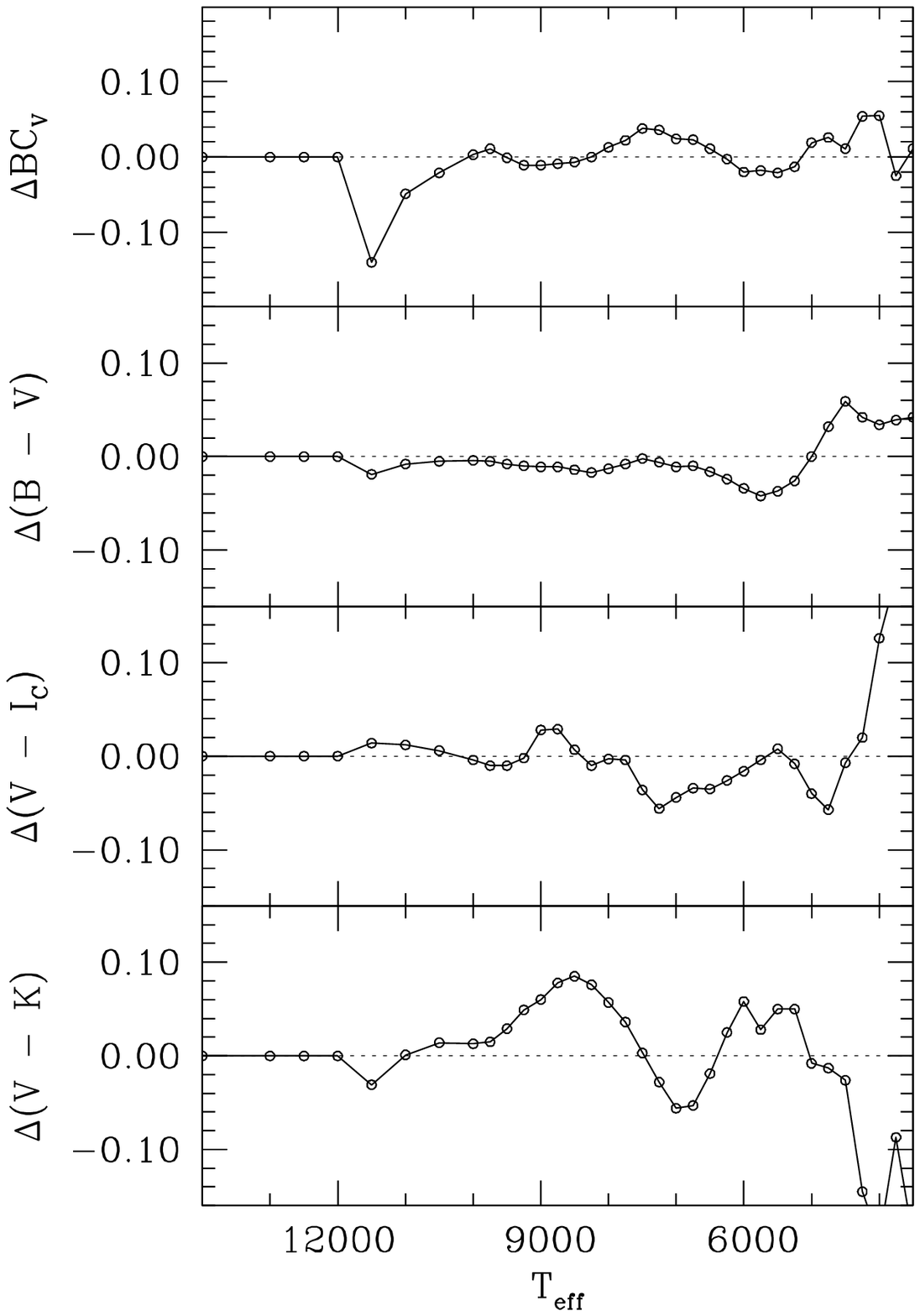}
\caption{Differences in the bolometric correction and colors between
``corrected'' and ``original'' LCB tables for MS stars ([M/H] = 0,
$\log{g}=4.5$).  Differences are in the sense of the former minus the
latter values.\label{fig:lj}}
\end{figure}

In the course of this work we became aware of some problematic features
in the LCB color tables.  They computed synthetic colors from both the
``original'' and the ``corrected'' model flux distributions (hereafter
original and corrected LCB tables, respectively).  In Figure~\ref{fig:lj}
we compare the corrected LCB table with the original LCB table for MS stars
($\log{g}=4.5$).  The LCB empirical color-$T_{\rm eff}$ relations
(corrected LCB table) are
defined only for $T_{\rm eff} < 11500~K$, which results in an artificial
jump in CMDs, particularly in the $V$-band bolometric correction.  In
addition, there are small scale structures in the color corrections
(especially in $V - I_C$) that cause interpolation noise in our isochrones
at $T_{\rm eff} \ga 8000$~K.  We therefore used the original LCB table above
8000~K, the corrected LCB table below 6500~K, and a linear ramp between
the two tables in order to produce smoother base isochrones.
Isochrones at an age of 100~Myr generated with these ``merged''
color-$T_{\rm eff}$ relations are shown as dot-dashed lines in
Figure~\ref{fig:plcmd}.  However, these still fail to match the
observed MS of the Pleiades and therefore require further corrections.

\begin{figure}
\epsscale{1.1}
\plotone{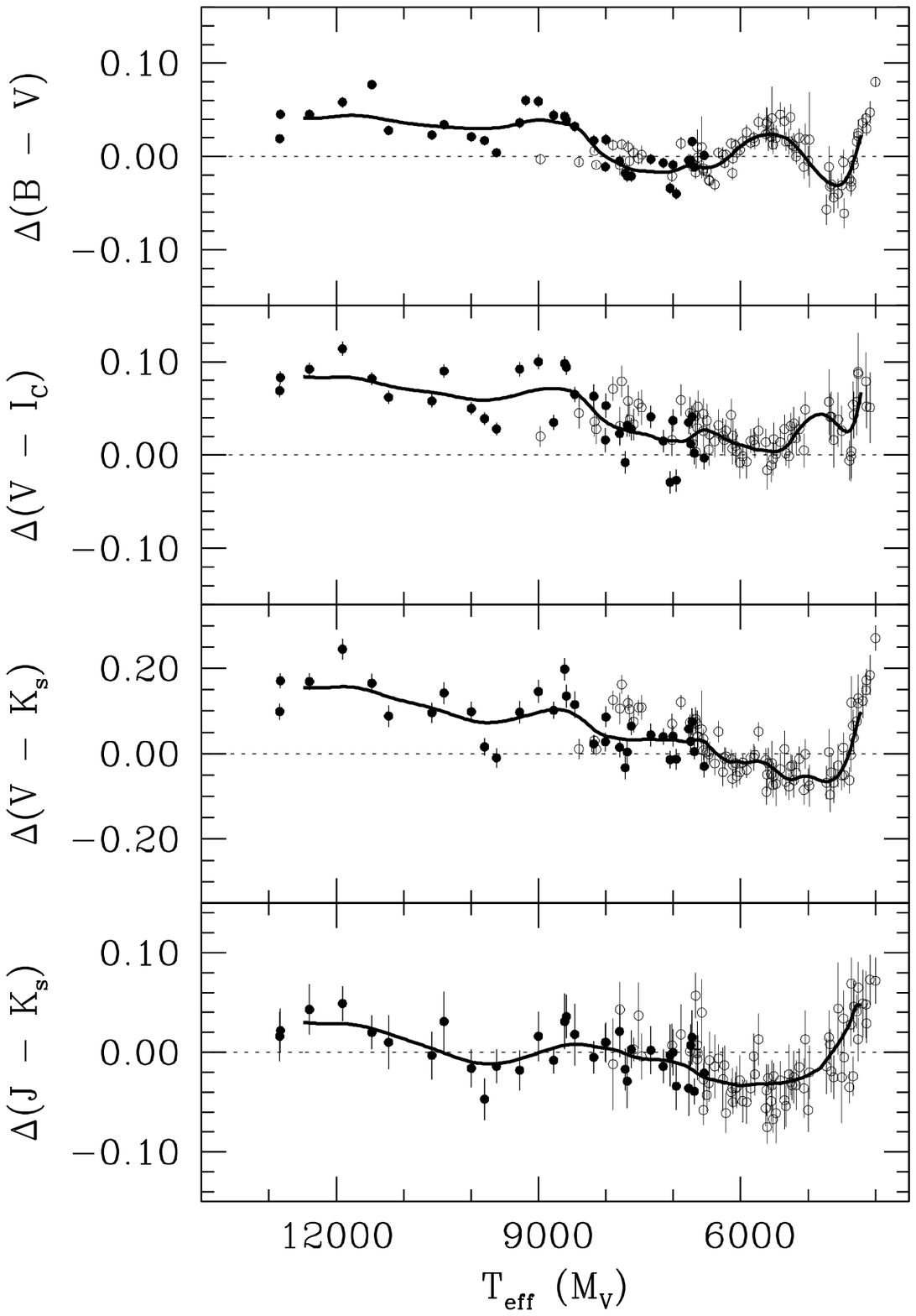}
\caption{Empirical color-$T_{\rm eff}$ corrections ({\it solid line}).
The filled circles show the differences in colors between the Pleiades
data and the uncorrected isochrone at a constant $M_V$.  The sense of
the difference is the former minus the latter values.  The open circles
are those for the Hyades stars.\label{fig:corr}}
\end{figure}

We derived empirical corrections to the isochrone in $B - V$, $V - I_C$,
$V - K_s$, and $J - K_s$ by forcing a match to the Pleiades photometry
as done for the Hyades (Paper II).  Figure~\ref{fig:corr} plots the
difference in color between individual stars and an isochrone generated
with the merged LCB color-$T_{\rm eff}$ relations.  The $T_{\rm eff}$
was calculated from $M_V$.  Filled circles are for the Pleiades stars,
and the open circles are for the Hyades stars used in the calibration of
the lower MS.  The Pleiades stars have smaller error bars than the
Hyades points because we assumed a negligible distance spread for the
Pleiades, while the Hyades ones reflect parallax errors for individual
stars \citep{debruijne01}.  The solid lines in Figure~\ref{fig:corr}
represent our empirical corrections, which were computed at a constant
$M_V$ in a seven-point moving window.  Given the modest sample size and
intrinsic scatter in the upper MS, we did not attempt to resolve out
smaller scale structures.  We defined the empirical color-$T_{\rm eff}$
corrections as those for the Pleiades at $T_{\rm eff} \geq 6838$~K and
for the Hyades at $T_{\rm eff} \leq 6500$~K.  A spline function was used
to bridge over the narrow $T_{\rm eff}$ gap between these two corrections.
We also used $H - K_s$ in MS fitting but without color corrections
because photometric errors were too large in this color to provide
a reliable calibration.

\begin{deluxetable*}{cccccccccc}
\tabletypesize{\scriptsize}
\tablewidth{6in}
\tablecaption{Empirically Calibrated Pleiades Isochrone\label{tab:pliso}}
\tablehead{
  \colhead{$T_{\rm eff}$} \nl
  \colhead{(K)} &
  \colhead{$M_{V,0}$} &
  \colhead{$(B - V)_0$} &
  \colhead{$(V - I_C)_0$} &
  \colhead{$(V - K_s)_0$} &
  \colhead{$(J - K_s)_0$} &
  \colhead{$\Delta (B - V)$} &
  \colhead{$\Delta (V - I_C)$} &
  \colhead{$\Delta (V - K_s)$} &
  \colhead{$\Delta (J - K_s)$}
}
\startdata
$12484$& $-0.034$  &$-0.073$  &$-0.023$  &$-0.128$  &$-0.056$  &$ 0.041$ &$0.084$ &$0.155$ &$ 0.030$\nl
$12286$& $ 0.100$  &$-0.068$  &$-0.020$  &$-0.115$  &$-0.053$  &$ 0.041$ &$0.083$ &$0.154$ &$ 0.029$\nl
$11769$& $ 0.414$  &$-0.053$  &$-0.012$  &$-0.079$  &$-0.046$  &$ 0.044$ &$0.083$ &$0.156$ &$ 0.028$\nl
$11153$& $ 0.723$  &$-0.040$  &$-0.008$  &$-0.052$  &$-0.045$  &$ 0.038$ &$0.073$ &$0.130$ &$ 0.017$\nl
$10468$& $ 1.053$  &$-0.017$  &$ 0.005$  &$-0.012$  &$-0.044$  &$ 0.032$ &$0.066$ &$0.102$ &$ 0.001$\nl
$9881 $& $ 1.335$  &$ 0.016$  &$ 0.026$  &$ 0.034$  &$-0.039$  &$ 0.030$ &$0.059$ &$0.074$ &$-0.011$\nl
$9404 $& $ 1.569$  &$ 0.056$  &$ 0.060$  &$ 0.112$  &$-0.021$  &$ 0.033$ &$0.063$ &$0.080$ &$-0.009$\nl
$9020 $& $ 1.769$  &$ 0.097$  &$ 0.101$  &$ 0.200$  &$ 0.002$  &$ 0.039$ &$0.070$ &$0.096$ &$-0.001$\nl
$8684 $& $ 1.956$  &$ 0.134$  &$ 0.142$  &$ 0.285$  &$ 0.024$  &$ 0.036$ &$0.071$ &$0.103$ &$ 0.006$\nl
$8383 $& $ 2.138$  &$ 0.166$  &$ 0.183$  &$ 0.358$  &$ 0.044$  &$ 0.028$ &$0.063$ &$0.084$ &$ 0.008$\nl
$8107 $& $ 2.316$  &$ 0.194$  &$ 0.224$  &$ 0.433$  &$ 0.063$  &$ 0.006$ &$0.042$ &$0.049$ &$ 0.005$\nl
$7853 $& $ 2.487$  &$ 0.221$  &$ 0.269$  &$ 0.527$  &$ 0.084$  &$-0.006$ &$0.030$ &$0.037$ &$ 0.002$\nl
$7606 $& $ 2.652$  &$ 0.254$  &$ 0.313$  &$ 0.629$  &$ 0.103$  &$-0.014$ &$0.025$ &$0.032$ &$-0.004$\nl
$7362 $& $ 2.819$  &$ 0.290$  &$ 0.349$  &$ 0.724$  &$ 0.123$  &$-0.016$ &$0.021$ &$0.034$ &$-0.007$\nl
$7128 $& $ 2.993$  &$ 0.324$  &$ 0.384$  &$ 0.807$  &$ 0.141$  &$-0.017$ &$0.016$ &$0.032$ &$-0.008$\nl
$6932 $& $ 3.167$  &$ 0.360$  &$ 0.424$  &$ 0.882$  &$ 0.157$  &$-0.014$ &$0.015$ &$0.031$ &$-0.011$\nl
$6838 $& $ 3.261$  &$ 0.380$  &$ 0.446$  &$ 0.925$  &$ 0.167$  &$-0.011$ &$0.015$ &$0.028$ &$-0.013$
\enddata
\end{deluxetable*}

Applying these corrections to the colors at each $T_{\rm eff}$ defines
the empirically calibrated isochrone for the Pleiades, which is tabulated
along with the color corrections in Table~\ref{tab:pliso}.  Isochrones
incorporating the corrections for the Pleiades are shown as solid lines
in Figure~\ref{fig:plcmd}.  Systematic errors in the color corrections
are $\sim0.01$~mag from the errors in the adopted distance, metallicity,
and reddening values for the Pleiades.  We assumed that our empirical
corrections in the upper MS are independent of metallicity and age
and applied them to all isochrones generated in Paper~III.  The isochrones
constructed in this way are available online
at\\{\tt http://www.astronomy.ohio-state.edu/iso/}.

There is one important functional difference between the calibration
for luminous stars and that for the lower MS stars.  The MS for FGK-type
stars is intrinsically very narrow.  As a result, for the low-mass stars
we could plausibly claim that the empirical cluster sequence could be
used to redefine the color-$T_{\rm eff}$ relationship.  On the other
hand, the dispersion about the mean trend for upper MS stars is
significantly larger.  For the Pleiades, we found a dispersion of 0.02~mag
in $B - V$ for $0.0 \la B - V \la 0.40$ (after excluding outliers),
which is significantly above the photometric precision.  A comparable
increase was seen in $V - I_C$ and $V - K_s$ (0.02 and 0.04~mag, respectively).
Although this could be due to differential reddening across the cluster
field \citep{breger86}, the persistence of rapid rotation in early-type
stars can impact both their evolution and the mapping of $T_{\rm eff}$
onto colors \citep[e.g.,][]{collins77} as a consequence of the
\citet{kraft65} break in rotational properties.  Furthermore, these
effects depend on both the rotation rate and the inclination to the line
of sight and can be as large as
0.1 mag in $B - V$.  Nonetheless, the mean cluster locus can still be
used to define a template for distance studies but should not be
interpreted as a direct change in the color-$T_{\rm eff}$ relationship.
In other words, our empirical corrections remove systematic trends from
the rotation-induced color anomalies, as well as from any intrinsic
problems in the adopted color-$T_{\rm eff}$ relationship.

\subsection{Main-Sequence Fitting}\label{sec:msfit}

We determined the distance, reddening, and the ratio of total to
selective extinction in $V$ [$R_V \equiv A_V / E(B - V)$] for each cluster
via an iterative approach, which fits an isochrone to individual CMDs of
various color indices.  All of these parameters are essential to
obtain self-consistent solutions for the absolute magnitude of a Cepheid
(\S~\ref{sec:pl}).  Our MS-fitting process also includes rejection of
stars that are too far away from an MS ridge-line, defined by the locus of
points with the highest density on each CMD.  Individual color excesses
in each cluster were used to derive $R_V$ according to the extinction law
described by \citet[][hereafter CCM89]{cardelli89}.  The rest of this
section describes the MS-fitting process in detail.

Our sample clusters do not have detailed radial velocity or proper-motion
membership studies.  Before isochrone fitting we therefore
applied the photometric filtering procedure described in Paper~III to
identify and reject stars that are likely foreground/background objects
or cluster binaries.  The procedure iteratively identifies a cluster MS
ridge-line independently of the isochrones and rejects stars if they are
too far away from the ridge-line.  This was done by computing $\chi^2$
in each CMD as
\begin{equation}
\chi^2 = \sum_{i=1}^N \chi_i^2
= \sum_{i=1}^N \frac {(\Delta X_i)^2 }
 {\sigma_{X,i}^2 + (\gamma_i^{-1} \sigma_{V,i})^2 + \sigma_0^2},
\label{eq:chifilter}
\end{equation}
where $\Delta X_i$ is the color difference between the $i^{th}$ data
point and MS at the same $V$; $\sigma_{X,i}$ and $\sigma_{V,i}$ are
photometric errors in color and $V$, respectively.  The error in
$V$ contributes to the error in $\Delta X_i$ by the inverse slope of
the MS, $\gamma_i^{-1}$.  We added $\sigma_0$ in quadrature to the
propagated photometric errors in the denominator to take into account
the presence of differential reddening, cluster binaries, non-cluster
members, and other effects that would increase the MS width above the
photometric precision.  We adjusted the value of $\sigma_0$ so that
the total $\chi^2$ is equal to the total number of stars $N$ used in
the filtering.

Initially we rejected all data points as outliers if $\chi_i^2$ (the
individual contribution to $\chi^2$) was greater than 9 (corresponding
to a $3\sigma$ outlier). We repeated adjusting $\sigma_0$ and rejecting
outliers with the reduced set of data points until there remained no
point with $\chi_i^2 > 9$.  We combined the results from all CMDs, and rejected
stars if they were tagged as an outlier at least in one of the CMDs.
Missing data in any colors was not a criterion for rejection.  For this
study we included $(J - K_s, K_s)$ and $(H - K_s, K_s)$ in the filtering
procedure.  We reduced the rejection threshold at each iteration until
it was limited to $2.0$--$2.4\sigma$, but our result is insensitive to
the final rejection threshold within the fitting errors.

Our filtering results are shown in Figures~\ref{fig:bvcmd}--\ref{fig:jkcmd}.
The bull's-eyes represent stars retained and used in the following MS fits,
and open circles are the ones rejected by the filtering.  At the end of the
iteration, $\sigma_0$ values in $(B - V, V)$ and $(V - K_s, V)$ were
(0.055~mag, 0.101~mag) in NGC~129, (0.050~mag, 0.102~mag) in VDB~1,
(0.023~mag, 0.064~mag) in NGC~5662, (0.048~mag, 0.102~mag) in Lyng{\aa}~6,
(0.031~mag, 0.078~mag) in NGC~6067, (0.033~mag, 0.058~mag) in NGC~6087,
(0.043~mag, 0.111~mag) in M25, and (0.011~mag, 0.026~mag) in NGC~7790.
These excess dispersions are most likely due to differential reddening
of heavily reddened, young open clusters.  Note that the ratio of $\sigma_0$
values from $(B - V, V)$ and $(V - K_s, V)$ is also approximately equal to
the color-excess ratio $E(V - K_s)/E(B - V) \approx 2.8$ (see below).
We discuss in detail the effects of differential reddening on {\it P-L}
relations in \S~\ref{sec:pl}.
In $(J - K_s, K_s)$ and $(H - K_s, K_s)$, photometric errors were usually
large enough that $\sigma_0$ values were automatically set to zero.

The fitting process took advantage of the changing slope of the MS with
absolute magnitude.  Near the MS turnoff ($M_V \leq 1.5$), the MS is
nearly vertical, allowing a precise determination of the reddening.
Further down ($1.5 \leq M_V \leq 3.5$), the slope of the MS is still
relatively steep [e.g., $\Delta M_V / \Delta (B - V) \approx 5$], but
it nevertheless provides sufficient leverage for determining the distance.
We divided each CMD into two zones, shown as boxes in
Figures~\ref{fig:bvcmd}--\ref{fig:jkcmd}, which define the regions used
for determining the color excess and distance.  The color excess was determined
for stars with  $-1.0 \leq M_V \leq 2.0$ in $(B - V, V)$, $(V - I_C, V)$,
and $(V - K_s, V)$, and at $-0.8 \leq M_K \leq 1.7$ in $(J - K_s, K_s)$
and $(H - K_s, K_s)$.  Distance determinations came from stars with
$-0.1 \leq (B - V)_0 \leq 0.5$ and the corresponding color ranges in other
colors with the same $M_V$.  For Lyng{\aa}~6 and NGC~7790, $(V - K_s, V)$
and $(J - K_s, K_s)$ were not used in the distance estimation because
the lower parts of the MS are below the 2MASS completeness limit
($K_s \approx 14.3$).

\begin{deluxetable*}{lccccc}
\tabletypesize{\scriptsize}
\tablewidth{4in}
\tablecaption{Color-Excess Ratios for Zero-Color Stars\label{tab:ebvlaw}}
\tablehead{
  \colhead{Color-Excess Ratios} &
  \colhead{CCM89\tablenotemark{a}} &
  \colhead{BCP98\tablenotemark{a}} &
  \colhead{This Paper\tablenotemark{b}} &
  \colhead{Others\tablenotemark{c}} &
  \colhead{ref} 
}
\startdata
$E(V-I_C)/E(B-V)$ & $1.30$ & $1.32$ & $1.26\pm0.06$ & $1.25\pm0.01$ & 1\nl
$E(V-K_s)/E(B-V)$ & $2.88$ & $2.91$ & $2.82\pm0.06$ & $2.78\pm0.02$ & 2,3,4\nl
$E(J-K_s)/E(B-V)$ & $0.56$ & $0.58$ & $0.53\pm0.04$ & $0.52\pm0.03$ & 4\nl
$E(H-K_s)/E(B-V)$ & $0.20$ & $0.22$ & $0.16\pm0.02$ & $0.20\pm0.04$ & 4\nl
\enddata
\tablerefs{(1) Dean et~al. 1978; (2) Schultz \& Wiemer 1975;
(3) Sneden et~al. 1978; (4) Rieke \& Lebofsky 1985.}
\tablecomments{Color excesses involving 2MASS $JHK_s$ bandpasses are
corrected for the difference in filter effective wavelengths.}
\tablenotetext{a}{Estimated at $R_{V,0} = 3.26$.}
\tablenotetext{b}{Weighted mean and rms deviation.}
\tablenotetext{c}{Revised for zero-color stars using color-dependent
reddening relations in BCP98.}
\end{deluxetable*}

Since the color excess and $R_V$ depend on the intrinsic color of
stars, we adopted the color-dependent relations from \citet[][
hereafter BCP98]{bessell98} and M. S. Bessell (2007, private communication).
Their formulae were
based on the theoretical stellar spectral energy distributions with
the extinction law from \citet{mathis90}.  Specifically,
their formulation yields
\begin{equation}
R_V = R_{V,0} + 0.22 (B - V)_0,
\label{eq:rv}
\end{equation}
\begin{equation}
E(B - V) = E(B - V)_0 \{1 - 0.083 (B - V)_0 \},
\label{eq:ebv}
\end{equation}
where $(B - V)_0$ is an intrinsic color of each star.  The $R_{V,0}$
and $E(B - V)_0$ are the values for zero-color stars, where $R_{V,0}$
is $3.26$ in their model.  For other colors we used the following
color-excess ratios:
\begin{eqnarray}
\frac{E(V - I_C)}{E(B - V)} &=& 1.30 + 0.06 (V - I_C)_0 \\
\frac{E(V - K_s)}{E(B - V)} &=& 2.88 + 0.07 (V - K_s)_0 \\
\frac{E(J - K_s)}{E(B - V)} &=& 0.56 + 0.06 (J - K_s)_0 \\
\frac{E(H - K_s)}{E(B - V)} &=& 0.20 + 0.15 (H - K_s)_0.
\label{eq:color}
\end{eqnarray}
The first terms represent the color-excess ratios from CCM89 at
$R_{V,0} = 3.26$, and the second terms are color-dependent relations
from BCP98.  Differences
between the original coefficients in BCP98 and those for the 2MASS
$JHK_s$ system using transformation equations in \citet{carpenter01}\footnote{
Updated color transformations for 2MASS All-sky Data Release can be found at\\
http://www.ipac.caltech.edu/2mass/releases/allsky/doc/sec6\_4b.html.}
were found to be negligible.  Table~\ref{tab:ebvlaw} lists color-excess
ratios for zero-color stars from CCM89 and BCP98, as well as observationally
derived values from \citet{schultz75}, \citet{dean78}, \citet{sneden78},
and \citet{rieke85}.  Color-excess ratios determined from our cluster
sample are also shown in the fourth column, which is discussed in
the next section.

The fitting process begins with a guess at the average distance.  At
this distance, individual color excesses for zero-color stars [$E(B - V)^{\rm obs}_0$,
$E(V - I_C)^{\rm obs}_0$, etc.] were determined using stars on the upper MS.
These were used to determine the best-fit $R_{V,0}$ and $E(B - V)_0$
from the CCM89 extinction law by minimizing
\begin{equation}
\chi^2 = \sum_{i=1}^5 \biggr[\frac {E(\lambda_i - V)^{\rm obs}_0
- E(\lambda_i - V)^{\rm CCM89}_0}{\sigma_{E(\lambda - V)}} \biggr]^2,
\label{eq:rvchi}
\end{equation}
where $\lambda_i$ represents effective wavelengths in $BI_C$
\citep{bessell98} and isophotal wavelengths in $JHK_s$ \citep{cohen03}.
The index in the summation runs over $B$, $I_C$, $J$, $H$, and $K_s$.
The $\sigma_{E(\lambda - V)}$ is the fitting error in the upper MS,
which was determined from the scatter of the points around the isochrone
and the individual errors in color.  The CCM89 color excess,
$E(\lambda - V)^{\rm CCM89}_0$, can be expressed as
\begin{equation}
E(\lambda - V)^{\rm CCM89}_0
= R_{V,0} E(B - V)_0 \left( \frac {A_{\lambda}}{A_V} - 1 \right),
\label{eq:ccm}
\end{equation}
where ${A_{\lambda}} / {A_V}$ is a function of $R_{V,0}$
from polynomial relations in CCM89.  Note that $R_{V,0}$ and $E(B - V)_0$
in equation~(\ref{eq:rvchi}) are two parameters to be determined and
that $E(B - V)_0$ in equation~(\ref{eq:ccm}) is the normalization factor,
which is different from the input color excess in $(B - V, V)$.

The fitting procedure then found weighted median distances on each CMD by
using the stars in the lower part of the MS.  The weighted average distance
from all CMDs was computed, and the process was repeated until convergence,
which occurred when the difference in the average distance from the previous
iteration became smaller than its propagated error.

\begin{figure}
\epsscale{1.1}
\plotone{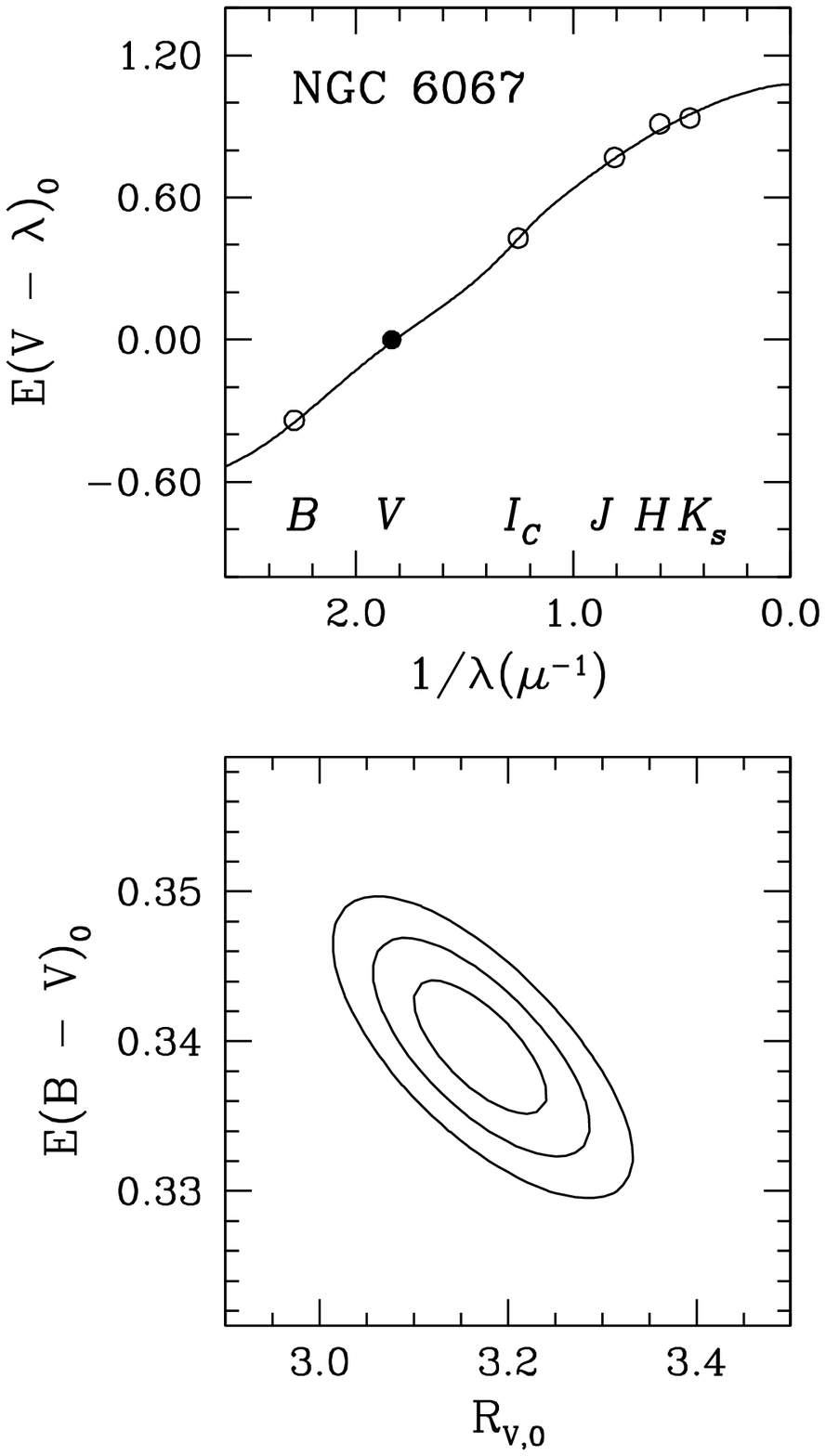}
\caption{Example of determining $R_{V,0}$ and $E(B - V)_0$ for NGC~6067.
{\it Top:} Solid line represents the best-fitting CCM89 extinction
curve to color excesses ({\it open circles}).
{\it Bottom:} Likelihood contours in $R_{V,0}$ and $E(B - V)_0$ shown at
$\Delta \chi^2$ = 2.30, 6.17, and 11.8 (68.3\%, 95.4\%, and 99.73\%
confidence levels for 2 degrees of freedom).\label{fig:6067.ebv}}
\end{figure}

An example of the fit is shown in Figure~\ref{fig:6067.ebv} for NGC~6067.
The solid line in the top panel is the best-fitting CCM89 extinction
curve to the data ({\it open circles}).  The bottom panel shows likelihood
contours in $R_{V,0}$ and $E(B - V)_0$ drawn at $\Delta \chi^2$ = 2.30, 6.17, and 11.8
from the minimum $\chi^2 = 4.5$ (68.3\%, 95.4\%, and 99.73\% confidence
levels for 2 degrees of freedom).

\section{Cluster Properties}

In this section we present our results on the distance, reddening, and
$R_V$ of the sample clusters and evaluate various systematic errors in
these parameters.  In particular, we assess the accuracy of MS-fitting
based on internal consistency of distances from multi-color CMDs.  We
show that our distance estimates are generally shorter than
previous ZAMS-fitting results and that this is mainly due
to our lower reddening estimates.

\subsection{Distance, Reddening, and $R_V$}\label{sec:par}

\begin{deluxetable*}{lrcrcrcrcrc}
\tabletypesize{\scriptsize}
\tablewidth{6.5in}
\tablecaption{MS-Fitting Distance\label{tab:dist}}
\tablehead{
  \colhead{Cluster} &
  \multicolumn{8}{c}{$(m -  M)_0$} &
  \colhead{$(m - M)_{0,\rm ave}$\tablenotemark{a}} &
  \colhead{$\sigma_{(m - M)}$\tablenotemark{b}}\nl
  \cline{2-3}\cline{4-5}\cline{6-7}\cline{8-9}
  \colhead{} &
  \colhead{$(B - V, V)$} &
  \colhead{$N_{\rm fit}$} &
  \colhead{$(V - I_C, V)$} &
  \colhead{$N_{\rm fit}$} &
  \colhead{$(V - K_s, V)$} &
  \colhead{$N_{\rm fit}$} &
  \colhead{$(J - K_s, K_s)$} &
  \colhead{$N_{\rm fit}$} &
  \colhead{} &
  \colhead{}
}
\startdata
NGC~129    \dotfill & $11.030\pm0.041$  &79& \nodata & \nodata    & $11.078\pm0.044$ &58& $11.007\pm0.037$ &53& $11.035\pm0.023$ & $0.027$ \nl
VDB~1      \dotfill & $11.067\pm0.097$  &20& $10.744\pm0.033$  &13& $10.576\pm0.062$ &13& $10.706\pm0.082$ &18& $10.734\pm0.026$ & $0.209$ \nl
NGC~5662   \dotfill &  $9.351\pm0.025$  &52& \nodata & \nodata    &  $9.290\pm0.037$ &46&  $9.201\pm0.043$ &37&  $9.307\pm0.019$ & $0.075$ \nl
Lyng{\aa}~6\dotfill & $11.604\pm0.114$  &21& $11.415\pm0.117$  &21& \nodata & \nodata   & \nodata & \nodata   & $11.512\pm0.082$ & $0.134$ \nl
NGC~6067   \dotfill & $10.991\pm0.019$ &141& $11.125\pm0.026$ &120& $11.004\pm0.033$ &43& $10.996\pm0.059$ &35& $11.029\pm0.013$ & $0.064$ \nl
NGC~6087   \dotfill &  $9.708\pm0.048$  &22& \nodata & \nodata    &  $9.601\pm0.045$ &20&  $9.649\pm0.049$ &19&  $9.650\pm0.027$ & $0.054$ \nl
M25        \dotfill &  $8.961\pm0.030$  &45& \nodata & \nodata    &  $8.899\pm0.054$ &36&  $8.885\pm0.052$ &35&  $8.934\pm0.023$ & $0.040$ \nl
NGC~7790   \dotfill & $12.474\pm0.014$  &63& $12.427\pm0.019$  &57& \nodata & \nodata   & \nodata & \nodata   & $12.457\pm0.011$ & $0.033$ \nl
\enddata
\tablecomments{Formal fitting errors are shown for individual distance moduli with a number of points used in the fit.}
\tablenotetext{a}{Weighted average distance and its propagated error, not including systematic uncertainties.}
\tablenotetext{b}{Standard deviation of individual distance moduli.}
\end{deluxetable*}

\begin{deluxetable*}{lrcrcrcrcrcc}
\tabletypesize{\scriptsize}
\tablewidth{6.5in}
\tablecaption{MS-Fitting Color Excess\label{tab:ebv}}
\tablehead{
  \colhead{Cluster} &
  \colhead{$E(B - V)_0$} &
  \colhead{$N_{\rm fit}$} &
  \colhead{$E(V - I_C)_0$} &
  \colhead{$N_{\rm fit}$} &
  \colhead{$E(V - K_s)_0$} &
  \colhead{$N_{\rm fit}$} &
  \colhead{$E(J - K_s)_0$} &
  \colhead{$N_{\rm fit}$} &
  \colhead{$E(H - K_s)_0$} &
  \colhead{$N_{\rm fit}$} &
  \colhead{$\sigma$\tablenotemark{a}}
}
\startdata
NGC~129    \dotfill & $0.500\pm0.008$  &53& \nodata    &\nodata& $1.431\pm0.020$ &45& $0.280\pm0.006$ &43& $0.069\pm0.006$ &41& $0.002$\nl
VDB~1      \dotfill & $0.573\pm0.014$  &12& $0.611\pm0.023$ & 9& $1.458\pm0.031$ &11& $0.358\pm0.013$ &10& $0.066\pm0.012$ &10& $0.075$\nl
NGC~5662   \dotfill & $0.277\pm0.003$  &47& \nodata    &\nodata& $0.795\pm0.012$ &41& $0.172\pm0.006$ &38& $0.050\pm0.005$ &37& $0.018$\nl
Lyng{\aa}~6\dotfill & $1.326\pm0.012$  &23& $1.698\pm0.017$ &22& $3.659\pm0.056$ & 9& $0.635\pm0.020$ & 8& $0.200\pm0.016$ & 9& $0.087$\nl
NGC~6067   \dotfill & $0.340\pm0.003$ &107& $0.428\pm0.007$ &67& $0.935\pm0.014$ &31& $0.166\pm0.008$ &26& $0.025\pm0.008$ &22& $0.019$\nl
NGC~6087   \dotfill & $0.132\pm0.007$  &30& \nodata    &\nodata& $0.410\pm0.010$ &25& $0.084\pm0.009$ &24& $0.019\pm0.007$ &24& $0.009$\nl
M25        \dotfill & $0.424\pm0.009$  &60& \nodata    &\nodata& $1.160\pm0.022$ &52& $0.233\pm0.007$ &47& $0.077\pm0.005$ &47& $0.011$\nl
NGC~7790   \dotfill & $0.509\pm0.002$  &35& $0.634\pm0.005$ &33& $1.457\pm0.021$ &27& $0.248\pm0.010$ &28& $0.110\pm0.011$ &28& $0.030$\nl
\enddata
\tablecomments{Color excess values are those for zero-color stars (see text).  Formal fitting
errors are shown for individual color excesses with a number of points used in the fit.}
\tablenotetext{a}{Standard deviation of $E(B - V)$ estimates from all CMDs  except $(H - K_s, K_s)$,
after transforming color-excess values to $E(B - V)$ using the CCM89 color-excess ratios.}
\end{deluxetable*}

Our MS-fitting results on the distance and color excess from individual
CMDs are shown in Tables~\ref{tab:dist} and \ref{tab:ebv}, respectively,
along with a number of stars used in the fit.  The last two columns in
Table~\ref{tab:dist} show the weighted average distance modulus from
all available CMDs with its propagated error and the standard deviation
($\sigma$) of individual distances.  The last column in Table~\ref{tab:ebv}
lists a standard deviation of $E(B - V)$ from all CMDs, after transforming
individual color-excess values in $V - I_C$, $V - K_s$, and $J - K_s$ to
$E(B - V)$ using the CCM89 reddening law (Table~\ref{tab:ebvlaw}).  We
excluded $E(H - K_s)$ from this computation since we found a large
systematic deviation of our color-excess estimates from the CCM89 law
as discussed below.

\begin{figure}
\epsscale{1.1}
\plotone{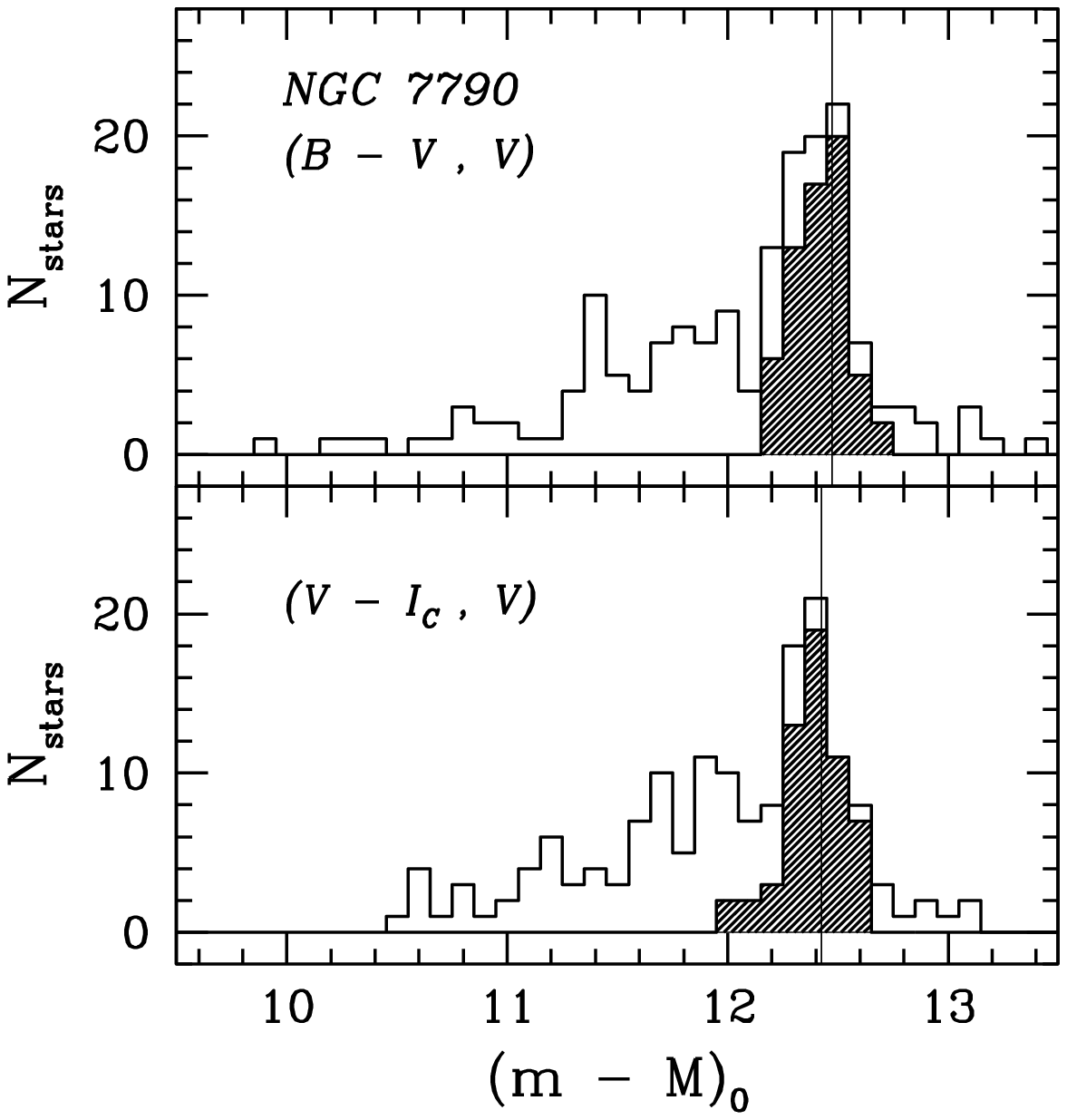}
\caption{Distribution of individual distance moduli for NGC~7790 with
respect to the isochrones shown in Fig.~\ref{fig:bvcmd} and \ref{fig:vicmd}.
The open histogram is the distribution of all stars in the isochrone
fitting range where the distance was determined.  The hatched histogram
shows only for those that passed the photometric filtering.  The vertical
line indicates the weighted median distance modulus.\label{fig:7790res}}
\end{figure}

In Figures~\ref{fig:bvcmd}--\ref{fig:jkcmd} the best-fitting isochrones
are overlaid on the cluster CMDs.  While there is some variation in
the quality of the data, we could divide the sample clusters into three
groups based on how tight the MS is on optical CMDs.  The first group
contains NGC~6067 and NGC~7790, which have well-populated CMDs and
small fitting errors on each CMD.  They also have a well-defined peak on
the distribution of an individual star's distance modulus as shown in
Figure~\ref{fig:7790res} for NGC~7790.  The second group is NGC~129,
NGC~5662, Lyng{\aa}~6, NGC~6087, and M25.  These clusters have less
well-defined peaks, either because of strong differential reddening
(\S~\ref{sec:pl}) or because of a bias in the sampling from photoelectric
observations at faint magnitudes.  The last group is VDB~1, which
has a sparse MS and correspondingly large distance uncertainties.

We can also divide the clusters into groups by examining the difference
in derived cluster properties from different colors.  Clusters where
the results from multi-color CMDs have a high internal consistency are
strong cases, while clusters with substantial differences indicate
problems.  For the above first and second groups of clusters except
Lyng{\aa}~6, the standard deviations of individual distances
[$\sigma_{(m - M)}$ in Table~\ref{tab:dist}] are 0.03--0.08~mag, which
are generally consistent with the fitting error on individual CMDs.
On the other hand, both Lyng{\aa}~6 and VDB~1 exhibit a larger
dispersion in distance [$\sigma_{(m - M)} = 0.21$ and $0.13$~mag,
respectively], and they also have a larger standard deviation of $E(B - V)$
from different colors ($\sigma$ in Table~\ref{tab:ebv}).

However, the dispersion in Lyng{\aa}~6 is consistent with the statistical
scatter expected from the small sample size.  By contrast, VDB~1 has a
statistically unreliable result that demands its exclusion from our data
set.  For this cluster, the best distance estimates inferred from
different colors disagree by much more than their individual error
estimates would imply.  One possible explanation for the internal
inconsistency is that the cluster MS is contaminated by less reddened
field stars as shown by spectral-type study \citep{preston64}.
Because these stars happen to lie on the cluster MS, our photometric
filtering could have misidentified cluster members.  We therefore
exclude VDB~1 from the following analysis and leave it for future studies.

\begin{deluxetable}{llccc}
\tabletypesize{\scriptsize}
\tablewidth{3.2in}
\tablecaption{Error Budget for NGC~6067\label{tab:error}}
\tablehead{
  \colhead{Source of Error} &
  \colhead{$\Delta$Quantity} &
  \colhead{$\Delta R_{V,0}$} &
  \colhead{$\Delta E(B - V)_0$} &
  \colhead{$\Delta (m - M)_0$}
}
\startdata
${\rm [M/H]}$\dotfill    & $\pm0.05$  & $\pm0.005$ & $\mp0.001$ & $\pm0.056$ \nl
$\log{t}$ (Myr)\dotfill  & $\pm0.2$   & $\pm0.009$ & $\mp0.013$ & $\mp0.020$ \nl
Helium ($\Delta Y$)\dotfill & $\pm0.010$ & \nodata    & \nodata    & $\mp0.030$ \nl
$\Delta (B-V)$\dotfill   & $\pm0.025$ & $\pm0.264$ & $\mp0.022$ & $\mp0.007$ \nl
$\Delta (V-I)_C$\dotfill & $\pm0.025$ & $\mp0.030$ & $\mp0.002$ & $\pm0.027$ \nl
$\Delta V$\dotfill       & $\pm0.025$ & $\mp0.085$ & $\pm0.000$ & $\mp0.003$ \nl
$\Delta J$\dotfill       & $\pm0.011$ & $\pm0.009$ & $\pm0.000$ & $\mp0.001$ \nl
$\Delta H$\dotfill       & $\pm0.007$ & $\pm0.009$ & $\pm0.000$ & $\mp0.002$ \nl
$\Delta K_s$\dotfill     & $\pm0.007$ & $\pm0.011$ & $\pm0.000$ & $\mp0.005$ \nl
Fitting\dotfill          & \nodata    & $\pm0.061$ & $\pm0.004$ & $\pm0.032$ \nl
Total\dotfill            & \nodata    & $\pm0.286$ & $\pm0.026$ & $\pm0.079$ \nl
\enddata
\tablecomments{$\Delta (B-V)$, $\Delta (V-I)_C$, $\Delta V$, $\Delta J$, $\Delta H$,
and $\Delta K_s$ represent zero-point errors in the cluster photometry.}
\end{deluxetable}

\begin{deluxetable}{lccc}
\tabletypesize{\scriptsize}
\tablewidth{3in}
\tablecaption{Summary of Cluster Parameters\label{tab:cluster}}
\tablehead{
  \colhead{Cluster}  &
  \colhead{$R_{V,0}$} &
  \colhead{$E(B - V)_0$} &
  \colhead{$(m - M)_0$}
}
\startdata
NGC~129     & $3.31\pm0.22$ & $0.49\pm0.03$ & $11.04\pm0.05$ \nl
NGC~5662    & $3.30\pm0.38$ & $0.27\pm0.03$ & $ 9.31\pm0.06$ \nl
Lyng{\aa}~6 & $3.16\pm0.10$ & $1.32\pm0.03$ & $11.51\pm0.13$ \nl
NGC~6067    & $3.18\pm0.29$ & $0.34\pm0.03$ & $11.03\pm0.08$ \nl
NGC~6087    & $3.59\pm0.85$ & $0.13\pm0.03$ & $ 9.65\pm0.06$ \nl
M25         & $3.16\pm0.21$ & $0.42\pm0.03$ & $ 8.93\pm0.08$ \nl
NGC~7790    & $3.18\pm0.26$ & $0.51\pm0.02$ & $12.46\pm0.11$ \nl
\enddata
\end{deluxetable}

Table~\ref{tab:error} shows the error budget for NGC~6067 as
an example.  The first column displays individual sources of error,
and the second column shows the size of the errors adopted for each
quantity.  The third through fifth columns list error contributions
to $R_{V,0}$, $E(B - V)_0$, and $(m - M)_0$, respectively.  The size of
systematic errors in cluster metallicity, age, and photometry was
discussed in \S~\ref{sec:data}.  The quantity $\Delta Y$ represents
the uncertainty in the helium abundance at fixed metallicity and comes
from the consideration of the initial solar abundance and those of the Hyades
and the Pleiades (Paper~I and Paper~III).  The fitting errors represent
the internal precision of the fit and were taken as the larger value
of the propagated error or $\sigma_{(m-M)}/\sqrt{N}$ in Table~\ref{tab:dist}.
We estimated the fitting errors in $R_{V,0}$ and $E(B - V)_0$ from
the size of the $1\sigma$ contours on the $\Delta \chi^2$ distribution
(e.g., Fig.~\ref{fig:6067.ebv}).  The total systematic error is the
quadrature sum of these errors.  The $R_{V,0}$, $E(B - V)_0$, and
$(m - M)_0$ of our cluster sample are listed in Table~\ref{tab:cluster}
with total systematic errors in these quantities.

\begin{figure}
\epsscale{1.2}
\plotone{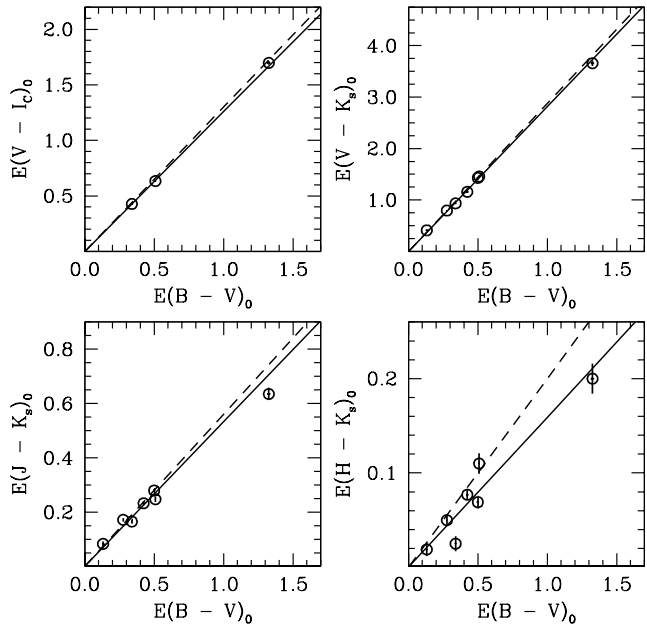}
\caption{Mean color-excess ratios from the cluster sample.  The error
bars are from the isochrone fitting only.  The solid line is a fit to
the data, constrained to pass through the origin.  The dashed line is
the color-excess ratio from the CCM89 reddening law.\label{fig:ebv}}
\end{figure}

Independently of the CCM89 reddening law, we also derived color-excess ratios
based on our cluster sample.  Figure~\ref{fig:ebv} shows a correlation
between $E(B - V)$ and color-excess values in other color indices.
Individual clusters are represented as a point, and fitting errors are
shown only.  The solid line is a linear fit to the data, constrained to
pass through the origin.  Table~\ref{tab:ebvlaw} summarizes our best-fit
slopes of these lines or color-excess ratios.  Our values are generally
in good agreement with the CCM89 law and other observational estimates.
However, our method yields a lower $E(H - K_s) / E(B - V)$ than these
studies, which may reflect a zero-point offset in $H - K_s$ of our
isochrones.  Note that we did not derive color corrections for $H - K_s$.

\subsection{Comparison with Previous Studies}\label{sec:comp}

\begin{deluxetable*}{lccccc}
\tablewidth{4in}
\tabletypesize{\scriptsize}
\tablecaption{Comparison of Cluster Distance with Published Results\label{tab:comp.dist}}
\tablehead{
  \colhead{Cluster} &
  \colhead{This Study} &
  \colhead{FW87\tablenotemark{a}} &
  \colhead{TSR03\tablenotemark{a}} &
  \colhead{HST03} &
  \colhead{G05\tablenotemark{b}}
}
\startdata
NGC~129    \dotfill & $11.04\pm0.05$ & $11.30$ & $11.24$ & $10.94\pm0.14$ & \nodata        \nl
NGC~5662   \dotfill & $ 9.31\pm0.06$ &  $9.19$ &  $9.19$ & \nodata        &  $9.33\pm0.06$ \nl
Lyng{\aa}~6\dotfill & $11.51\pm0.13$ & $11.48$ & $11.49$ & $11.33\pm0.18$ & \nodata        \nl
NGC~6067   \dotfill & $11.03\pm0.08$ & $11.19$ & $11.19$ & $11.18\pm0.12$ & $11.25\pm0.19$ \nl
NGC~6087   \dotfill & $ 9.65\pm0.06$ &  $9.87$ &  $9.87$ & \nodata        & $10.02\pm0.03$ \nl
M25        \dotfill & $ 8.93\pm0.08$ &  $9.09$ &  $9.09$ &  $9.08\pm0.18$ &  $8.98\pm0.02$ \nl
NGC~7790   \dotfill & $12.46\pm0.11$ & $12.45$ & $12.71$ & $12.58\pm0.14$ & \nodata        \nl
\enddata
\tablecomments{FW87, \citet{feast87}; TSR03, \citet{tammann03};
HST03, \citet{hoyle03}; G05, \citet{gieren05}.}
\tablenotetext{a}{Distances revised assuming the same Pleiades distance, $(m - M)_0 = 5.63$,
as in this paper.}
\tablenotetext{b}{Distances to individual Cepheids from the surface brightness technique.}
\end{deluxetable*}

\begin{figure}
\epsscale{1.1}
\plotone{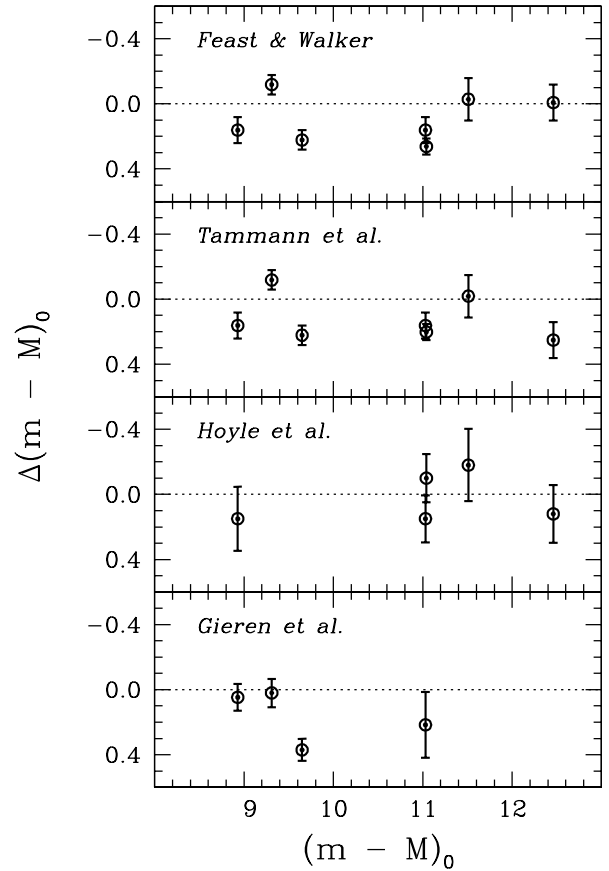}
\caption{Comparison of cluster distances with those in \citet{feast87},
\citet{tammann03}, \citet{hoyle03}, and \citet{gieren05}.  The
differences are in the sense of their distances minus those in this
paper.  The error bars are quadrature sum from the two studies in the
lower two panels.  Individual distances in the top two panels
were revised assuming our Pleiades distance scale.\label{fig:comp.dist}}
\end{figure}

The most up-to-date compilation of Cepheid parameters in open clusters
and associations can be found in \citet{tammann03}.  Their distances are
from the earlier compilation by \citet{feast99}.  These ZAMS-fitting
distances are the same as in \citet{feast87}, but include more recent
distance estimates for NGC~129 \citep{turner92} and NGC~7790 \citep{romeo89}.
All of our Cepheids are listed in these tables, and the comparisons with
distances in \citeauthor{feast87} and \citeauthor{tammann03} are shown in the upper
two panels of Figure~\ref{fig:comp.dist} and listed in
Table~\ref{tab:comp.dist}.  Both of these studies estimated distances
relative to the Pleiades but assumed different Pleiades distance of
$(m - M)_0 = 5.57$ and $5.61$~mag, respectively.  We revised their
distances assuming our Pleiades distance, $(m - M)_0 = 5.63$.
Since they did not explicitly present errors, we plotted our errors only.
The average differences in distance are $\Delta \langle (m - M)_0
\rangle = 0.09\pm0.05$ and $0.12\pm0.05$ with \citeauthor{feast87} and
\citeauthor{tammann03}, respectively.  The sense of the differences
is that our new distance estimates are shorter on average.  The standard
deviations of the differences are $0.14$~mag for both
comparisons, while the expected size from our error estimates is
$0.07$~mag.  If our error estimates are correct, the dispersion
indicates a random error of $0.12$~mag in distance estimates from
\citeauthor{feast87} and \citeauthor{tammann03}.

In addition to these studies, \citet{hoyle03} derived ZAMS-fitting
distances to 11 open clusters with new photometry in $UBVK$.  Our
comparison with this study is shown in Figure~\ref{fig:comp.dist}
and tabulated in Table~\ref{tab:comp.dist}.  The error bars represent
quadrature sums of our and their reported error estimates.  Their
distances are marginally consistent with our estimates, and the average
difference is $\Delta \langle (m - M)_0 \rangle = 0.04\pm0.07$, our
distances being shorter on average.  However, we did not attempt to revise
their distances assuming our Pleiades distance scale.  They used a ZAMS from
\citet{allen73} for the optical data, but they also claimed that the
same results were obtained from the ZAMS in \citet{turner79} and
\citet{mermilliod81}.  Since the Pleiades distance modulus of 5.56~mag
was adopted in \citeauthor{turner79}, the distance moduli in \citeauthor{hoyle03}
would then become longer by $\approx0.07$~mag on average than ours when
they are on the same Pleiades distance scale as in this paper.

The bottom panel in Figure~\ref{fig:comp.dist} shows the comparison
with distances from the surface brightness technique by \citet{gieren05}.
Their distances are also larger on average than our estimates by
$0.15\pm0.09$~mag as in the previous MS-fitting studies.  However,
they recalibrated the projection factor, which was used to convert
observed velocities to pulsational velocities, using MS-fitting
distances to cluster Cepheids.  Therefore, their estimates are not
completely independent from the above MS-fitting studies.

\begin{deluxetable*}{lccccc}
\tablewidth{4in}
\tabletypesize{\scriptsize}
\tablecaption{Comparison of Cepheid $E(B - V)$ with Published Results\label{tab:comp.ebv}}
\tablehead{
  \colhead{Cluster} &
  \colhead{This Study\tablenotemark{a}} &
  \colhead{FW87} &
  \colhead{F95} &
  \colhead{TSR03} &
  \colhead{HST03}
}
\startdata
NGC~129    \dotfill & $0.46\pm0.03$ & $0.49$ & $0.52\pm0.01$ & $0.48$ & $0.52\pm0.05$ \nl
NGC~5662   \dotfill & $0.26\pm0.03$ & $0.29$ & $0.28\pm0.02$ & $0.26$ & \nodata       \nl
Lyng{\aa}~6\dotfill & $1.23\pm0.03$ & $1.22$ & $1.27\pm0.02$ & $1.21$ & $1.24\pm0.08$ \nl
NGC~6067   \dotfill & $0.32\pm0.03$ & $0.32$ & $0.33\pm0.01$ & $0.32$ & $0.33\pm0.03$ \nl
NGC~6087   \dotfill & $0.12\pm0.03$ & $0.17$ & $0.19\pm0.01$ & $0.18$ & \nodata       \nl
M25        \dotfill & $0.39\pm0.03$ & $0.44$ & $0.43\pm0.01$ & $0.40$ & $0.45\pm0.04$ \nl
NGC~7790\tablenotemark{b}\dotfill & $0.48\pm0.02$ & $0.59$ & $0.59\pm0.02$ & $0.55$ & $0.55\pm0.05$ \nl
\enddata
\tablecomments{FW87, \citet{feast87}; F95, \citet{fernie95};
TSR03, \citet{tammann03}; HST03, \citet{hoyle03}.}
\tablenotetext{a}{Errors from differential reddening are not included.}
\tablenotetext{b}{Average value for CEa Cas, CEb Cas, and CF Cas.}
\end{deluxetable*}

\begin{figure}
\epsscale{1.1}
\plotone{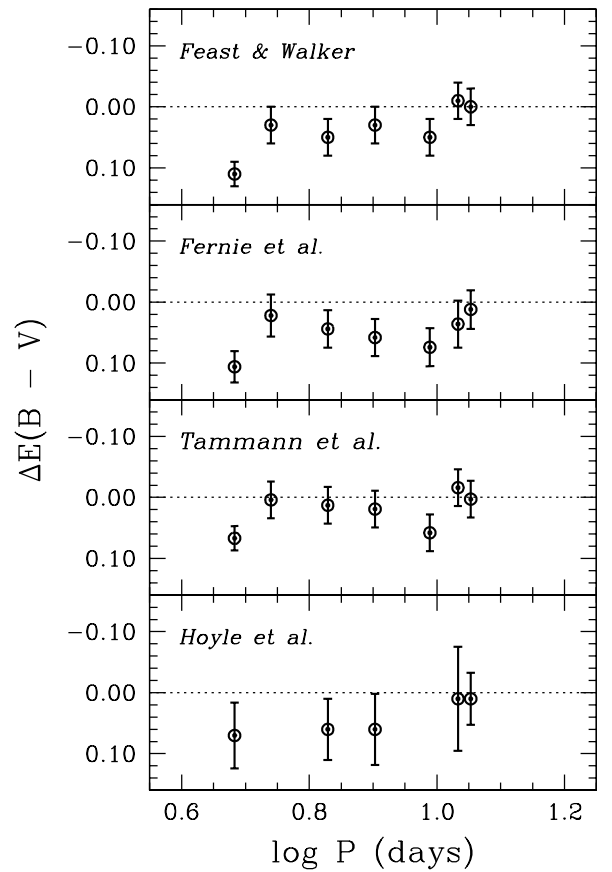}
\caption{Comparison of Cepheid $E(B - V)$ with those in \citet{feast87},
\citet{fernie95}, \citet{tammann03}, and \citet{hoyle03}.  The differences
are in the sense of their values minus those in this paper.  Error bars
represent our error estimates in the top three panels, but they show
a quadrature sum from two studies in the bottom panel (Hoyle et~al.).
\label{fig:comp.ebv}}
\end{figure}

Our short distances are mainly due to our lower $E(B - V)$ values than
those in the previous work.  This is because the MS-fitting distances
are correlated with reddening by $\Delta (m - M)_0 / E(B - V) \sim 2$
(Paper~III).  The top panel in Figure~\ref{fig:comp.ebv} compares
the Cepheid reddening with those in \citet{feast87}, and individual
estimates are listed in Table~\ref{tab:comp.ebv}.  They estimated the OB
star reddening at the location of the Cepheids and then transformed it to
the value appropriate for the Cepheid using a color-dependent reddening
law (e.g., eqs.~[\ref{eq:rv}] and [\ref{eq:ebv}]).  The mean difference from
our study is $\langle E(B - V) \rangle = 0.04\pm0.02$~mag, our values
being smaller on average.  The difference in $E(B - V)$ then
approximately matches the size expected to produce the difference in
distance modulus ($\sim0.1$~mag).

Individual reddening estimates from \citet{fernie95}, \citet{tammann03},
and \citet{hoyle03} are also listed in Table~\ref{tab:comp.ebv}, and
comparisons with our estimates are shown in Figure~\ref{fig:comp.ebv}.
Many studies used an extensive compilation of Cepheid reddening
in the David Dunlap Observatory (DDO) database of Galactic classical
Cepheids,\footnote{http://www.astro.utoronto.ca/DDO/research/cepheids/}
which is from heterogeneous methods \citep{fernie90,fernie94,fernie95}.
The mean difference with their data is $\Delta \langle E(B - V) \rangle
= 0.05$ in the sense that our reddening is smaller.  \citeauthor{tammann03}
adopted these reddening estimates for their cluster Cepheids but made
slight corrections to remove a mild correlation of the color excess
with residuals from the period-color relation.  This results in a reduced
difference with our estimates by $\Delta \langle E(B - V) \rangle = 0.02$.
However, the results from \citeauthor{tammann03} are internally less
consistent, in the sense
that the extinction corrections on the Cepheid magnitudes and the
reddening values for the MS-fitting distance estimates were not derived
from the same methods.  Finally, we compared our reddening values with those
in \citeauthor{hoyle03}, which were based on the cluster $UBV$
color-color diagram.  Their values are also systematically
larger than ours by $\Delta \langle E(B - V) \rangle = 0.04\pm0.01$.

There are a limited number of reddening estimates from the spectral
types in the literature.  \citet{kraft58} estimated $E(B - V) = 0.45$
from one B-type star in NGC~129, while \citet{arp59} found $\langle
E(B - V) \rangle = 0.53\pm0.01$ from seven stars in the vicinity of its
Cepheid (DL~Cas).  In M25, \citet{sandage60} estimated $\langle E(B - V)
\rangle = 0.43\pm0.05$ from the spectral types given by \citet{feast57}
and $0.43\pm0.06$ by \citet{wallerstein57}, but found a slightly larger
reddening, $0.46\pm0.04$, in the vicinity of its Cepheid (U~Sgr).  Finally,
\citet{kraft58} estimated $\langle E(B - V) \rangle = 0.49\pm0.02$ from
six stars in NGC~7790.  These estimates are generally found between
the previous estimates and our new values.  Because of low Galactic
latitudes of the clusters ($|b| < 5\arcdeg$), the dust map of
\citet{schlegel98} could not provide meaningful upper limits on
the color excess.

We contend that our reddening estimates are more reliable than the previous
values, which mostly relied on the solar metallicity ZAMS-fitting on
$UBV$ color-color diagrams.  In contrast to these studies, we took into
account metallicity and age effects on MS fitting, and our estimates are
based on simultaneous fits over a wider range of spectral bands,
including near-infrared $JHK_s$.  We also note that photometric zero-point
errors in $U$-band \citep[e.g.,][]{bessell05} could introduce significant
uncertainties in color excesses derived from $UBV$ color-color diagrams.

\section{Galactic Period-Luminosity Relations}\label{sec:pl}

With cluster parameters derived in the previous section, we are now
in a position to estimate absolute magnitudes of Cepheids and to
redetermine the Galactic {\it P-L} relations.  We show that our new
{\it P-L} relation in $V$ is generally fainter than in previous studies.
However, it is in good agreement with recent parallax studies
\citep{benedict07,vanleeuwen07} when the Wesenheit magnitudes are employed.

\subsection{Absolute Magnitudes of Cepheids}\label{sec:mv}

\begin{deluxetable*}{lcccccccc}
\tablewidth{5in}
\tabletypesize{\scriptsize}
\tablecaption{Cepheid Absolute Magnitudes\label{tab:cepheid}}
\tablehead{
  \colhead{Cepheids} &
  \colhead{$\log{P}$} &
  \colhead{$\langle M_B \rangle$} &
  \colhead{$\langle M_V \rangle$} &
  \colhead{$\langle M_{I_C} \rangle$} &
  \colhead{$\langle M_J \rangle$} &
  \colhead{$\langle M_H \rangle$} &
  \colhead{$\langle M_{K_s} \rangle$} &
  \colhead{$\langle M_{W(VI)} \rangle$\tablenotemark{a}}
}
\startdata
DL~Cas  \dotfill &$0.903$ & $-2.988$ & $-3.673$ & $-4.364$ & \nodata  & \nodata  & \nodata  & $-5.285$ \nl
V~Cen   \dotfill &$0.740$ & $-2.754$ & $-3.371$ & $-4.042$ & $-4.557$ & $-4.848$ & $-4.951$ & $-4.974$ \nl
TW~Nor  \dotfill &$1.033$ & $-3.187$ & $-3.957$ & $-4.716$ & $-5.287$ & $-5.580$ & $-5.775$ & $-5.676$ \nl
V340~Nor\dotfill &$1.053$ & $-2.879$ & $-3.720$ & $-4.505$ & $-5.142$ & $-5.500$ & $-5.641$ & $-5.604$ \nl
S~Nor   \dotfill &$0.989$ & $-2.864$ & $-3.685$ & $-4.518$ & $-5.128$ & $-5.470$ & $-5.593$ & $-5.688$ \nl
U~Sgr   \dotfill &$0.829$ & $-2.829$ & $-3.535$ & $-4.270$ & $-4.794$ & $-5.086$ & $-5.196$ & $-5.294$ \nl
CEa~Cas \dotfill &$0.711$ & $-2.458$ & $-3.129$ & $-3.951$ & \nodata  & \nodata  & \nodata  & $-5.090$ \nl
CEb~Cas \dotfill &$0.651$ & $-2.306$ & $-2.998$ & $-3.731$ & \nodata  & \nodata  & \nodata  & $-4.739$ \nl
CF~Cas  \dotfill &$0.688$ & $-2.189$ & $-2.911$ & $-3.667$ & \nodata  & \nodata  & \nodata  & $-4.707$ \nl
\enddata
\tablenotetext{a}{Wesenheit index from \citet{freedman01}: $\langle W (VI) \rangle \equiv
\langle V \rangle - 2.45 (\langle V \rangle - \langle I_C \rangle)$.}
\end{deluxetable*}

The absolute $V$ magnitude was computed for each Cepheid from
\begin{eqnarray}
\langle M_V \rangle
= \langle m_V \rangle - (m - M)_0 - R_V E(B - V),\label{eq:mv}
\label{eq:mv}
\end{eqnarray}
where $\langle m_V \rangle$ is the observed intensity-mean magnitude and
$(m - M)_0$ is the extinction-corrected distance modulus.  We used the
same color-dependent prescriptions for $R_V$ and $E(B - V)$ as those for
MS dwarfs (eqs.~[\ref{eq:rv}] and [\ref{eq:ebv}]).  The absolute
magnitudes in other filter passbands were computed from the CCM89
reddening law at each cluster's $R_{V,0}$.  Table~\ref{tab:cepheid}
lists absolute magnitudes in $BVI_CJHK_s$ for all Cepheids in this
study.  In the last column we also show the reddening-insensitive
Wesenheit magnitude, $\langle W_{VI} \rangle \equiv \langle V \rangle -
2.45 (\langle V \rangle - \langle I_C \rangle)$, which was adopted in the
{\it HST} Key Project.\footnote{We simply adopted $A_V / E(V - I_C) =
2.45$ from \citet{freedman01} to directly compare our results with those
of other studies, which also employed the same relation.   We would
derive $A_V / E(V - I_C) = 2.56$ for the Cepheids with $\langle B_0
\rangle - \langle V_0 \rangle = 0.7$ at $R_V = 3.35$, computed
from the average $R_{V,0}$ of our cluster sample.  This difference in
$A_V / E(V - I_C)$ has a negligible effect on $W_{VI}$ of Cepheids.
It is noted that \citet{macri01b} tested the CCM89 reddening law using
$VIH$ photometry for 70 extragalactic Cepheids in 13 galaxies.  They
showed that a mean color-excess ratio of $E(V - H) / E(V - I)$ is in
good agreement with the CCM89 predicted value.}

\begin{deluxetable*}{lcccccccc}
\tablewidth{6in}
\tabletypesize{\scriptsize}
\tablecaption{Error Budget in $\langle M_V \rangle$\label{tab:error2}}
\tablehead{
  \colhead{Source of Error} &
  \colhead{$\Delta$Quantity} &
  \colhead{NGC~129} &
  \colhead{NGC~5662} &
  \colhead{Lyng{\aa}~6} &
  \colhead{NGC~6067} &
  \colhead{NGC~6087} &
  \colhead{M25} &
  \colhead{NGC~7790} \\
  \colhead{} &
  \colhead{} &
  \colhead{DL~Cas} &
  \colhead{V~Cen} &
  \colhead{TW~Nor} &
  \colhead{V340~Nor} &
  \colhead{S~Nor} &
  \colhead{U~Sgr} &
  \colhead{CEa, CEb, CF~Cas}
}
\startdata
${\rm [M/H]}$\tablenotemark{a}\dotfill & \nodata & $\mp0.032$ & $\mp0.028$ & $\mp0.056$ & $\mp0.054$ & $\mp0.028$ & $\mp0.045$ & $\mp0.054$ \nl
$\log{t}$ (Myr)\dotfill  & $\pm0.2$   & $\pm0.037$ & $\pm0.007$ & $\mp0.014$ & $\pm0.057$ & $\pm0.061$ & $\pm0.051$ & $\pm0.021$ \nl
Helium ($\Delta Y$)\dotfill & $\pm0.010$ & $\mp0.030$ & $\mp0.030$ & $\mp0.030$ & $\mp0.030$ & $\mp0.030$ & $\mp0.030$ & $\mp0.030$ \nl
$\Delta (B-V)$\dotfill   & $\pm0.025$ & $\mp0.009$ & $\mp0.001$ & $\pm0.014$ & $\mp0.007$ & $\mp0.002$ & $\mp0.002$ & $\pm0.001$ \nl
$\Delta (V-I)_C$\dotfill & $\pm0.025$ & \nodata    & \nodata    & \nodata    & $\mp0.012$ & \nodata    & \nodata    & $\mp0.011$ \nl
$\Delta V$\dotfill       & $\pm0.025$ & $\pm0.048$ & $\pm0.025$ & $\pm0.028$ & $\pm0.030$ & $\pm0.022$ & $\pm0.013$ & $\pm0.026$ \nl
$\Delta J$\dotfill       & $\pm0.011$ & $\mp0.014$ & $\mp0.004$ & $\mp0.002$ & $\mp0.003$ & $\mp0.005$ & $\mp0.005$ & $\pm0.000$ \nl
$\Delta H$\dotfill       & $\pm0.007$ & $\pm0.001$ & $\mp0.001$ & $\mp0.002$ & $\mp0.001$ & $\mp0.002$ & $\mp0.002$ & $\pm0.001$ \nl
$\Delta K_s$\dotfill     & $\pm0.007$ & $\pm0.010$ & $\pm0.002$ & $\mp0.002$ & $\pm0.002$ & $\pm0.004$ & $\pm0.004$ & $\mp0.001$ \nl
Fitting\dotfill          & \nodata    & $\pm0.023$ & $\pm0.044$ & $\pm0.094$ & $\pm0.032$ & $\pm0.031$ & $\pm0.040$ & $\pm0.023$ \nl
$\Delta m_V$\tablenotemark{b}\dotfill & $\pm0.025$ & $\pm0.025$ & $\pm0.025$ & $\pm0.025$ & $\pm0.025$ & $\pm0.025$ & $\pm0.025$ & $\pm0.025$ \nl
$\sigma_{E(B-V)}$\tablenotemark{c}\dotfill & \nodata    & $\pm0.347$ & $\pm0.066$ & $\pm0.316$ & $\pm0.159$ & $\pm0.197$ & $\pm0.379$ & $\pm0.041$ \nl
Total\dotfill & \nodata    & $\pm0.357$ & $\pm0.097$ & $\pm0.338$ & $\pm0.187$ & $\pm0.216$ & $\pm0.390$ & $\pm0.089$ \nl
\enddata
\tablecomments{$\Delta (B-V)$, $\Delta (V-I)_C$, $\Delta V$, $\Delta J$, $\Delta H$,
and $\Delta K_s$ represent zero-point errors in the cluster photometry.}
\tablenotetext{a}{Individual errors taken from Table~\ref{tab:list}.}
\tablenotetext{b}{Photometric zero-point error in the intensity mean magnitude of a Cepheid.}
\tablenotetext{c}{Differential reddening over a cluster field.}
\end{deluxetable*}

Table~\ref{tab:error2} shows systematic errors in $\langle M_V \rangle$
of the Cepheids.  As in Table~\ref{tab:error}, we estimated
the size of errors by performing MS fits for alternate values of
systematic error quantities.  These are shown in the first and
second columns, which are the same as those in Table~\ref{tab:error}
except the last two rows:  $\Delta m_V$ and $\sigma_{E(B-V)}$.
The $\Delta m_V$ represents a systematic error in the Cepheid photometry,
which was treated as independent of the cluster photometry and any
of the MS-fitting process.  Note that $(m - M)_0$, $R_V$, and
$E(B - V)$ in equation~(\ref{eq:mv}) are correlated with each other
via MS fitting.

The $\sigma_{E(B-V)}$ in Table~\ref{tab:error2} represents the error
in the reddening for individual Cepheids.  In equation~(\ref{eq:mv})
we adopted the Cepheid reddening as the mean cluster reddening value.
However, reddening for individual stars becomes more uncertain as
we deal with more differentially reddened clusters.  For these
clusters, the colors for individual stars scatter along a reddening
vector, and this can be detected by a large size of the MS width above
photometric precision.

\begin{deluxetable}{lcc}
\tablewidth{2in}
\tablecaption{Results from Artificial Cluster Tests\label{tab:art}}
\tablehead{
  \colhead{Cluster}  &
  \colhead{$\sigma_{E(B - V)}$\tablenotemark{a}} &
  \colhead{$\Delta \langle M_V \rangle$\tablenotemark{b}}
}
\startdata
NGC~129    \dotfill & $0.105$ & $+0.04\pm0.14$ \nl
NGC~5662   \dotfill & $0.020$\tablenotemark{c} & $-0.06\pm0.05$ \nl
Lyng{\aa}~6\dotfill & $0.100$ & $+0.04\pm0.14$ \nl
NGC~6067   \dotfill & $0.050$ & $-0.06\pm0.07$ \nl
NGC~6087   \dotfill & $0.055$ & $-0.06\pm0.08$ \nl
M25        \dotfill & $0.120$ & $+0.02\pm0.15$ \nl
NGC~7790   \dotfill & $0.013$\tablenotemark{c} & $-0.05\pm0.04$ \nl
\enddata
\tablenotetext{a}{Size of differential reddening inferred from
artificial cluster tests, assuming a normal distribution of $E(B - V)$.}
\tablenotetext{b}{This value should be subtracted from the estimated
$\langle M_V \rangle$.  Errors are semi-interquartile ranges.}
\tablenotetext{c}{Not corrected from the artificial cluster tests.}
\end{deluxetable}

To estimate the size of differential reddening for each cluster, we
first computed the standard deviation of $B - V$ colors in the upper MS
from Table~\ref{tab:ebv}.  Since the photometric filtering likely
reduces the size of the dispersion, we corrected for it from artificial
cluster CMD tests (Paper~III).  We generated solar metallicity, 80~Myr
isochrones with photometric errors of $0.02$~mag in colors and magnitudes.
We used the \citet{salpeter55} mass function for the primaries and
a flat mass function for the secondaries for a 40\% binary
fraction.\footnote{The binary fraction is defined as the number of binaries
divided by the total number of systems in the considered fitting range.}
Single stars and binaries were then randomly displaced from the MS
assuming a normal distribution of individual reddening with a standard deviation
$\sigma_{E(B-V)}$.  After applying the photometric filtering, we then estimated
the size of the reduction in dispersion as a function of $\sigma_{E(B-V)}$.
In this way, we inferred $\sigma_{E(B-V)}$ for each cluster
from the observed dispersion in $B - V$, which are shown in the second
column of Table~\ref{tab:art}.  We made no corrections to those for NGC~5662
and NGC~7790 because the dispersions in $B - V$ were equal to or smaller
than our assumed photometric precision ($\sigma_{B-V} = 0.02$).

In addition to previously known clusters with differential reddening
(NGC~129, NGC~5662, and M25), we also found that the remaining clusters
except NGC~7790 have larger dispersions than photometric errors.  While
this could be due to remaining binaries, background stars, rapid
rotators, or an underestimation of the photometric errors, we took
the excess dispersion in Table~\ref{tab:art} as a characteristic
measure of differential reddening as a conservative error estimate.
The $\sigma_{E(B-V)}$ in Table~\ref{tab:art} was then multiplied by
$R_V$ to estimate the error in $\langle M_V \rangle$, which is shown in
Table~\ref{tab:error2}.  For all clusters except NGC~7790, differential
reddening is the largest error source in $\langle M_V \rangle$.

In previous studies, local Cepheid reddening was often determined
using photometry of neighboring stars.  We experimented with this
approach for M25.  We estimated individual Cepheid's reddening
on $(B - V, V)$ and assumed that the dispersion around the best-fitting
isochrone is solely due to differential reddening.  However, we could
not place a strong constraint on the local color excess for U~Sgr
because there were not a sufficient number of stars with good $E(B - V)$.
In addition, cluster binaries cannot be easily distinguished from
highly reddened single stars on CMDs, which could lead to an overestimation
of reddening.

\subsection{The Galactic {\it P-L} Relations}\label{sec:galpl}

Our Cepheid sample does not span a wide range of periods, which leads to
a large error in {\it P-L} slopes.  Instead, we adopted the {\it P-L} slopes
derived from LMC Cepheids, assuming that the {\it P-L} relations for Cepheids
in the Galaxy and in the LMC have the same slopes.  Some investigators have
claimed that the Galactic {\it P-L} relations are steeper than those of
the LMC Cepheids, and that the LMC {\it P-L} relations have a break at
$P\approx10$~days, possibly due to different metal contents of Cepheids
\citep{kanbur04,sandage04,ngeow05}.  However, \citet{macri06} recently
showed that {\it P-L} relations from the LMC Cepheids in the OGLE-II
catalog \citep{udalski99a} are a good fit to Cepheids in the two
fields of NGC~4258, which have approximately the LMC and the Galactic
metal abundances (see \S~\ref{sec:4258}).

We derived LMC {\it P-L} slopes using fundamental mode Cepheids in the
OGLE-II catalog \citep{udalski99a} and those in \citet{persson04}.
The OGLE-II Cepheids have an average period of $\langle \log{P} {\rm (days)}
\rangle \approx 0.6$ with about half of them being in the same
period range as our Galactic Cepheid sample.  The {\it P-L} relations in
the LMC from the OGLE-II database were originally derived by \citet{udalski99b}
and then revised according to the new photometric calibration \citep{udalski00}.
We rederived the period-{\it apparent} magnitude relations in $BVI_C$
using the revised data from the OGLE-II Internet
archive\footnote{http://www.astrouw.edu.pl/\~{}ogle/ogle2/cep\_lmc.html}
after iterative $2.5\sigma$ rejection from 680 Cepheids with $\log{P} >
0.4$ \citep{udalski99b}.  We also derived similar relations in $JHK_s$
on the LCO system from the photometry in \citet{persson04}.  As in their
analysis, we excluded four stars (HV~883, HV~2447, and HV~2883 have
periods longer than 100 days, and HV~12765 is about 0.2~mag brighter
than the mean {\it P-L} relation), leaving 88 Cepheids with $\log{P} > 0.4$.

We then derived the Galactic {\it P-L} relations in $BVI_CJHK_s$ after
correcting for interstellar extinction:
\begin{eqnarray}
\langle M_B     \rangle = (-3.169\pm0.085) - (2.503\pm0.048) (\log{P} - 1),\nonumber\\
\langle M_V     \rangle = (-3.932\pm0.066) - (2.819\pm0.032) (\log{P} - 1),\nonumber\\
\langle M_{I_C} \rangle = (-4.712\pm0.057) - (3.004\pm0.021) (\log{P} - 1),\nonumber\\
\langle M_J     \rangle = (-5.271\pm0.076) - (3.148\pm0.053) (\log{P} - 1),\nonumber\\
\langle M_H     \rangle = (-5.593\pm0.069) - (3.233\pm0.044) (\log{P} - 1),\nonumber\\
\langle M_{K_s} \rangle = (-5.718\pm0.064) - (3.282\pm0.040) (\log{P} - 1).\nonumber
\label{eq:galpl}
\end{eqnarray}
The slope errors are those estimated from the LMC Cepheids.  The zero-point
errors were estimated from the magnitude dispersion among the Cepheids,
assuming that the magnitude errors are uncorrelated with each
other.  We took the average magnitudes and periods of the three Cepheids
in NGC~7790 as one data point.  Cepheids in the two northern clusters
(NGC~129 and NGC~7790) have no photometry in \citet{laney92}, so the
{\it P-L} relations in $JHK_s$ were derived from five clusters.

\begin{figure}
\epsscale{1.1}
\plotone{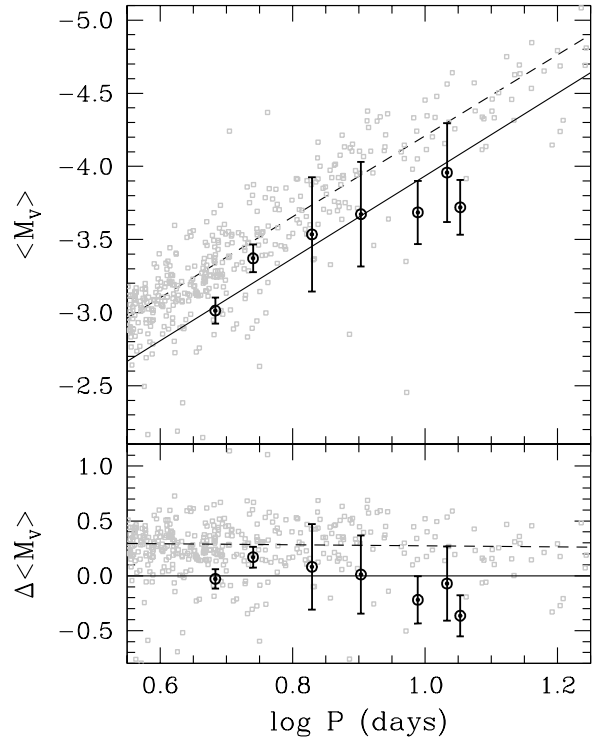}
\caption{Galactic {\it P-L} relation in $V$.  The bull's-eyes are our
Galactic calibrators with their $1\sigma$ errors, and the solid line is
a fit to these points assuming the LMC {\it P-L} slope.  The small boxes
are LMC fundamental mode Cepheids from the OGLE-II survey, assuming
$(m - M)_0 = 18.50$ and the OGLE-II reddening map.  The dashed line is
a fit to these points.\label{fig:pl}}
\end{figure}

\begin{figure*}
\epsscale{0.8}
\plotone{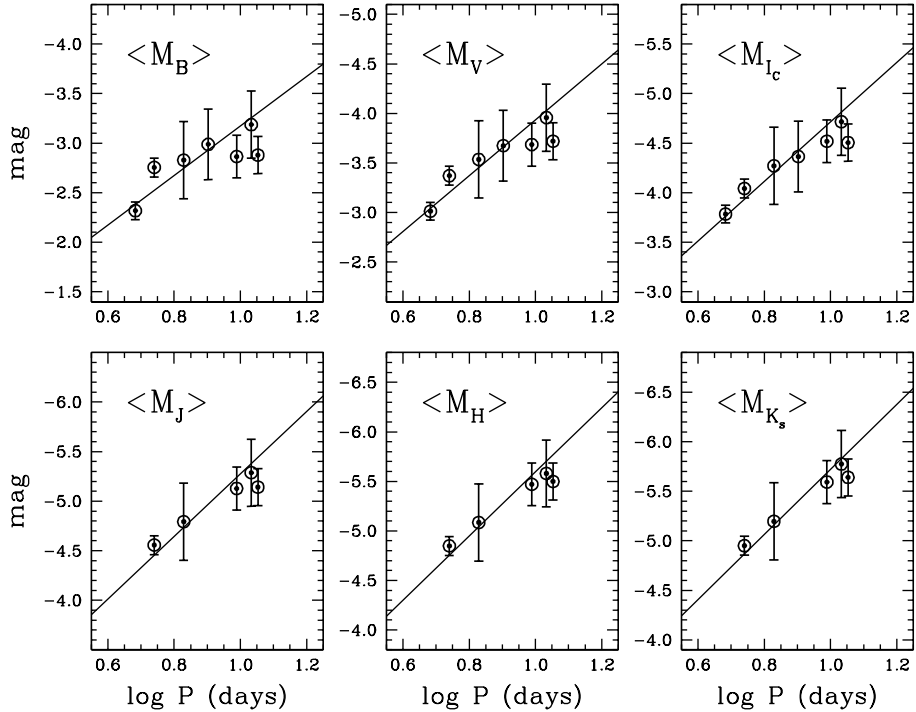}
\caption{Same as Fig.~\ref{fig:pl}, but in all $BVI_CJHK_s$.
\label{fig:pl2}}
\end{figure*}

Figures~\ref{fig:pl} and \ref{fig:pl2} display {\it P-L} relations for
our Galactic Cepheids ({\it bull's-eyes}) in all filter passbands.  In
$V$-band the $\chi^2$ of the fit is $8.0$ for 6 degrees of freedom,
which indicates our reasonable error estimation.  The standard deviation
of the Galactic Cepheids around the {\it P-L} relation is $0.18$~mag, which
is comparable to that of the LMC counterparts \citep{udalski99b}.  This
value is smaller than previous estimates for the Galactic Cepheids
from MS fitting: \citet{tammann03} found 0.26~mag for 25 Cepheids
\citep[see also][]{sandage04}, and \citet{fouque03} estimated 0.27~mag
from 24 Cepheids.

In Figure~\ref{fig:pl} the OGLE-II Cepheids are also shown at
a distance modulus of 18.50 mag, which was adopted in the {\it HST} Key
Project ({\it small boxes}).  Magnitudes of these Cepheids were corrected
for extinction based on the OGLE-II reddening map, which was derived
from the $I_C$-band magnitudes of red clump stars \citep{udalski99a}.
As seen in the figure, the LMC {\it P-L} relation ({\it dashed line}) is
brighter than our Galactic {\it P-L} relation ({\it solid line}) by
$\sim0.3$~mag.  This is due to metallicity and reddening effects, as well
as the LMC distance as discussed in \S~\ref{sec:lmc}.

\subsection{Comparison with Previous Studies}\label{sec:plcomp}

\begin{deluxetable*}{lccccc}
\tablewidth{4in}
\tabletypesize{\scriptsize}
\tablecaption{Comparison of Cepheid $\langle M_V \rangle$ with Published Results\label{tab:comp.mv}}
\tablehead{
  \colhead{Cluster} &
  \colhead{This Study} &
  \colhead{FW87\tablenotemark{a}} &
  \colhead{TSR03\tablenotemark{a}} &
  \colhead{HST03} &
  \colhead{G05}
}
\startdata
DL Cas  \dotfill & $-3.67\pm0.36$ & $-3.92$ & $-3.77$ & $-3.69$ & \nodata  \nl
V Cen   \dotfill & $-3.37\pm0.10$ & $-3.30$ & $-3.19$ & \nodata & $-3.45$ \nl
TW Nor  \dotfill & $-3.96\pm0.34$ & $-3.85$ & $-3.72$ & $-3.79$ & \nodata  \nl
V340 Nor\dotfill & $-3.72\pm0.19$ & $-3.86$ & $-3.85$ & $-3.92$ & $-3.91$ \nl
S Nor   \dotfill & $-3.69\pm0.22$ & $-4.01$ & $-4.00$ & \nodata & $-4.21$ \nl
U Sgr   \dotfill & $-3.54\pm0.39$ & $-3.83$ & $-3.66$ & $-3.88$ & $-3.62$ \nl
CEa,CEb,CF Cas\tablenotemark{b}\dotfill & $-3.01\pm0.09$ & $-3.36$ & $-3.37$ & $-3.34$ & \nodata \nl
\enddata
\tablecomments{FW87, \citet{feast87}; TSR03, \citet{tammann03};
HST03, \citet{hoyle03}; G05, \citet{gieren05}.}
\tablenotetext{a}{Magnitudes revised assuming the same Pleiades distance,
$(m - M)_0 = 5.63$, as in this paper.}
\tablenotetext{b}{Average value for CEa Cas, CEb Cas, and CF Cas.}
\end{deluxetable*}

\begin{figure}
\epsscale{1.1}
\plotone{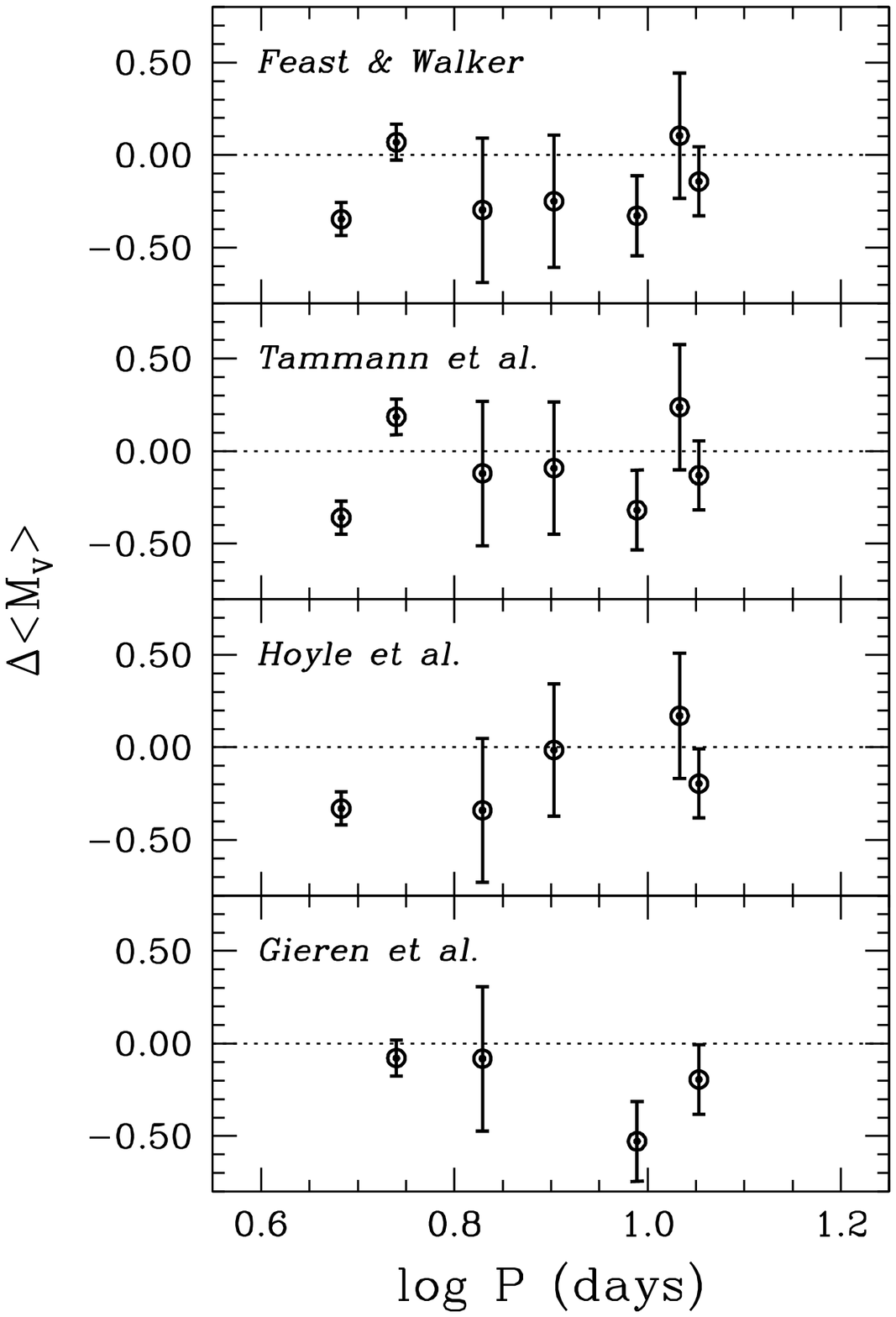}
\caption{Comparison of Cepheid absolute magnitudes with those in
\citet{feast87}, \citet{tammann03}, \citet{hoyle03}, and \citet{gieren05}.
The differences are in the sense of their magnitudes minus those in this
paper.  The error bars represent our error estimates only.  Individual
$\langle M_V \rangle$ estimates in the top two panels were revised
assuming our Pleiades distance scale.\label{fig:comp.mv}}
\end{figure}

Figure~\ref{fig:comp.mv} compares our $\langle M_V \rangle$ with those of
\citet{feast87}, \citet{tammann03}, \citet{hoyle03}, and \citet{gieren05}.
Error bars represent our error
estimates only.  We revised absolute magnitudes in \citeauthor{feast87}
and \citeauthor{tammann03} assuming our Pleiades distance ($5.63$~mag in
distance modulus).  Individual $\langle M_V \rangle$ estimates are shown
in Table~\ref{tab:comp.mv}.  We restricted our comparison to $\langle M_V
\rangle$ since absolute magnitudes in other passbands are correlated with
each other.  The weighted average differences are $\Delta \langle M_V
\rangle = -0.17\pm0.07$, $-0.09\pm0.09$, $-0.14\pm0.10$, and $-0.22\pm0.11$
from \citeauthor{feast87}, \citeauthor{tammann03}, \citeauthor{hoyle03},
and \citeauthor{gieren05}, respectively, in the sense that our estimates
are fainter on average.  This is mainly due to the fact that our distance
and reddening estimates are smaller on average than those in the previous
studies (\S~\ref{sec:comp}).

\begin{figure*}
\epsscale{1.0}
\plottwo{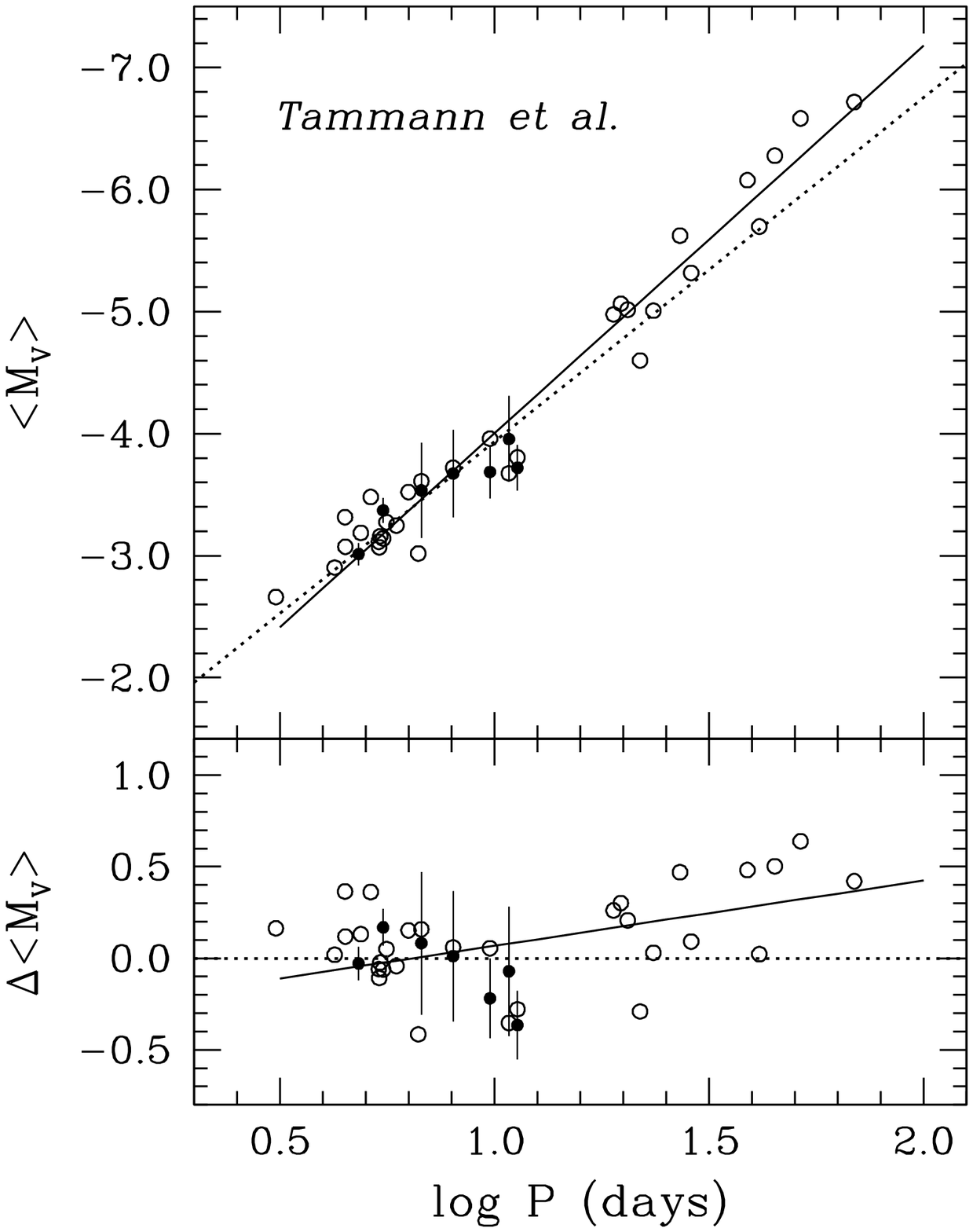}{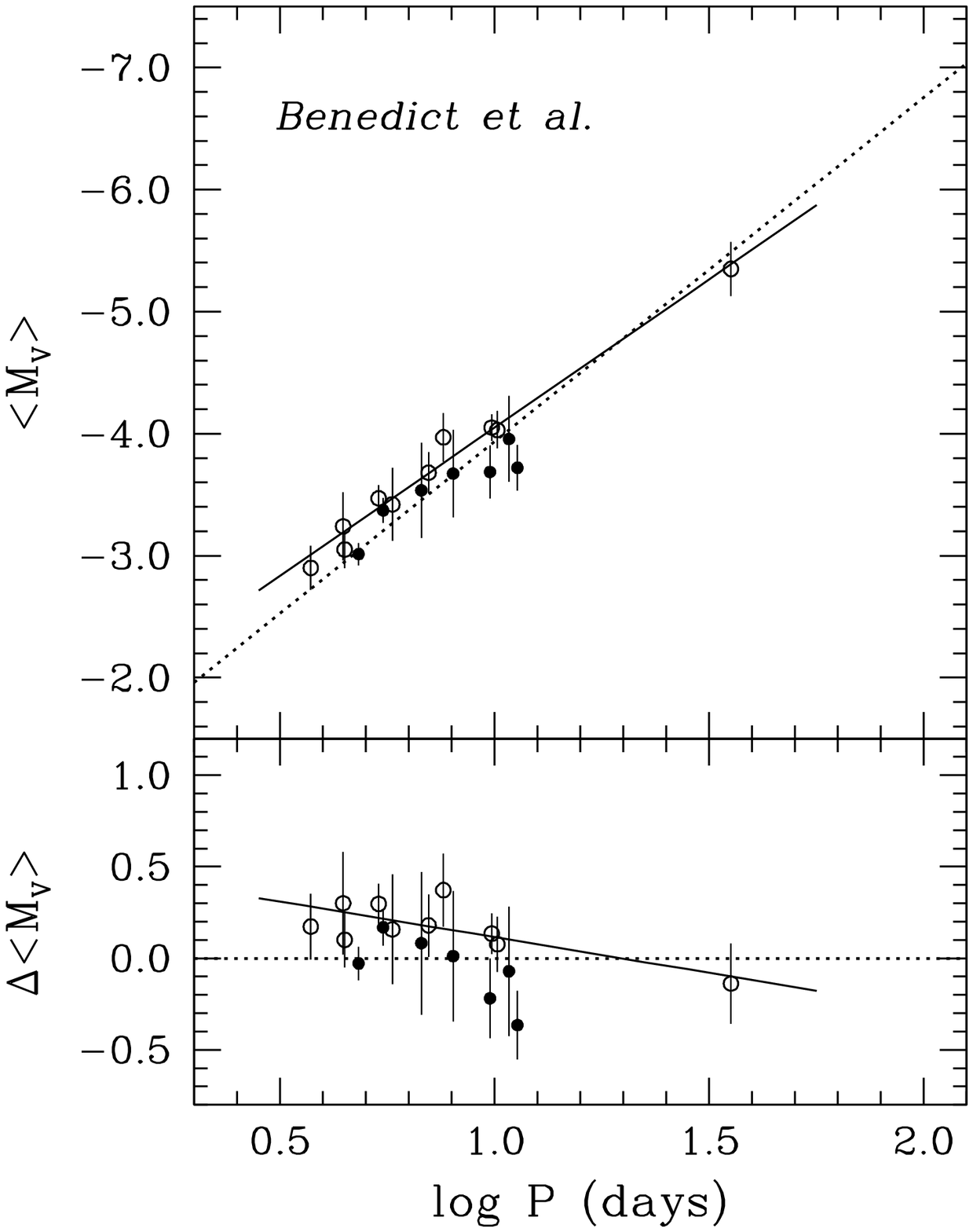}
\caption{Comparison of the $V$-band Galactic {\it P-L} relation with
those in \citet[][{\it left}]{tammann03} and \citet[][{\it right}]{benedict07}.
The open circles are Cepheids in these studies,
and the solid lines are their best-fit {\it P-L} relations.  Our
Galactic calibrators are shown as filled circles, and the dotted lines
represent our {\it P-L} relation.  Magnitudes in \citeauthor{tammann03}
were revised assuming the same Pleiades distance scale as in this paper.
\label{fig:plcomp}}
\end{figure*}

In addition, we also made a comparison with all cluster Cepheids in
\citet{tammann03} as shown in the left panel of Figure~\ref{fig:plcomp}.
As found in the above Cepheid-to-Cepheid comparison, the difference with
their $\langle M_V \rangle$ is statistically insignificant for Cepheids
with $\log{P} \la 1.1$.  However, their {\it P-L} relation is steeper
than ours because their long-period Cepheids at $\log{P} \ga 1.1$ are
brighter than expected from our {\it P-L} relation.  If we assume their
{\it P-L} slope for our sample, we would derive $\langle M_V \rangle =
-4.01$ for 10-day period Cepheids, which is $0.25$~mag fainter than
their {\it P-L} zero point.

In the right panel of Figure~\ref{fig:plcomp}, we compare our Galactic
{\it P-L} relation in $V$ with that of the recent {\it HST} parallax study
\citep{benedict07}, which provided accurate parallaxes (average $8\%$)
for nine Galactic Cepheids using the {\it HST} Fine Guidance Sensor
\citep[see also][]{benedict02b}.  Their Cepheid parallaxes were estimated
with respect to photometric parallaxes of reference stars, which set the
absolute distance scale of each reference frame.  The magnitudes were
then corrected for interstellar extinction from either color-color relations
or the DDO Galactic Cepheid database.  We have no Cepheids in common with
their study since their distance measurements were focused on nearby
field Cepheids.  If we assume their {\it P-L} slope, our 10-day period
Cepheids would be $0.20$~mag fainter than their zero point.  If the
difference is solely due to the error in our reddening estimates,
our estimates would have been underestimated by $\Delta \langle E(B - V)
\rangle \sim 0.04$.  However, it is noted that reddening can be
determined far more accurately for cluster Cepheids than it can for
field stars.  Our error bars in Figure~\ref{fig:plcomp} are slightly
larger than those of the {\it HST} study because of our conservative
error estimates for the extinction correction.

\begin{figure}
\epsscale{1.1}
\plotone{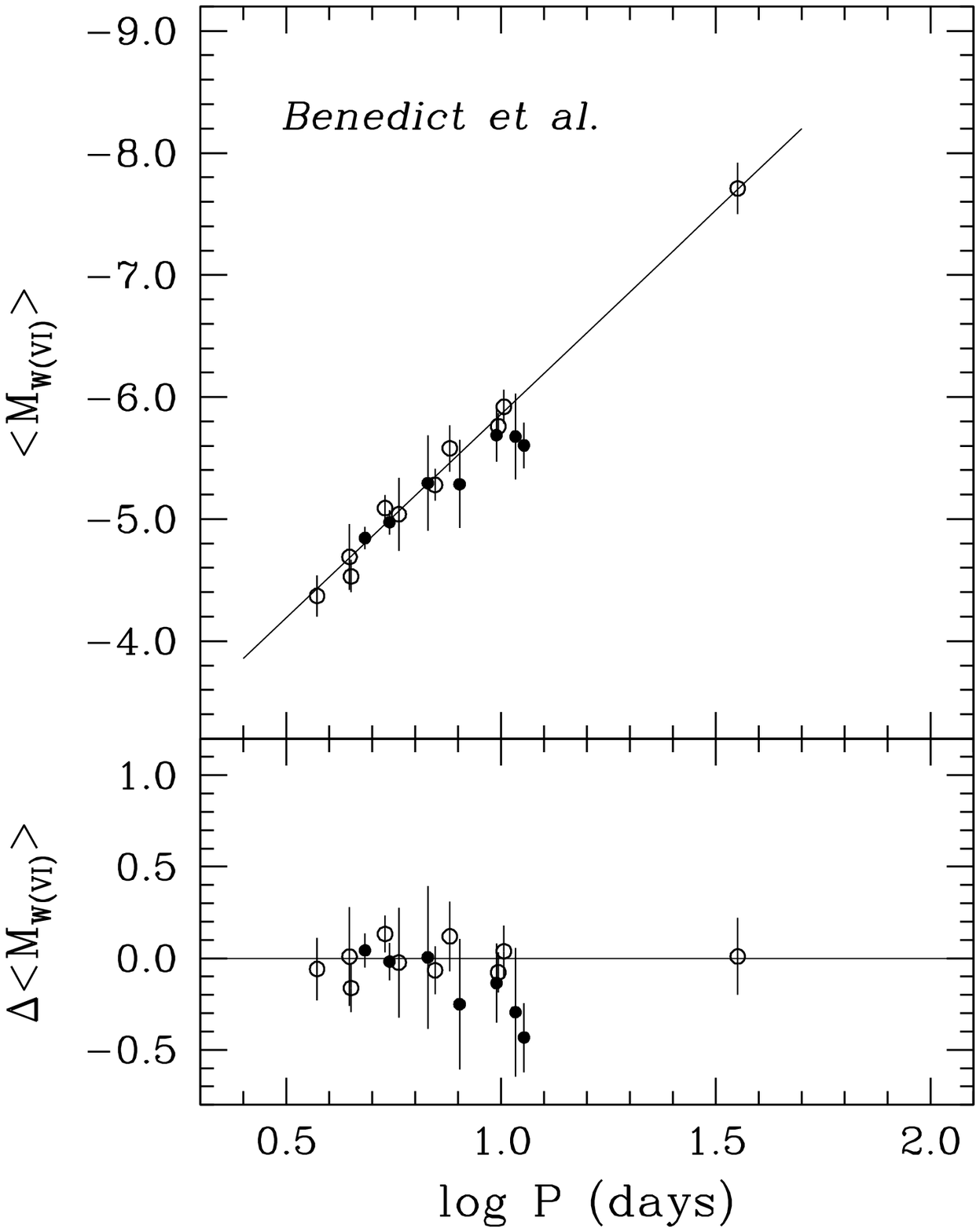}
\caption{Same as Fig.~\ref{fig:plcomp}, but a comparison of the {\it
P-L} relation in $M_{W(VI)}$ with that of \citet[][{\it solid line}]
{benedict07}.  Residuals are shown with respect to their {\it P-L}
relation in the bottom panel.\label{fig:plmw}}
\end{figure}

On the other hand, the {\it HST}-based {\it P-L} relation is in better
agreement with ours when the Wesenheit magnitude $W(VI)$ is used as
shown in Figure~\ref{fig:plmw}.  The weighted mean difference of our
Galactic Cepheids from their best-fitting {\it P-L} relation ({\it solid
line}) is $\Delta \langle M_{W(VI)} \rangle = 0.05\pm0.06$ in the sense
that our {\it P-L} relation is fainter.  Since the Wesenheit magnitude
was designed to minimize the effect of errors in the reddening, the
agreement may indicate that the large difference in $\langle M_V \rangle$
is due to the difference in the distance estimation.

Previous studies based on {\it Hipparcos} data typically yielded
$\langle M_V \rangle \approx -4.2$ for 10-day period Cepheids
\citep{feast97,feast98,lanoix99,groenewegen00}, while \citet{luri98}
obtained $\langle M_V \rangle \approx -3.9$ from statistical parallaxes.
Recently, \citet{vanleeuwen07} reexamined the Galactic {\it P-L} relations
based on the new reduction of the {\it Hipparcos} astrometric data
\citep{vanleeuwen05b}, where improvements of up to a factor of 2 in
parallax accuracy have been achieved by reconstructing the satellite's
attitude \citep{vanleeuwen05a}.  They analyzed their data together with
the {\it HST} parallaxes to derive a {\it P-L} relation in $W(VI)$.
If we adopt their {\it P-L} slope in $W(VI)$, our {\it P-L} relation
would be $\Delta \langle M_{W(VI)} \rangle = 0.07\pm0.06$ fainter than
their best-fitting {\it P-L} relation.

\subsection{Systematic Errors in {\it P-L} relations}\label{sec:plsys}

\begin{deluxetable}{lc}
\tabletypesize{\scriptsize}
\tablewidth{2.5in}
\tablecaption{Error Budget in $\langle M_V \rangle$
for 10-day Period Cepheids from MS Fitting\label{tab:plerr}}
\tablehead{
  \colhead{Source of Error} &
  \colhead{$\Delta \langle M_V \rangle$}
}
\startdata
Isochrone Calibration\dotfill               & $\pm0.02$ \nl
Metallicity Scale ($\pm0.1$ dex)\dotfill    & $\pm0.11$ \nl
Age Scale [$\Delta \log{t} {\rm (Myr)} = \pm0.2$] \dotfill & $\pm0.03$ \nl
Reddening Laws\dotfill                      & $\pm0.05$ \nl
Differential Reddening + Binary             & $\pm0.05$ \nl
Total Systematic\dotfill                    & $\pm0.14$ \nl
Statistical\dotfill                         & $\pm0.07$ \nl
Combined (Systematic + Statistical)\dotfill & $\pm0.16$ \nl
\enddata
\end{deluxetable}

When determining the {\it P-L} relations in the previous section, we
assumed that the errors in $\langle M_V \rangle$ of individual Cepheids
(Table~\ref{tab:error2}) are uncorrelated with each other.  However,
there are a number of error sources that could systematically change
$\langle M_V \rangle$.  These are shown in Table~\ref{tab:plerr} with
their estimated size of error contributions to the {\it P-L} zero point,
which is described below in detail.  In short, the calibration error is
from the uncertainty in the adopted distance of the Pleiades (Paper~III).
Errors in the metallicity scale of $\pm0.1$~dex and the age scale of $\Delta
\log{t {\rm (Myr)}} = 0.2$ were adopted.  The error from the reddening
laws represents the case of using different color-dependent prescriptions
for $E(B - V)$ and $R_V$ from \citet{feast87} and \citet{laney93}.
We performed artificial cluster CMD tests to estimate a probable size
of a bias in the MS-fitting technique from differential reddening and
unresolved cluster binaries.

From the comparison with \citet{yong06} and \citet{andrievsky02},
we inferred that our adopted metallicity scale \citep{fry97} has
a $0.1$~dex systematic error (\S~\ref{sec:mehage}).  Metallicity changes
of $\pm0.1$~dex would vary the $\langle M_V \rangle$ of 10-day period
Cepheids from $-3.84$ to $-4.04$~mag.

In this paper, only the relative age scale is relevant with respect
to the Pleiades.  Our assumed age range in \S~\ref{sec:mehage} reflects
the cluster-to-cluster variations with respect to the adopted
age of $80$~Myr.  However, it is possible that the Cepheid clusters
are all systematically younger or older than the Pleiades.  This would
change the reddening values since the mean color of the upper MS
varies with the cluster age.  In fact, some of the CMDs have stars
that extend far above the top end of the $80$~Myr isochrone.  If
these bright stars are cluster members, they would suggest younger
ages than $80$~Myr.  Nevertheless, the {\it P-L} zero point is relatively
insensitive to the age.  If the overall age scale is uncertain at
the $1\sigma$ level of individual cluster ages ($50$--$120$~Myr), we
would have $\langle M_V \rangle = -3.90$ and $-3.96$~mag for 10-day
period Cepheids at $50$ and $120$~Myr, respectively.

We estimated the error from the reddening laws using different
prescriptions for the color-dependent $E(B - V)$ and $R_V$ in
\citet{feast87} and \citet{laney93}.  For the same $E(B - V)_0$ and
$R_{V,0}$ values derived from the cluster MS fitting, we found that
these prescriptions result in a fainter {\it P-L} relation by $\Delta
\langle M_V \rangle \approx 0.05$ than our default case
(eqs.~[\ref{eq:rv}] and [\ref{eq:ebv}]).

The average $R_{V,0}$ that we inferred from the clusters in this study is
$3.19$ (Table~\ref{tab:cluster}), which is in good agreement with the
average Galactic value for the diffuse interstellar medium, $R_V \sim
3.1$.  The formal standard deviation of $R_{V,0}$ is $0.17$, but
individual errors are too large to detect a cluster-to-cluster
variation.  Even if we adopt a different $R_{V,0}$, it would have
a reduced impact on $\langle M_V \rangle$.  This is because a larger
$R_V$ makes an MS-fitting distance shorter (Paper~III), but it makes the
extinction correction larger at the same time (eq.~[\ref{eq:mv}]).

In \S~\ref{sec:mv} we considered the case where the presence of
differential reddening makes it difficult to estimate individual
extinctions for Cepheids.  This is independent of the MS-fitting
technique in the sense that it does not affect any of the cluster
parameters.  However, differential reddening could also modify
cluster parameters because of cluster binaries.  Unresolved binaries
are typically brighter and redder than a single-star MS, which would make
a distance apparently shorter.  If there exists differential reddening
across the cluster field, the distribution of these binaries on CMDs
would be modified, and the MS-fitting parameters would subsequently change.

\begin{figure}
\epsscale{1.1}
\plotone{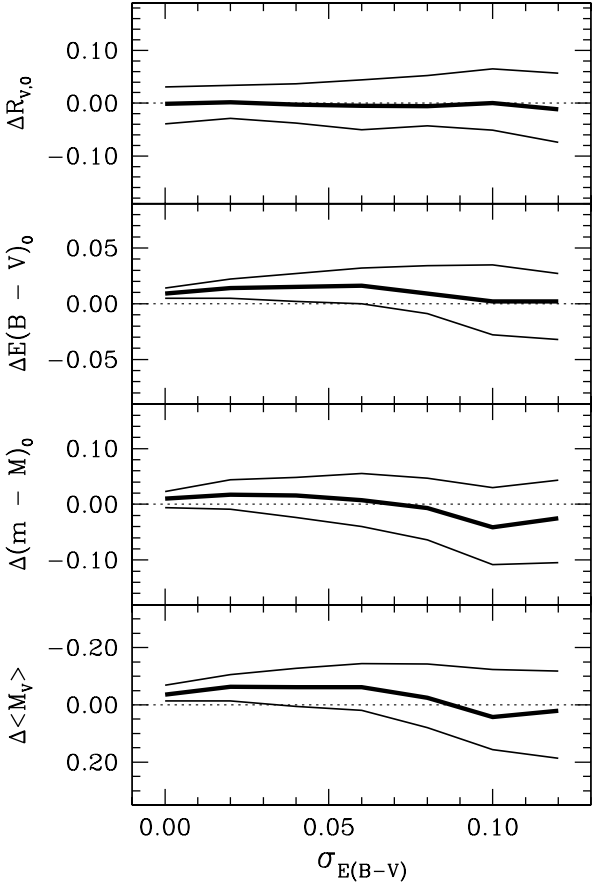}
\caption{Effects of differential reddening from artificial cluster CMD
tests (see text for details).  A bias in each quantity is shown assuming
40\% binary fraction.  The thick solid line is the median of these values
from 200 artificial cluster CMDs as a function of differential reddening,
and thin lines on either side are the first and third quartiles.  The dashed
line indicates a zero bias.  The differences are in the sense that Cepheids
look brighter at $\sigma_{E(B-V)} \approx 0.05$ with an overestimation of
$E(B - V)$.\label{fig:diff}}
\end{figure}

To estimate the effect of differential reddening, we performed artificial
cluster CMD tests as described earlier in \S~\ref{sec:mv}.  We assumed
that $E(B - V)$ has a normal distribution with a standard deviation
$\sigma_{E(B-V)}$.  For each set of artificial cluster CMDs we applied
the photometric filtering and estimated $R_{V,0}$, $E(B - V)_0$, $(m - M)_0$,
and $\langle M_V \rangle$ as for the actual cluster data, assuming $\langle
B_0 \rangle - \langle V_0 \rangle = 0.7$ for Cepheids. Each panel in
Figure~\ref{fig:diff} shows the bias in each of these quantities as
a function of $\sigma_{E(B-V)}$.  The thick solid line shows the median
of the bias from 200 artificial cluster CMDs, and each set of CMDs contained
60 single stars and 40 unresolved binaries in our MS-fitting range for
the distance determination.  Thin lines on either side are the first and
third quartiles of these estimates.  The dashed line indicates a zero bias.

Even in the presence of a small differential reddening, low mass-ratio
binaries cannot be easily detected from the photometric filtering.
Because these unresolved binaries are redder than a single-star MS at
a given $M_V$, the reddening would be overestimated as shown in
Figure~\ref{fig:diff} for $\sigma_{E(B-V)} \la 0.05$.  A higher $E(B - V)$
then results in a longer distance, and a Cepheid would look brighter
by several hundredths of a magnitude in $\langle M_V \rangle$.  As
differential reddening becomes stronger, binaries would be more scattered
around the MS, and it would become further difficult to distinguish
them from highly reddened single stars.  As a result, both cluster
reddening and distance estimates decrease at $\sigma_{E(B-V)} \ga 0.05$,
while semi-interquartile ranges continuously increase.  We also
experimented with 30\% and 50\% binary fractions but found insignificant
changes within the semi-interquartile ranges.

The last column in Table~\ref{tab:art} shows the size of the bias in
$\langle M_V \rangle$ for each Cepheid, which was determined at
$\sigma_{E(B-V)}$ in the second column.  Applying these corrections to
$\langle M_V \rangle$ in Table~\ref{tab:cepheid} leads to $0.05$~mag
reduction in $\langle M_V \rangle$ for 10-day period Cepheids, again
making worse the agreement with the brighter calibration in the previous
studies.  The total $\chi^2$ of the fit becomes smaller ($6.9$ for 6
degrees of freedom) if we add in quadrature the semi-interquartile range
to the error in $\langle M_V \rangle$.

In summary, our Galactic {\it P-L} relation in $V$-band has a statistical
error of 0.07~mag and a systematic error of 0.14~mag in the zero point
(Table~\ref{tab:plerr}).  If we add these quantities in quadrature, the
combined error would be 0.16~mag.  Since we determined the Galactic
{\it P-L} relations in other filter passbands with the same cluster
distance, reddening, and $R_V$ estimates as in $V$, their zero-point
errors are correlated with each other.  Therefore, we took $0.16$~mag
as a systematic error in the Cepheid distance scale in the following section.

\section{Extragalactic Distance Scales}

Cepheids have long served as our most commonly used standard candles
for extragalactic distance studies.  The conventional way of estimating
extragalactic Cepheid distances is to use a fiducial {\it P-L} relation
derived from the LMC Cepheids with an adopted distance to the LMC.
However, geometric distances to a few galaxies have become available
in recent years, providing an opportunity to compare with the Cepheid
distance scale.  In this section we therefore apply our Galactic
{\it P-L} relations to estimate a distance to NGC~4258 (M106) that has
an accurate geometric distance measurement from its water maser sources
\citep{herrnstein99}.  We then infer a distance to the LMC and compare
our result with those adopted by the {\it HST} Key Project and the
SN~Ia calibration program \citep{sandage06b}, from which we consider
a possible increase in their Hubble constant $H_0$ measurements.  We
finish with a discussion of the potentially interesting case of M33
(NGC~598).  Our most fundamental conclusion is that the distance to
NGC~4258 is in good agreement with the geometric measurement when
using the {\it P-L} relations inferred from open clusters.
We also compare our LMC distance with those from the {\it HST}
\citep{benedict07} and the {\it Hipparcos} parallax studies
\citep{vanleeuwen07}.

\subsection{The Maser-host Galaxy NGC~4258}\label{sec:4258}

\citet{herrnstein99} inferred a geometric distance to NGC~4258 from
the orbital motions of water maser sources on its nucleus
and found $(m - M)_0 = 29.29\pm0.09{\rm\ (statistical)}\pm0.12
{\rm\ (systematic)}$ \citep[see also][]{argon07}.  Many Cepheids
were also observed in this galaxy, providing an opportunity to check
the Cepheid distance scale \citep{maoz99,newman01}.  Recently,
\citet{macri06} discovered 281 Cepheids and provided accurate $BVI$
photometry using the Advanced Camera for Surveys/Wide Field Camera
(ACS/WFC) onboard the {\it HST}.  In particular, they observed two
fields in the galaxy, located at two different galactocentric distances
with significantly different gas-phase metal abundances, and derived
the metallicity sensitivity of the Cepheid luminosity.

We reexamined the \citeauthor{macri06} data set in light of our revised
{\it P-L} relationship.  Following their selection procedure, we used
the ``restricted'' sample of 69 Cepheids and estimated distance moduli
for the two groups of Cepheids in the inner (with a period cut of
$P > 12$~days) and outer fields ($P > 6$~days).  For each group of this
sample we corrected for extinction using our {\it P-L} relations in
$BVI_C$.  The absolute distance to NGC~4258 was then estimated by
anchoring our Galactic sample to a reference gas-phase metal abundance
within the galaxy.

\begin{deluxetable}{lccc}
\tabletypesize{\scriptsize}
\tablewidth{2.5in}
\tablecaption{Distance Modulus and Metallicity Sensitivity in NGC4258\label{tab:4258}}
\tablehead{
  \colhead{Filter} &
  \multicolumn{2}{c}{$(m - M)_0$} &
  \colhead{$\gamma$} \nl
  \cline{2-3}
  \colhead{} &
  \colhead{Inner Field} &
  \colhead{Outer Field} &
  \colhead{mag dex$^{-1}$}
}
\startdata
$B$    \dotfill & $29.25\pm0.05$ & $29.33\pm0.07$ & $-0.20$ \nl
$V$    \dotfill & $29.26\pm0.04$ & $29.32\pm0.05$ & $-0.16$ \nl
$I_C$  \dotfill & $29.26\pm0.03$ & $29.32\pm0.04$ & $-0.17$ \nl
Average\dotfill & $29.26\pm0.02$ & $29.32\pm0.03$ & $-0.17$ \nl
\enddata
\tablecomments{Distances are shown at the best-fit reddening from
the Galactic {\it P-L} relations (see text).  Errors represent fitting
uncertainties, and do not include the error from the correlation
with reddening.}
\end{deluxetable}

\begin{figure*}
\epsscale{0.75}
\plotone{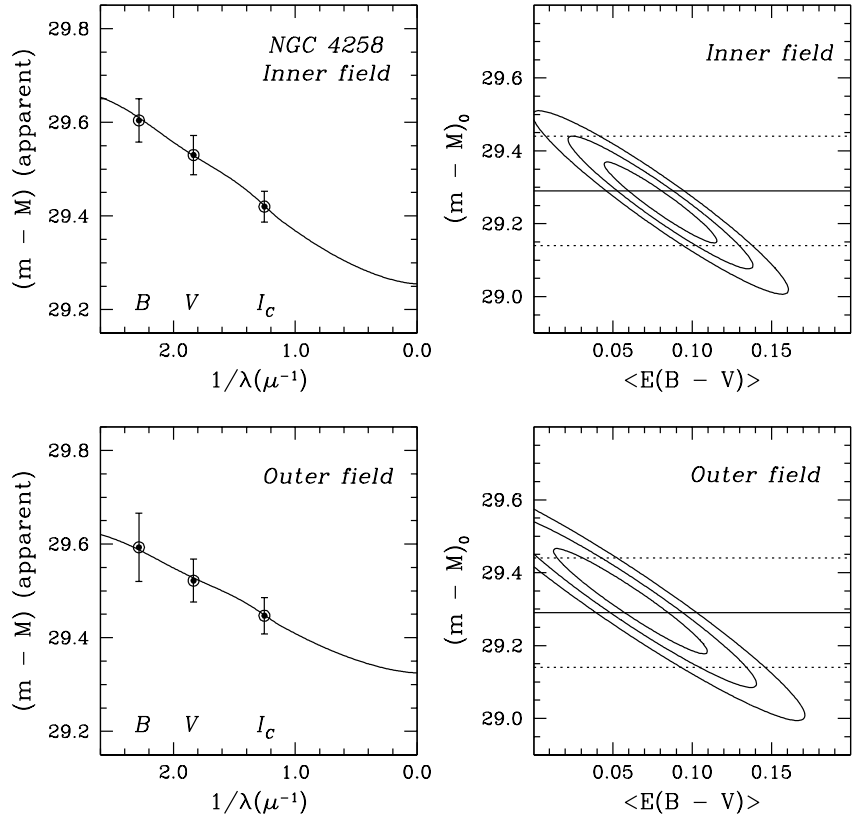}
\caption{{\it Top left}: Apparent distance moduli of inner field Cepheids
in NGC~4258 with the best-fitting CCM89 extinction curve ({\it solid line}).
{\it Top right}: Likelihood contours in the average reddening and distance
modulus of inner field Cepheids shown at $\Delta \chi^2$ = 2.30, 6.17, and
11.8 (68.3\%, 95.4\%, and 99.73\% confidence levels for 2 degrees of
freedom).  The horizontal lines represent a geometric distance from the
water maser sources and its $1\sigma$ error \citep{herrnstein99}.  {\it
Bottom left}: Same as the top left panel, but for outer field Cepheids.
{\it Bottom right}: Same as the top right panel, but for outer field
Cepheids.\label{fig:4258}}
\end{figure*}

Figure~\ref{fig:4258} shows how we determined the average distance
and reddening.  In the left two panels {\it apparent} distance moduli in
$BVI_C$ are shown for Cepheids in the inner and outer fields ({\it top} and
{\it bottom}, respectively).  Distances from shorter wavelength filters are
systematically longer, and the data are well fitted by a CCM89 extinction
curve ({\it solid line}).  Here we assumed $R_V = 3.35$ for the Cepheids,
which was computed from the average $R_{V,0}$ of our cluster sample with
equation~(\ref{eq:rv}) at $\langle B \rangle_0 - \langle V \rangle_0
= 0.72$.  At this value, the CCM89 extinction law yields absorption ratios
$A_{\lambda}/A_V$ of 1.30 and 0.61 for $B$ and $I_C$, respectively,
where we took effective wavelengths of these filters from BCP98.
We added 0.02~mag in quadrature to the fitting error to take into
account the photometric zero-point error.  At the limit of $\lambda^{-1}
\rightarrow 0$, an intercept on the ordinate is the true or unreddened
distance modulus \citep[e.g.,][]{madore91,madore98}.  The right two panels
in Figure~\ref{fig:4258} show likelihood contours in the average reddening
and distance modulus from these fits.  Contours are shown at $\Delta
\chi^2$ = 2.30, 6.17, and 11.8 (68.3\%, 95.4\%, and 99.73\% confidence
levels) for 2 degrees of freedom.  Horizontal lines represent the
distance estimate from the water maser sources and its combined $1\sigma$
error \citep{herrnstein99}.  Distance moduli at the best-fit reddening
are listed in Table~\ref{tab:4258} for each of the two galactic fields.

When the reddening is constrained in this way, distances in both fields
are in good agreement with the maser distance.  However, the distance
from the inner field is shorter than the distance from the outer field
by $\Delta (m - M)_0 = 0.07$.  The most natural interpretation of this
difference is that it measures the metallicity dependence of
the Cepheid luminosity.  This effect has been extensively discussed
in the literature and tested empirically \citep{freedman90,gould94,
sasselov97,kochanek97,kennicutt98,udalski00,kennicutt03,groenewegen04,
sakai04,storm04,romaniello05,macri06}.

To estimate the true distance modulus of the galaxy, we first estimated
metal abundances in the two NGC~4258 fields and then inferred a distance
at the metallicity of the Galactic Cepheids.  Stellar abundances cannot
be measured directly for most of the Cepheid galaxies.  Instead, the
{\it HST} Key Project adopted ``empirical'' gas-phase [O/H] abundances
from \citet{zaritsky94}.  \citeauthor{macri06} computed oxygen
abundances at each Cepheid's location from the \ion{H}{2} region
abundance gradient in \citeauthor{zaritsky94}.  The average abundances
from this table are $\langle 12 + \log{\rm (O/H)} \rangle = 8.94$ and
$8.54$~dex for the inner and outer field Cepheids, respectively, when
the LMC Cepheids have $12 + \log{\rm (O/H)} = 8.50$ on this scale
\citep{sakai04}.  To relate these abundances to those of
our Galactic Cepheid sample, we adopted a stellar abundance difference
between the LMC Cepheids and Galactic supergiants from \citet{luck98} of
$\Delta{\rm \langle [Fe/H] \rangle (Galaxy - LMC)} \approx 0.30$.  They
also found a similar $\alpha$-element enhancement between the LMC and
the Galactic samples.  In the following discussion we adopted $\Delta
\langle \log{\rm (O/H)} \rangle {\rm (Galaxy - LMC)} = 0.30\pm0.06$,
assuming that the stellar abundance difference is the same as the
difference in the gas-phase oxygen abundance.

Since the metallicity of the Galactic Cepheids is found between the metal
abundances for the inner and outer fields, the true distance would lie
between distance estimates from the two galactic fields.  From a linear interpolation,
we obtained $(m - M)_0 = 29.28\pm0.10\pm0.16$ ({\it P-L} zero point) for
a distance to NGC~4258, which is in good agreement with the current maser
distance.  The first error is from the fit to the {\it HST} ACS data
($0.10$~mag), an error in $R_V$ ($\mp0.03$~mag for $\pm0.5$ change in
$R_V$), and an error in the metallicity scale ($\mp0.02$~mag for
$\pm0.10$~dex).  We adopted an error in the metallicity scale of
$\pm0.10$~dex to account for a possible difference in the stellar and
gas-phase abundances, but it has only a moderate effect on the distance
determination.

Previous studies found somewhat longer Cepheid distances to NGC~4258
although these values are consistent within the errors with the maser
solution.  As a part of the {\it HST} Key Project, \citet{maoz99}
estimated $(m - M)_0 = 29.54\pm0.12\pm0.18$ using the Wide-Field
Planetary Camera 2 (WFPC2), and \citet{newman01} revised it
to be $(m - M)_0 = 29.47\pm0.09\pm0.15$, where the first errors represent
those unique to each determination and the second errors are systematic
uncertainties in the {\it HST} Key Project distances.

However, the reddening in the {\it HST} Key Project is significantly larger
than our estimates.  The final result in the {\it HST} Key Project lists
$\langle E(V - I)_C \rangle = 0.22\pm0.04$ or $\langle E(B - V) \rangle
= 0.17\pm0.03$ for Cepheids
located in the middle of galactocentric distances to the two ACS fields.
On the other hand, we found $\langle E(B - V) \rangle = 0.08\pm0.04$
and $0.06\pm0.05$ for the inner- and outer-field Cepheids, respectively,
where the errors are the $1\sigma$ contour sizes in Figure~\ref{fig:4258}.
\citeauthor{macri06} also estimated individual Cepheid reddenings using
the OGLE-II LMC {\it P-L} relations, and their average values are also
larger than ours, being $\langle E(B - V) \rangle = 0.18$ and $0.13$~mag
for the inner and outer fields, respectively.

In Table~\ref{tab:4258} we also listed the metallicity sensitivity
($\gamma$) in each band, which was computed from the distances and the average
abundance difference in the two galactic fields.  The magnitude of this
effect is smaller than the original estimate in \citeauthor{macri06},
$\gamma=-0.29\pm0.09\ {\rm (random)}\pm0.05\ {\rm (systematic)}$ mag dex$^{-1}$.
Since our estimate neglects the change in color with metallicity,
we employed the metallicity corrections from \citet{kochanek97} in
the next two sections.  \citeauthor{kochanek97} derived corrections in
$UBVRIJHK$ based on a simultaneous analysis of Cepheids in 17 galaxies
with a proper treatment of correlations between extinction, temperature,
and metallicity.  The corrections are $+0.20\pm0.18$, $-0.08\pm0.14$,
$-0.21\pm0.14$, $-0.26\pm0.13$, $-0.34\pm0.14$, and $-0.40\pm0.13$
mag dex$^{-1}$ in $BVI_CJHK$, respectively, in the sense that
metal-rich Cepheids have a decreased flux in $B$ but increased fluxes
in $VI_CJHK$, possibly due to line-blanketing and back-warming among
other effects. These estimates are based on $\langle E(B - V) \rangle
= 0.15$ for LMC Cepheids, but their dependence on the LMC reddening
is small.  If we adopt these corrections for NGC~4258, we would derive
$(m - M)_0 = 29.34\pm0.13$ for the inner-field and $29.17\pm0.19$~mag
for the outer-field Cepheids, where the errors represent the $1\sigma$
statistical uncertainty.

\subsection{LMC}\label{sec:lmc}

Our Galactic {\it P-L} relations can also be applied to estimate the
distance to the LMC and its mean reddening.  We included $JHK_s$
photometry of the LMC Cepheids from \citet{persson04}, as well as $BVI_C$
photometry from the OGLE-II catalog in the multi-wavelength fitting
technique.  This approach assumes that Cepheids in these two surveys
were drawn from the same population.\footnote{The OGLE-II Cepheids were
observed along the LMC bar, while the Cepheids observed in $JHK_s$
\citep{persson04} are more dispersed over the LMC.}

\begin{figure*}
\epsscale{1.15}
\plotone{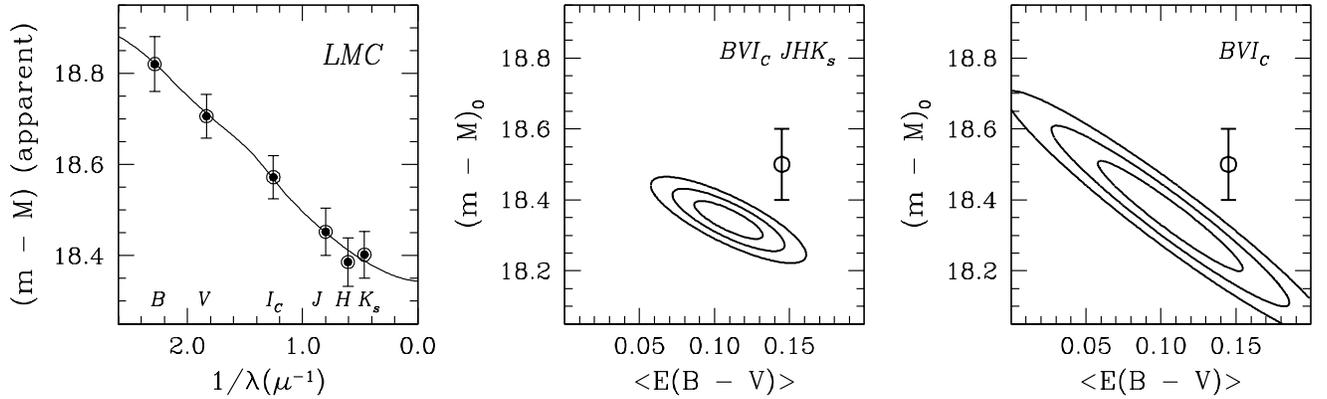}
\caption{{\it Left}: Apparent LMC distance moduli in $BVI_CJHK_s$ with
the best-fitting CCM89 extinction curve ({\it solid line}).  {\it Middle}:
Likelihood contours in the average LMC reddening and its distance modulus
in $BVI_CJHK_s$ shown at $\Delta \chi^2$ = 2.30, 6.17, and 11.8 (68.3\%,
95.4\%, and 99.73\% confidence levels for 2 degrees of freedom).  The
error bar represents the adopted distance and its error by {\it HST} Key
Project at the average reddening from the OGLE-II reddening map.  {\it
Right}: Same as the left panel, but likelihood contours when $BVI_C$
photometry is used only.\label{fig:lmc}}
\end{figure*}

Figure~\ref{fig:lmc} shows likelihood contours in the solution for
the average LMC reddening and its distance modulus ({\it middle panel})
from the fits on the apparent distance moduli in $BVI_CJHK_s$ ({\it left
panel}). We applied metallicity corrections for $\Delta \langle \log
{\rm (O/H) \rangle (Galaxy - LMC)} = 0.30$ and assumed that the
metallicity sensitivities on the Glass-Carter $JHK$ system in \citet{kochanek97}
are the same as in the LCO $JHK_s$ system.  As in the NGC~4258 case, we
assumed $R_V = 3.35$ for the Cepheids, which yields absorption ratios
$A_{\lambda}/A_V$ of 0.29, 0.18, and 0.12 for $JHK_s$, respectively,
from the CCM89 extinction law.  In this calculation we took the effective
filter wavelengths from \citet{persson04} for $JHK_s$ on the LCO system.
We added $0.02$~mag in $BVI_C$ \citep{udalski99a} and $0.03$~mag in
$JHK_s$ \citep{persson04} in quadrature to the fitting errors to account
for photometric zero-point errors.

\begin{deluxetable}{ccc}
\tabletypesize{\scriptsize}
\tablewidth{2.5in}
\tablecaption{Distance Modulus of the LMC\label{tab:lmc}}
\tablehead{
  \colhead{Filter} &
  \multicolumn{2}{c}{$(m - M)_0$} \nl
  \cline{2-3}
  \colhead{} &
  \colhead{OGLE-II $E(B - V)$\tablenotemark{a}} &
  \colhead{Best-fit $E(B - V)$\tablenotemark{b}}
}
\startdata
$B$    \dotfill & $18.19\pm0.06$ & $18.35\pm0.06$ \nl
$V$    \dotfill & $18.23\pm0.05$ & $18.34\pm0.05$ \nl
$I_C$  \dotfill & $18.29\pm0.05$ & $18.35\pm0.05$ \nl
$J$    \dotfill & \nodata          & $18.35\pm0.05$ \nl
$H$    \dotfill & \nodata          & $18.32\pm0.05$ \nl
$K_s$  \dotfill & \nodata          & $18.36\pm0.05$ \nl
Average\dotfill & $18.24\pm0.03$ & $18.34\pm0.02$ \nl
\enddata
\tablecomments{After applying metallicity corrections from \citet{kochanek97}.}
\tablenotetext{a}{Distances derived assuming the OGLE-II reddening map.}
\tablenotetext{b}{Distances shown at the best-fit reddening from the Galactic
{\it P-L} relations (see text).  Errors represent fitting uncertainties, and
do not include the error from the correlation with reddening.}
\end{deluxetable}

As shown in Figure~\ref{fig:lmc}, our best-fit solution yields
$(m - M)_0 = 18.34\pm0.06\pm0.16$ ({\it P-L} zero point).
Extinction-corrected distance moduli at the best-fit reddening are
listed in Table~\ref{tab:lmc}.  The first error in distance is a
quadrature sum of a fitting error ($\pm0.05$~mag) and an error from
$R_V$ ($\mp0.01$~mag for $\pm0.5$ in $R_V$).  We also added in
quadrature an error from the metallicity ($\mp0.03$~mag for
$\pm0.06$~dex).  As in the NGC~4258 case the zero-point error
of the Galactic {\it P-L} relations dominates the combined error
in the LMC distance modulus.  Without metallicity corrections, we would
derive $(m - M)_0 = 18.48$.

We also derived $\langle E(B - V) \rangle = 0.11\pm0.02$, where the
error is from the $1\sigma$ contour in Figure~\ref{fig:lmc}.  Our
average reddening is 0.03~mag smaller than the average reddening from
the OGLE-II, $\langle E(B - V) \rangle = 0.143$, which was adopted in
the {\it HST} Key Project.\footnote{The {\it HST} Key Project adopted
the OGLE-II LMC {\it P-L} relations and scaled them to $(m - M)_0 = 18.50$.
However, the {\it P-L} relations were not scaled to $\langle E(B - V)
\rangle = 0.10$ as they claimed \citep[see also][]{fouque03}.}  However,
our reddening estimate is closer to the value of $E(B - V) \sim 0.10$
\citep{walker99} from several different reddening estimates for LMC
stars \citep[see Table~II in][]{mcnamara80}.  It is also in agreement
with the reddening value determined for the LMC Cepheids, $\langle
E(B - V) \rangle = 0.076$ \citep[]{caldwell85,laney07}.  The zero point
of the OGLE-II reddening map was set by two LMC clusters (NGC~1850 and
NGC~1835) and one eclipsing binary (HV~2274), but the reddening for
HV~2274 has different values from various studies:  OGLE-II adopted
$E(B - V) = 0.149\pm0.015$ \citep{udalski98}, which is larger than
$E(B - V) = 0.120\pm0.009$ \citep{guinan98}, $0.088\pm0.023$
\citep{nelson00}, and $0.103\pm0.007$ \citep{groenewegen01}.
Furthermore, as shown in Table~\ref{tab:lmc}, extinction corrections
from the OGLE-II reddening map result in systematically shorter
distances at shorter wavelengths, which may indicate an overcorrection
for extinction \citep[see also][]{fouque03}.

Our distance modulus is $0.16$~mag smaller (or $7\%$ in distance) than
the adopted distance modulus by the {\it HST} Key Project ($18.50$~mag).
Formally, this implies an increase in their $H_0$ by $7\%\pm8\%$.  Our
LMC distance is also shorter by $0.20$~mag (or $9\%$ in distance) than
the distance modulus adopted by the \citet{sandage06b} program ($18.54$~mag).
In addition to our study, recent parallax studies from the {\it HST}
\citep{benedict07} and the revised {\it Hipparcos} catalog \citep{vanleeuwen07}
used the Wesenheit index $W_{VI}$ to derive $(m - M)_0 = 18.40\pm0.05$
and $18.39\pm0.05$~mag, respectively, also suggesting a downward revision
of the LMC distance.  Finally, \citet{macri06} estimated
a distance modulus of NGC~4258 relative to the LMC of $10.88\pm0.04
{\rm\ (statistical)}\pm0.05{\rm\ (systematic)}$.  Given the maser
distance, this was then translated into the LMC distance modulus of
$(m - M)_0 = 18.41\pm0.10{\rm\ (statistical)} \pm0.13{\rm\ (systematic)}$,
which is within $1\sigma$ of our estimate.

We note that stronger constraints on distance can be achieved when several
photometric bands are used including near-infrared data.  As illustrated
in Figure~\ref{fig:lmc}, the statistical error in distance becomes
larger by a factor of 3 when only $BVI_C$ information is used
({\it right panel}) rather than from $BVI_CJHK_s$ ({\it middle panel}).
The error would be even larger when the distance and reddening values
are estimated from only two bands, such as the Wesenheit index.

\subsection{M33}

As our final case, we estimated the average distance and reddening
for M33 Cepheids using the same method as in the previous section.
The distance to this nearby spiral galaxy has recently been measured
from a detached eclipsing binary, $(m - M)_0 = 24.92\pm0.12$
\citep{bonanos06}, and from two water maser sources in \ion{H}{2}
regions, $(m - M)_0 = 24.32^{+0.45}_{-0.57}$ \citep{brunthaler05}.

\begin{figure*}
\epsscale{0.75}
\plotone{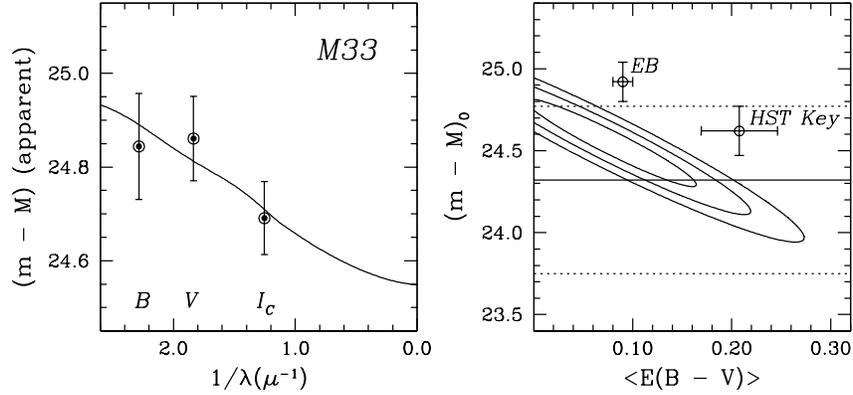}
\caption{{\it Left:} Apparent distance moduli in $BVI_C$ with the
best-fitting CCM89 extinction curve ({\it solid line}).  {\it Right:}
Likelihood contours in the average reddening and distance modulus of M33
Cepheids shown at $\Delta \chi^2$ = 2.30, 6.17, and 11.8 (68.3\%, 95.4\%,
and 99.73\% confidence levels for 2 degrees of freedom).  Two open
circles represent the distance and reddening estimates from an eclipsing
binary \citep{bonanos06} and those from the {\it HST} Key Project.
The horizontal lines represent a distance from the water maser sources
and its $1\sigma$ error \citep{brunthaler05}.\label{fig:m33}}
\end{figure*}

Figure~\ref{fig:m33} shows likelihood contours in the solution for
the average reddening and distance modulus.  As in the {\it HST} Key
Project, we used $BVI_C$ photometry in \citet{freedman91} for their
10 best observed Cepheids, which was obtained at the 3.6 m
Canada-France-Hawaii Telescope (CFHT), the Kitt Peak National
Observatory (KPNO) 4 m telescope, and the Palomar 1.5 m telescope.
We assumed $0.03$~mag for the photometric zero-point error in each
band.  On the \citet{zaritsky94} scale, M33 Cepheids have $\langle 12 +
\log{\rm (O/H)} \rangle = 8.82$, which was adopted in the {\it HST}
Key Project.  We therefore applied metallicity corrections from
\citet{kochanek97} for $\Delta {\rm \langle \log (O/H) \rangle
(Galaxy - M33}) = -0.02$.

At the best-fit reddening, we obtained an extinction-corrected distance
modulus of $24.51\pm0.11$, $24.60\pm0.09$, and $24.53\pm0.08$ in
$BVI_C$, respectively, where the uncertainties are fitting errors only.
The average distance is $(m - M)_0 = 24.55\pm0.28\pm0.16$
({\it P-L} zero point).  The first error was determined by adding in
quadrature the fitting error (0.28~mag), the error from $R_V$
($\mp0.03$~mag for $\Delta R_V = \pm0.5$), and the error from the
metallicity ($\mp0.08$~mag for $\pm0.15$~dex).  We adopted the error in
the metallicity from the {\it HST} Key Project.  If we adopt metal abundances
derived from electron temperatures \citep{sakai04}, the abundance
difference between the LMC and M33 becomes $\Delta \langle {\rm \log
(O/H) \rangle} = +0.21$, which leads to $\Delta {\rm \langle \log (O/H)
\rangle (Galaxy - M33)} = +0.09$.  We would derive $(m - M)_0 = 24.53$
without metallicity corrections.

The {\rm HST} Key Project used the same ground-based photometry for
this galaxy and derived a metallicity-corrected distance modulus of
$(m - M)_0 = 24.62\pm0.15$ based only on $VI$ as for the other
{\it HST} sample galaxies.  While this distance is in good
agreement with our estimate, our average reddening of $\langle E(B - V)
\rangle = 0.08\pm0.09$ is lower than their estimate,
$\langle E(V - I) \rangle = 0.27\pm0.05$ or $\langle E(B - V) \rangle
= 0.21\pm0.04$.\footnote{We note that \citet{freedman91} originally
derived $\langle E(B - V) \rangle = 0.10\pm0.09$ and $(m - M)_0 =
24.64\pm0.09$ using $BVRI$ photometry, assuming the LMC distance modulus
of 18.5~mag and its mean reddening of 0.10~mag.}  A part of the
difference is due to different zero points of {\it P-L} relations.
However, our reddening is in good agreement with the value of the
eclipsing binary, $E(B - V) = 0.09\pm0.01$, which was obtained from
$BVRJHK_s$ photometry \citep{bonanos06} and with previous estimates
based on several different methods (see Table~7 of \citeauthor{bonanos06}).

A similar result was obtained from a larger Cepheid sample in
\citet{macri01a}.  They provided $BVI_C$ photometry for 251 Cepheids
obtained at the Fred Lawrence Whipple Observatory (FLWO) 1.2 m telescope
and the MDM 1.3 m telescope \citep[see also][]{mochejska01a,mochejska01b}.
From the restricted sample of 90 Cepheids in $1.1 \leq \log{P} \leq
1.7$ similar to those of the \citeauthor{freedman91} sample, we derived
$(m - M)_0 = 24.67\pm0.18$ and $\langle E(B - V) \rangle = 0.04\pm0.06$,
where the uncertainties are fitting errors only.

Distances from these two independent sets of photometry are about
$0.3$~mag shorter than the distance from the eclipsing binary, although
our distance has a large error.  One potential explanation for the
short Cepheid-based distance is the unresolved blends by a strong
star-to-star correlation function.  \citet{mochejska01c} investigated
the influence of blending by comparing {\it HST} WFPC2 and
\citeauthor{macri01a} ground-based images for a sample of 102 Cepheids
\citep[see also][]{mochejska00}.  They found that the average flux
contribution from unresolved luminous companions on the ground-based
images could be on average $\sim20\%$ in $BVI$ for $\log{P}>1$, which
would systematically underestimate the Cepheid distance by $\sim10\%$.
In addition, \citet{lee02} obtained single-epoch $I$-band photometry
for a subset of the \citeauthor{macri01a} sample using {\it HST} WFPC2
and found that the {\it HST} photometry is on average $\approx0.2$~mag
fainter than \citeauthor{macri01a} photometry.

Follow-up observations with a higher spatial resolution
\citep[e.g.,][]{macri04} would be helpful for a definitive test of the
distance to M33, although there are indications that our estimated reddening
is more accurate than the value in the {\it HST} Key Project.  Note
that the effect of stellar blending is less significant in our distance
estimates for NGC~4258 and the LMC.  The ground-based photometry with
the full width at half-maximum (FWHM) of $\sim1.5\arcsec$ would
correspond to $\sim0.3$~pc at the distance of the LMC, compared to
$\sim7$~pc at the distance of M33.  The ACS/WFC FWHM of 0.09\arcsec
\citep{ford03} corresponds to $\sim3$~pc at the distance of NGC~4258.

\section{SUMMARY AND DISCUSSION}

We have continued our effort to improve the accuracy of isochrones and
distances derived from MS fitting.  We extended the previous Hyades-based
calibration to the upper MS by constructing empirical color-$T_{\rm eff}$
corrections to match the observed Pleiades MS at the cluster's accurately
known distance.  We applied these empirically calibrated sets of isochrones
to Cepheid-bearing Galactic open clusters to derive distances, reddenings,
and $R_V$ at the spectroscopic metal abundances and obtained Galactic
{\it P-L} relations based on nine Cepheids in seven clusters.  Our distance
modulus of individual Cepheids has an accuracy of $\sim0.08$~mag (or
$4\%$ in distance), which is compatible to those of recent parallaxes
from the {\it HST} \citep{benedict07} and the revised {\it Hipparcos}
catalog \citep{vanleeuwen07}.

Using these {\it P-L} relations, we derived a distance to NGC~4258 and
found that our Cepheid distance is in excellent agreement with the maser
distance, supporting our distance scale from the MS-fitting technique.
From $BVI_CJHK_s$ photometry we derived the LMC
distance modulus of $(m - M)_0 = 18.34\pm0.06\pm0.16$ ({\it P-L} zero
point) after applying metallicity corrections.  This is also in close
agreement with distance estimates from the recent {\it HST} and {\it
Hipparcos} parallax studies.  However, our revised LMC distance is lower
than the distance adopted by the {\it HST} Key Project, which formally
implies an increase in their $H_0$ by $7\%\pm8\%$.  A similar size of an
increase in $H_0$ is expected from the SN~Ia calibration program
by \citet{sandage06b}.

Our reddening estimates are systematically lower than those in the
{\it HST} Key Project.  Part of the reason is the {\it P-L} relations
with different assumptions about the reddening.  They adopted the color
excess for LMC Cepheids from the OGLE-II reddening map, but its average
value is $\Delta \langle E(B - V) \rangle \approx 0.03$ larger
than our LMC reddening estimate.  Nonetheless, relative distances
between the LMC and target galaxies in the {\it HST} Key Project would
remain less affected because the reddening-free or Wesenheit index
adopted in the {\it HST} Key Project is designed to avoid
the problem of knowing the absolute zero point of the reddening scale.
In other words, the absolute reddening value would not be important
as long as the difference in reddening between target Cepheids and
the calibrating system is well-defined.  Therefore, the Wesenheit index
can be effectively used to build the cosmic distance scale once highly
accurate distances to calibrating Cepheids are available.

However, our $E(B - V)$ estimates for NGC~4258 and M33 are larger by
$\langle E(B - V) \rangle \sim0.1$ than those in the {\it HST} Key
Project, more than expected from the difference in the LMC reddening.
This is due to two reasons.  First, the Cepheid's color is not only
affected by reddening, but it can be also affected by the metal content.
The {\it HST} Key Project corrected a distance modulus for a metallicity
effect after deriving a color excess.  On the other hand, we applied
metallicity corrections to apparent distance moduli from
\citet{kochanek97}, and then derived a reddening value.
Because the sign and size of the metallicity corrections could not be
the same in different broadband filters, the difference in the Cepheid
metallicity could lead to different reddening estimates from those in
the {\it HST} Key Project.  Second, the Wesenheit magnitude is
based on only two broadband filters.  However, reddening and distance
estimates are naturally correlated in the standard procedure to estimate
these parameters using {\it P-L} relations, so a small zero-point error
in the photometry, for example, can be translated into a large error in
reddening.  Therefore, obtaining Cepheid photometry at least in three
or more passbands including near-infrared filters is of great importance
in the distance and reddening estimation \citep[e.g.,][]{gieren06,soszynski06}.

In the determination of individual Cepheids' absolute magnitudes, the extinction
is one of the largest sources of error because of strong differential
reddening in Cepheid-bearing young open clusters.  Although we found
an error of $\sim0.03$~mag in the mean $E(B - V)$ for a cluster, most
Cepheid-bearing clusters have patchy reddening that makes the
individual Cepheid reddening uncertain by up to $\sigma_{E(B-V)} \sim
0.12$.  An average reddening of cluster members located near a Cepheid
was often used to determine the Cepheid's reddening, but differential
reddening over a small scale could not be resolved unless we have a sufficient
number of nearby stars with well-determined reddening.  However, it is
noted that the most secure reddening estimates for Cepheids are considered
to be given by the cluster Cepheids.  In addition, our LMC reddening is
in close agreement with the previous standard value, $\langle E(B - V) \rangle
\sim0.10$ \citep{walker99}, and our average reddening for M33 Cepheids
is in good agreement with the value inferred from $BVRJHK_s$ photometry
of an eclipsing binary \citep{bonanos06}.

In terms of the zero point of the Galactic {\it P-L} relations, the
uncertainty in the metallicity scale of the Galactic open clusters is
the largest error source, with a change of $\Delta {\rm [M/H]}\sim0.1$~dex
producing a $\sim0.1$~mag shift in distance modulus.  No other source
is unusually large.  In Paper~III we have shown that several color
indices can be effectively used in MS fitting to determine a cluster
metallicity because of differential sensitivities of these colors on
metal abundance.  Although the noise in the current photometry prevented
the estimation of the photometric metal abundance, deep multi-color
photometry combined with near-infrared data will be useful to determine
$R_V$, $E(B - V)$, $(m - M)_0$, and [M/H], simultaneously.  A more
accurate metallicity scale could then be used to reduce the size of the systematic
error in the {\it P-L} zero point.  In addition, the zero-point error
can be further reduced by deeper photometry of other Cepheid-bearing clusters,
which were not included in this paper because of poor photometric quality.
These are C1814-191, Collinder~394, Mon~OB2, NGC~1647, NGC~6649,
NGC~6664, Platais~1, Trumpler~35, Vel~OB5, and Vul~OB1.

\acknowledgements
We are very grateful to Krzysztof Stanek for his valuable comments
and suggestions.  We gratefully acknowledge Michael Bessell for
answering a number of questions about his computations of
color-excess ratios, Fiona Hoyle for providing her photometry,
and Imants Platais for pointing out some problematic features of
the isochrones, which has led us to investigate the LCB color correction
tables more carefully.  We also would like to thank Andrew Gould,
Lucas Macri, Fritz Benedict, Christopher Kochanek, and Jennifer Johnson
for useful discussions.  We thank Richard Pogge, Subo Dong, and
Kristen Sellgren for their help.
The work reported here was partially supported by grants AST-0206008
and AST-0205789 from the National Science Foundation to the Ohio
State Research Foundation, and with funds provided by the Ohio State
University Department of Astronomy.
This publication makes use of data products from the Two Micron All Sky
Survey, which is a joint project of the University of Massachusetts
and the Infrared Processing and Analysis Center/California Institute of
Technology, funded by the National Aeronautics and Space Administration
and the National Science Foundation.

\end{document}